\documentclass[conference,compsoc]{IEEEtran}

\pdfoutput=1

\ifCLASSOPTIONcompsoc
  \usepackage[nocompress]{cite}
\else
  \usepackage{cite}
\fi

\ifCLASSINFOpdf
\else
\fi
\usepackage{amsmath}
\usepackage{amssymb}

\usepackage{textcomp}
\usepackage{wasysym}
\usepackage{amsthm}
\usepackage{multirow}
\usepackage{balance}
\usepackage{tabularx}
\usepackage{booktabs}
\usepackage{algpseudocode}
\usepackage{hyperref}
\usepackage{array}
\usepackage{boldline}
\usepackage{makecell}
\usepackage{pifont}
\usepackage{threeparttable}
\usepackage{graphicx}
\usepackage{subcaption}
\usepackage{svg}
\usepackage{amsthm}
\usepackage{diagbox}
\usepackage{booktabs}
\usepackage{lipsum}
\usepackage{caption}
\usepackage{xcolor}
\usepackage[ruled,vlined]{algorithm2e}
\usepackage{xspace,verbatim,cleveref,colortbl}
\usepackage[inline]{enumitem}
\usepackage{url}

\PassOptionsToPackage{hyphens}{url}
\usepackage[hyphens]{url}

\newcommand{\ComMark}{\textsc{DeMark}\xspace}
\newcommand{\ICS}{\text{ICS}\xspace}
\newcommand{\DistortionAttack}{Distortion\xspace}
\newcommand{\RegenVAE}{\text{RegenVAE}\xspace}
\newcommand{\RegenDiffusion}{\text{RegenDM}\xspace}
\newcommand{\MBRS}{\text{MBRS}\xspace}
\newcommand{\CIN}{\text{CIN}\xspace}
\newcommand{\TrustMark}{\text{TrustMark}\xspace}
\newcommand{\PIMoG}{\text{PIMoG}\xspace}
\newcommand{\VINEB}{\text{VINE-B}\xspace}
\newcommand{\VINER}{\text{VINE-R}\xspace}
\newcommand{\StableSignature}{\text{SS}\xspace}
\newcommand{\PTW}{\text{PTW}\xspace}
\newcommand{\SparseEncoder}{\(\mathcal{T}_\text{CNN}\)\xspace}
\newcommand{\Reconstructor}{\(\mathcal{R}_\text{CNN}\)\xspace}
\newcommand{\BitAcc}{BitAcc\xspace}
\newcommand{\DetectAcc}{DetectAcc\xspace}
\newcommand{\OpenImage}{OpenImage\xspace}
\newcommand{\COCO}{COCO\xspace}
\newcommand{\PSNR}{PSNR\xspace}
\newcommand{\SSIM}{SSIM\xspace}
\newcommand{\FID}{FID\xspace}
\newcommand{\ESRGAN}{ESRGAN\xspace}
\newcommand{\ISR}{ISR\xspace}
\newcommand{\AT}{AT\xspace}
\newcommand{\SW}{SW\xspace}
\newcommand{\LPIPS}{LPIPS\xspace}

\definecolor{mygreen}{RGB}{0,130,0}
\definecolor{myred}{RGB}{170,0,0}
\newcommand{\cmark}{\textcolor{mygreen}{\ding{51}}}
\newcommand{\xmark}{\textcolor{myred}{\ding{55}}}

\newcommand{\appref}[1]{Appendix~\ref{#1}}

\begin{document}

\title{{\ComMark: A Query-Free Black-Box Attack on Deepfake Watermarking Defenses}}

\author{%
\parbox{\textwidth}{\centering
\begin{tabular}{ccc}
\parbox[t]{0.30\textwidth}{\centering
Wei Song\\
\textit{UNSW Sydney}\\
\textit{wei.song1@unsw.edu.au}}
&
\parbox[t]{0.30\textwidth}{\centering
Zhenchang Xing\\
\textit{CSIRO's Data61}\\
\textit{zhenchang.xing@data61.csiro.au}}
&
\parbox[t]{0.30\textwidth}{\centering
Liming Zhu\\
\textit{CSIRO's Data61}\\
\textit{liming.zhu@data61.csiro.au}}
\end{tabular}

\vspace{0.8em}

\begin{tabular}{cc}
\parbox[t]{0.30\textwidth}{\centering
Yulei Sui\\
\textit{UNSW Sydney}\\
\textit{y.sui@unsw.edu.au}}
&
\parbox[t]{0.30\textwidth}{\centering
Jingling Xue\\
\textit{UNSW Sydney}\\
\textit{j.xue@unsw.edu.au}}
\end{tabular}
}%
}

\maketitle


\begin{abstract}
The rapid proliferation of realistic deepfakes has raised urgent concerns over their misuse, motivating the use of defensive watermarks in synthetic images for reliable detection and provenance tracking. However, this defense paradigm assumes such watermarks are inherently resistant to removal. 
We challenge this assumption with \textbf{\ComMark}, a query-free black-box attack framework that targets defensive image watermarking schemes for deepfakes. 
\ComMark exploits latent-space vulnerabilities in encoder–decoder watermarking models through a compressive-sensing-based sparsification process, suppressing watermark signals while preserving perceptual and structural realism appropriate for deepfakes.
Across eight state-of-the-art watermarking schemes, \ComMark reduces watermark detection accuracy from 100\% to 32.9\% on average while maintaining natural visual quality, outperforming existing attacks. 
We further evaluate three defense strategies—image super-resolution, sparse watermarking, and adversarial training—and find them largely ineffective. 
These results demonstrate that current encoder–decoder watermarking schemes remain vulnerable to latent-space manipulations, underscoring the need for more robust watermarking methods to safeguard against deepfakes.
\end{abstract}


%
\IEEEpeerreviewmaketitle

\section{Introduction}
\label{sec:introduction}

Generative AI (GenAI) models such as Stable Diffusion~\cite{LDM} and ChatGPT~\cite{GPT} have revolutionized media creation, driving advances in art, design, and productivity~\cite{art_creation, text_generation}. 
However, these same technologies have also accelerated the spread of deepfakes~\cite{deepfake-cvpr, copyright-mm, copyright-cvpr}. 
Their growing realism makes them difficult for both humans and automated systems to distinguish from authentic content, posing serious security and societal risks.

To mitigate the growing threat of deepfakes, defensive watermarking has emerged as a practical solution for attribution and detection~\cite{sok_aigc, unmark, synID, OpenAI-WM, DeepMind-WM}. 
These schemes embed invisible signals into generated images, allowing dedicated detectors to later verify their authenticity~\cite{sok_aigc, unmark}. 
Defensive watermarking has gained substantial traction in both research and industry~\cite{unmark, synID, OpenAI-WM, DeepMind-WM}: Google has deployed SynthID~\cite{synID}, while Microsoft~\cite{Microsoft-WM} and OpenAI~\cite{OpenAI-WM} have announced plans to integrate similar mechanisms into their GenAI products.

Most existing defensive image watermarking methods~\cite{MBRS, CIN, TrustMark, PIMoG, VINE, StableSignature, PTW} focus on robustness against distortions~\cite{WAVES} and regeneration attacks~\cite{WatermarkAttacker, Regeneration-Attack-Diff1, Regeneration-Attack-Diff2}. 
Although adversarial perturbation–based attacks~\cite{WM-adversarial-CCS, WM-adversarial-ICLR, WM-adversarial-TCSVT} can be effective, they generally require access to model parameters or extensive querying, limiting their practicality in real-world scenarios~\cite{unmark, WatermarkAttacker, meta-says}.

Despite significant engineering progress, defensive image watermarking schemes remain poorly understood under advanced adversaries. 
Modern watermark-based deepfake defenses~\cite{MBRS, CIN, TrustMark, PIMoG, VINE, StableSignature, PTW} typically rely on encoder--decoder architectures that embed watermark signals in latent representations, exposing structural weaknesses. 
We exploit these for the first time through \textbf{\ComMark}, a query-free black-box attack that leverages image compressive sensing theory~\cite{ICS-CVPR, ICS-TIP, ICS-TPAMI} within a tailored neural architecture to weaken—and often remove—embedded watermarks while preserving visual fidelity.

Our key observation is that these latent representations act as carriers for watermark attributes. 
During encoding, watermarks are fused with input images or prompts so that the decoded outputs covertly preserve the embedded signals~\cite{WAVES, unmark, PTW}. 
During verification, detectors re-encode images into latent space to extract and validate the watermarks. 
This reliance on latent attributes creates a critical attack surface: carefully targeted manipulations in latent space can suppress watermark evidence across images.

\ComMark constructs its attack on the foundation of image compressive sensing (\ICS) theory~\cite{ICS-CVPR, ICS-TIP, ICS-TPAMI}. 
It enforces sparsity in latent space to retain only essential components while eliminating redundancies through a \textit{sparse encoding} module with a sparsity-inducing loss that drives most elements toward zero. 
This induces a dispersal effect characterized by Sparsity Change (SC)~\cite{CS-Sparsity-Change-TIP19}, Intensity Redistribution (IR)~\cite{Intensity-Distribution-TIP17}, and Positional Redistribution (PR)~\cite{Positional-Redistribution-CVPR22}, which collectively weaken watermark signals by reducing their density (SC), diluting their energy (IR), and displacing their positions (PR). 
Rooted in compressive sensing theory~\cite{CS-Signal-Processing, CS-Theory-Book, CS-Theory-Survey} and instantiated in \ICS~\cite{ICS-CVPR, ICS-TIP, ICS-TPAMI}, this dispersal is the key mechanism that enables \ComMark to neutralize embedded watermarks. 
A subsequent \textit{image reconstruction} module restores visual consistency, optimized by pixel-level structural similarity~\cite{SSIM-TIP15} and feature-level perceptual alignment~\cite{LPIPS}, thereby preserving image integrity while suppressing watermark evidence. 
Since our focus is on deepfake imagery rather than authentic digital artworks, the objective is to maintain perceptual and structural realism rather than exact pixel-level fidelity.

We validate \ComMark's effectiveness through extensive evaluations across eight state-of-the-art defensive watermarking schemes, including six post-processing methods (\MBRS~\cite{MBRS}, \CIN~\cite{CIN}, \TrustMark~\cite{TrustMark}, \PIMoG~\cite{PIMoG}, \VINEB~\cite{VINE}, and \VINER~\cite{VINE}) and two in-processing methods (\PTW~\cite{PTW} and  \StableSignature~\cite{StableSignature}). 
\ComMark reduces average watermark detection accuracy from 100\% to 32.9\%, outperforming prior attacks such as Distortion~\cite{WAVES}, \RegenVAE~\cite{WatermarkAttacker}, and \RegenDiffusion~\cite{Diffusion-Attack-2, Diffusion-Attack, WatermarkAttacker}. 
It preserves image integrity comparable to \RegenVAE~\cite{WatermarkAttacker, VAE} and substantially better than \DistortionAttack~\cite{WAVES} and \RegenDiffusion~\cite{Diffusion-Attack-2, Diffusion-Attack, WatermarkAttacker}. 
We further evaluate three mitigation strategies—image super-resolution, sparse watermarking, and adversarial training~\cite{AT-1, AT-2, AT-3}—none of which prove effective, underscoring the urgent need to develop stronger watermark-based defenses against deepfake manipulation.

In summary, we make the following main contributions:
\begin{itemize}[leftmargin=*]

\item We propose \textbf{\ComMark}, the first query-free black-box attack against defensive image watermarking for deepfakes.

\item \ComMark establishes the first connection between image compressive sensing  theory and attacks on deepfake watermarking defenses, formalizing the dispersal effect of sparse encoding through three properties:  Sparsity Change, Intensity Redistribution, and Positional Redistribution.

\item We demonstrate \ComMark's effectiveness across eight state-of-the-art defensive watermarking schemes, showing substantial reductions in watermark detection accuracy under query-free and black-box settings.

\item \ComMark also proves highly practical, reducing memory usage by $2.2\times$–$14.9\times$ and improving computational efficiency by $1.3\times$–$5.7\times$ over \RegenVAE and \RegenDiffusion.

\item We evaluate three mitigation strategies—image super-resolution (\ISR), sparse watermarking (SW), and adversarial training (\AT)~\cite{AT-1, AT-2, AT-3}. 
\ISR inadvertently amplifies \ComMark's impact, lowering detection accuracy from 29.7\% to 12.1\%, while SW and \AT modestly raise it from 25.1\% to 35.8\% on average. 
Despite these efforts, \ComMark remains more effective than \DistortionAttack, \RegenVAE, and \RegenDiffusion, underscoring the need for more resilient watermark-based deepfake defenses.

\end{itemize}

\section{Background}  
\label{sec:background}  

\subsection{Defensive Image Watermarking}
\label{subsec:invisible_image_watermarking}

Modern watermark-based deepfake defenses commonly employ deep encoder--decoder architectures~\cite{sok_aigc, unmark, VINE}. 
These approaches fall into two categories: \textit{post-processing} methods~\cite{MBRS, CIN, PIMoG, TrustMark, VINE, StegaStamp}, which embed watermarks after image generation, and \textit{in-processing} methods~\cite{StableSignature, TreeRing, PTW, Diffusion-Attack-2}, which integrate watermarking directly into the generation process.

\noindent
\emph{\underline{Post-Processing vs. In-Processing:}}
\emph{Post-processing} approaches~\cite{MBRS, CIN, PIMoG, TrustMark, VINE, StegaStamp} employ encoder--decoder models where the encoder~\(\mathcal{E}\) embeds a watermark~\(m\) into a GenAI-generated image~\(x\), producing a latent representation \(z = \mathcal{E}(x, m)\) that combines watermark and image features. 
The decoder~\(\mathcal{D}\) then reconstructs the watermarked image \(x_m = \mathcal{D}(z)\), while the watermark detector~\(\mathcal{W}\) authenticates it by extracting the watermark \(m = \mathcal{W}(x_m)\). 
In contrast, \emph{in-processing} schemes~\cite{StableSignature, TreeRing, PTW, Diffusion-Attack-2} embed watermarks directly during image generation. 
GenAI models such as Stable Diffusion~\cite{LDM} encode both the watermark and text prompt into a shared latent representation~\cite{StableSignature, PTW, Diffusion-Attack-2}, ensuring the generated image inherently carries watermark information. 
Here, the encoder~\(\mathcal{E}\) and decoder~\(\mathcal{D}\) are integrated within the generator~\(\mathcal{G}\), which is trained to produce verifiable watermarked outputs authenticated by dedicated detectors.

\subsection{Compressive Sensing}
\label{subsec:compressive_sensing}

\ICS (Image Compressive Sensing)~\cite{ICS-CVPR, ICS-TIP, ICS-TPAMI} aims to reconstruct a high-dimensional signal \(x \in \mathbb{R}^N\) from a low-dimensional measurement \(y = \Phi x\), where \(y \in \mathbb{R}^M\), \(\Phi \in \mathbb{R}^{M \times N}\) is the \emph{measurement (or sensing) matrix}, and \(M \ll N\). 
Because \(\Phi\) projects \(x\) into a much lower-dimensional space, directly recovering \(x\) from \(y\) is impossible—many different signals can yield the same measurement \(y\). 
To enable reconstruction, compressive sensing assumes that \(x\) is sparse under a known invertible \emph{sparsifying basis} (or \emph{transform matrix}) \(\Psi \in \mathbb{R}^{N \times N}\), such that \(x = \Psi \mathcal{Z}\), where \(\mathcal{Z} \in \mathbb{R}^N\) is the sparse representation. 
Although natural images are not sparse in the pixel domain, they exhibit sparsity in appropriate transform domains, such as DCT or wavelet, which enables effective reconstruction from compressed measurements. 
Substituting yields \(y = \Phi \Psi \mathcal{Z}\), an underdetermined system~\cite{ICS-CVPR, ICS-TIP, ICS-TPAMI} with infinitely many feasible solutions. 
The core objective is therefore to identify the \emph{sparsest} representation \(\mathcal{Z}\) that satisfies the measurement constraint \(y = \Phi \Psi \mathcal{Z}\):
\begin{equation}
\label{eq:compressive_sensing}
\min_{\mathcal{Z}} \Vert \mathcal{Z} \Vert_1 \quad \text{s.t. } y = \Phi \Psi \mathcal{Z}
\end{equation}
This optimization seeks the most compact latent representation capable of reproducing the observed measurement \(y\), ensuring accurate reconstruction despite the dimensionality gap between \(y\) and \(x\).

\section{Attacks on Defensive Image Watermarking}
\label{sec:watermarking_attacks}

\begin{table*}[t]
\centering
\caption{Comparison of \ComMark with prior attack approaches against defensive watermarking schemes.}
\label{table:prior_attack_comparison}
\scalebox{1.0}{
\addtolength{\tabcolsep}{-1.3ex}
\begin{tabular}{l|c|c|c|c|c|c}
\toprule
\hline
\makecell{Attack Approach} 
& \makecell{Black-Box} 
& \makecell{Query-Free} 
& \makecell{Integrity-Preserving} 
& \makecell{No Pre-trained \\ Model Dependence} 
& \makecell{Latent-Representation \\ Targeted}
& \makecell{Cross-Scheme \\ Transferability} \\

\hline
Distortion Attacks \cite{WAVES} & \cmark & \cmark & \xmark & \cmark & \xmark & \xmark \\

\hline
Regeneration Attacks \cite{WatermarkAttacker, Regeneration-Attack-Diff1, Regeneration-Attack-Diff2, Regeneration-Attack-Diff3, Regeneration-Attack-Diff4} & \cmark & \cmark & \cmark & \xmark & \xmark & \xmark \\

\hline
Adversarial Attacks \cite{WM-adversarial-CCS, WM-adversarial-ICLR, WM-adversarial-TCSVT} & \xmark & \xmark & \cmark & \cmark & \cmark & \cmark \\

\hline
\rowcolor{red!10}
\ComMark [This Paper] & \cmark & \cmark & \cmark & \cmark & \cmark & \cmark \\

\hline
\bottomrule
\end{tabular}
}
\end{table*}

Prior efforts to attack defensive image watermarking can be broadly classified into three categories—distortion, regeneration, and adversarial attacks—summarized in \Cref{table:prior_attack_comparison}.

\begin{description}[leftmargin=0pt, itemsep=3pt]
\setlength{\parindent}{1em}
\setlength{\parskip}{0pt}

\item[Distortion Attacks.] Watermarked images may undergo distortions during transmission~\cite{WAVES}, which can inadvertently weaken embedded watermarks. Attackers exploit this by applying compression, blurring, or brightness adjustments to degrade watermark integrity~\cite{WAVES, WatermarkAttacker}. However, these approaches often introduce visible artifacts~\cite{WAVES, PTW, StableSignature}. Unmark~\cite{unmark} further perturbs spectral amplitudes for stronger disruption but still causes noticeable quality loss~\cite{sok_aigc} and requires up to five minutes per image~\cite{unmark}.

\item[Regeneration Attacks.] These methods~\cite{WatermarkAttacker, Regeneration-Attack-Diff1, Regeneration-Attack-Diff2, Regeneration-Attack-Diff3, Regeneration-Attack-Diff4} use pre-trained generative models, such as VAEs~\cite{VAE} (\RegenVAE) or diffusion models~\cite{Diffusion-Model} (\RegenDiffusion), to reconstruct watermarked images. \RegenVAE encodes and decodes the image, while \RegenDiffusion iteratively denoises it to produce a clean version. Both primarily optimize pixel-level fidelity (e.g., $\ell_2$ loss) rather than disrupting latent representations that carry watermark information~\cite{VINE, WAVES, unmark}, leaving watermarks partially intact. Their reliance on large generative models also makes them computationally expensive~\cite{unmark}.

\item[Adversarial Attacks.] Adversarial methods~\cite{WM-adversarial-CCS, WM-adversarial-ICLR, WM-adversarial-TCSVT} adapt standard neural network attacks to subvert watermark detectors. Given an encoder \(\mathcal{E}\) that maps an image to latent features~\cite{WAVES}, they generate adversarial examples \(x_{\text{adv}}\) from watermarked inputs \(x_m\) by maximizing the divergence \(\|\mathcal{E}(x_{\text{adv}}) - \mathcal{E}(x_m)\|_2\) under an \(\ell_\infty\) constraint \(\|x_{\text{adv}} - x_m\|_\infty \leq \epsilon\). These methods typically require access to the watermark model or query capability for the detector \(\mathcal{W}\), limiting their practicality in real-world settings (\Cref{sec:threat_model}). Even surrogate-based variants~\cite{WM-adversarial-ICLR} exhibit poor transferability~\cite{WAVES, unmark, Hu-Transfer}.

\item[\ComMark.] \ComMark is a query-free black-box attack framework that exploits latent-space vulnerabilities in encoder--decoder watermarking schemes through image compressive sensing (\ICS) theory~\cite{ICS-CVPR, ICS-TIP, ICS-TPAMI}. 
Unlike adversarial attacks~\cite{WM-adversarial-CCS, WM-adversarial-ICLR, WM-adversarial-TCSVT} requiring model knowledge or regeneration-based approaches~\cite{WatermarkAttacker, Regeneration-Attack-Diff1, Regeneration-Attack-Diff2} relying on pre-trained generative models, \ComMark enforces latent-space sparsity to suppress watermark-bearing components while preserving image fidelity through structural–perceptual optimized reconstruction. 
This design introduces a principled latent-space attack paradigm that bridges compressive sensing and deep watermark removal.

\end{description}

\section{\ComMark}
\label{sec:problem_formulation}

We  outline our threat model (\Cref{sec:threat_model}), design principles (\Cref{sec:design_principle}), and attack framework (\Cref{sec:commark}).

\subsection{Threat Model}
\label{sec:threat_model}

\ComMark targets defensive watermarking for deepfake detection, involving two key parties: \emph{\underline{GenAI providers}}~\cite{OpenAI-WM, synID, DeepMind-WM}, who manage model outputs, and \emph{\underline{adversaries}}, who exploit GenAI to generate malicious content.

\emph{\underline{GenAI providers}} embed watermarks into generated images to mitigate misuse. Leveraging vast resources and training data, they fine-tune models for performance and secure APIs to ensure only watermarked outputs are produced. When misuse is suspected, watermark detectors verify authenticity and attribute content to GenAI models.

\emph{\underline{Adversaries}} seek to disrupt embedded watermarks, generating images that evade detection while preserving quality. Following prior work~\cite{unmark, WatermarkAttacker, WAVES}, we assume attackers have access to watermarked images but lack knowledge of watermarking models (\emph{black-box}) and cannot query GenAI providers’ detection algorithms (\emph{query-free}), as access is typically restricted to trusted entities~\cite{unmark, WatermarkAttacker, meta-says}.

\ComMark targets advanced encoder--decoder watermarking schemes, including six post-processing methods—\MBRS~\cite{MBRS}, \CIN~\cite{CIN}, \TrustMark~\cite{TrustMark}, \PIMoG~\cite{PIMoG}, \VINEB~\cite{VINE}, and \VINER~\cite{VINE}—and two in-processing methods—\PTW~\cite{PTW} and \StableSignature~\cite{StableSignature}. We further explore mitigations against \ComMark, including two new approaches—image super-resolution~\cite{Image-SR-ICCV2023, Image-SR-TPAMI2020, Image-SR-TPAMI2022} and sparse watermarking—alongside traditional adversarial training~\cite{AT-1, AT-2, AT-3}. Further details appear in \Cref{sec:evaluation}.

\subsection{Design Principles}
\label{sec:design_principle}

Traditional \ICS~\cite{ICS-CVPR, ICS-TIP, ICS-TPAMI} samples via a fixed measurement matrix \(\Phi\) to obtain \(y = \Phi x\), and assumes sparsity under a predefined invertible transform \(\Psi\) (e.g., DCT or wavelet), recovering \(x\) by solving for the sparsest \(\mathcal{Z}\) in \(y = \Phi \Psi \mathcal{Z}\) (\Cref{subsec:compressive_sensing}). 
However, fixed \(\Phi\) and \(\Psi\) offer limited adaptability and cannot selectively preserve semantically meaningful features in deep latent representations.

To address this limitation, we introduce a \emph{learnable} encoder–decoder architecture that replaces these fixed operators with data-driven ones. 
The encoder produces a sparse latent representation \(\mathcal{Z}\) directly from the input \(x\), acting as an adaptive sensing/analysis operator that retains only task-relevant components, while the decoder serves as a learned synthesis operator that reconstructs the image from \(\mathcal{Z}\). 
In effect, we reformulate the classical \ICS\ objective—finding the sparsest \(\mathcal{Z}\) consistent with \(y = \Phi \Psi \mathcal{Z}\)—into a \emph{learnable sparse encoding} on deep latent features, enforced during training rather than through hand-crafted matrices \(\Phi\) and \(\Psi\).

Training employs two coupled losses: 
(i) a \emph{sparse encoding} loss that promotes sparsity in \(\mathcal{Z}\) (e.g., \(\ell_1\)-regularization), and 
(ii) a \emph{structural–perceptual} reconstruction loss (SSIM~\cite{SSIM-TIP15} + LPIPS~\cite{LPIPS}) that preserves both structural similarity and perceptual realism, providing feedback that guides sparsification toward suppressing watermark-bearing components while maintaining visual fidelity.

\subsubsection{Sparse Encoding Loss (SEL)} 

\(\mathcal{L}_\text{SEL}\) enforces sparsity in latent representations by preserving only salient features essential for reconstruction. It is defined as:
\begin{equation}
\label{eq:sparse_encoding_loss}
    \mathcal{L}_\text{SEL}(\mathcal{Z}) = \|\mathcal{Z}\|_1
\end{equation}
This minimizes the \(\ell_1\)-norm of \(\mathcal{Z}\), reducing the sum of absolute values in the sparse latent representation—consistent with the principles of ICS~\cite{CS-Signal-Processing, CS-Theory-Book, CS-Theory-Survey}. 
Unlike traditional \ICS, where sparsity strength depends on a fixed sparsifying basis \(\Psi\), we introduce a weighting coefficient \(\alpha\) (\Cref{eq:total_loss}) to control sparsity. 
A larger \(\alpha\) imposes stronger sparsity, emphasizing key components while suppressing irrelevant ones. 
An analysis of \(\alpha\)’s effect on sparsity and watermark removal is presented in \Cref{subsec:RQ4}.

Sparsity lies at the core of \ICS, ensuring efficient signal reconstruction by retaining only the most informative coefficients. 
In \ComMark, this principle is retained but operationalized differently: instead of solving an optimization problem for the sparsest representation (\Cref{eq:compressive_sensing}), the model learns to enforce sparsity directly through \(\mathcal{L}_\text{SEL}\). 
This learnable formulation preserves the \ICS\ \textbf{dispersal effect}—the redistribution of sparsity, intensity, and positional attributes in latent space (discussed below)—allowing targeted suppression of watermark-bearing components while maintaining perceptual fidelity.

\smallskip
\noindent
\textbf{Dispersal Effect.} Given a watermarked image \(x_m\), derived from the latent representation \(z = \mathcal{E}(x, m)\) that fuses image features with watermark attributes (\Cref{subsec:invisible_image_watermarking}), its sparse latent representation \(\mathcal{Z}\) is obtained via sparse transformation \(\mathcal{T}(x_m)\). 
The dispersal effect of \(\mathcal{T}\) is defined by three properties: \emph{sparsity change (SC)}, \emph{intensity redistribution (IR)}, and \emph{positional redistribution (PR)}~\cite{CS-Signal-Processing, CS-Theory-Book, CS-Theory-Survey}.

\emph{\underline{Sparsity Change (SC)}} quantifies the variation in the number of significant coefficients after sparsification, where a coefficient is considered \emph{significant} if its absolute value exceeds a small threshold \(\tau\):
\begin{equation}
\label{eq:SC}
SC_{\mathcal{T}} = \frac{|\{i \mid |\mathcal{Z}_i| > \tau\}|}{N_{\mathcal{Z}}} 
 - \frac{|\{i \mid |z_{i}| > \tau\}|}{N_{z}}
\end{equation}
where \(N_{\mathcal{Z}}\) and \(N_{z}\) denote the total number of coefficients in \(\mathcal{Z}\) and \(z\), respectively.

\emph{\underline{Intensity Redistribution (IR)}} measures the change in energy distribution between the sparse representation \(\mathcal{Z}\) and the original latent \(z\):
\begin{equation}
\label{eq:IR}
IR_{\mathcal{T}} = \frac{1}{N_{\mathcal{Z}}} \sum_{i=1}^{N_{\mathcal{Z}}} |\mathcal{Z}_i|^2 
  - \frac{1}{N_z} \sum_{i=1}^{N_z} |z_i|^2
\end{equation}

\emph{\underline{Positional Redistribution (PR)}} measures positional shifts of non-zero elements in the sparse representation using the Wasserstein distance \(W\)~\cite{wasserstein-distance-compressive-sensing}, which quantifies the minimal ``work'' required to transform one distribution into another. 
It is computed over the sets \(P_{\mathcal{Z}}\) and \(P_{z}\), representing the non-zero elements in \(\mathcal{Z}\) and \(z\), respectively:
\begin{equation}
\label{eq:PR}
PR_{\mathcal{T}} = W(P_{\mathcal{Z}}, P_{z})
\end{equation}
The overall dispersal effect in \ICS\ thus comprises SC, IR, and PR~\cite{CS-Signal-Processing, CS-Theory-Book, CS-Theory-Survey}.

Building on the design principles in \Cref{sec:design_principle}, sparse encoding via \ICS serves as the key mechanism through which \ComMark\ suppresses embedded watermark signals. 
By enforcing sparsity in the latent domain, it efficiently compresses structural carriers of watermark information, weakening their robustness and reducing detectability. 
As described in \Cref{subsec:invisible_image_watermarking}, a watermarked image \(x_m\) is decoded from the latent representation \(z = \mathcal{E}(x, m)\), which fuses image semantics with watermark attributes. 
We analyze how the sparse encoding transformation \(\mathcal{T}\) disperses these attributes, compromising watermark integrity through measurable changes in SC, IR, and PR.

The sparsity change \(SC_{\mathcal{T}}\) quantifies reductions in the density of significant coefficients in latent representations~\cite{CS-Signal-Processing, CS-Theory-Book, CS-Theory-Survey}. 
This reduction indicates that watermark attributes—reliant on dense activation patterns for reliable detection—are dispersed across fewer active positions, hindering watermark extraction.

Likewise, \(IR_{\mathcal{T}}\) typically shows a decrease in latent-space energy, reflecting a more uniform energy distribution. 
Since defensive watermarks are commonly embedded in low-frequency components to maintain imperceptibility~\cite{VINE, unmark}, this redistribution further diminishes their robustness and detectability.

Finally, \(PR_{\mathcal{T}}\) captures positional shifts of non-zero elements, revealing disrupted alignment between watermark patterns and their original latent positions. 
The resulting increase in Wasserstein distance signifies stronger displacement of watermark features, highlighting the effectiveness of sparsification in breaking their structured embeddings.

Collectively, the variations in \(SC_{\mathcal{T}}\), \(IR_{\mathcal{T}}\), and \(PR_{\mathcal{T}}\) induce a structural breakdown of watermark signals in the sparse latent domain, significantly reducing their detectability and undermining the resilience of current defensive watermarking schemes for deepfakes.

\begin{figure*}[t]
\centering
\begin{subfigure}[t]{1.0\columnwidth}
\includegraphics[width=\linewidth]{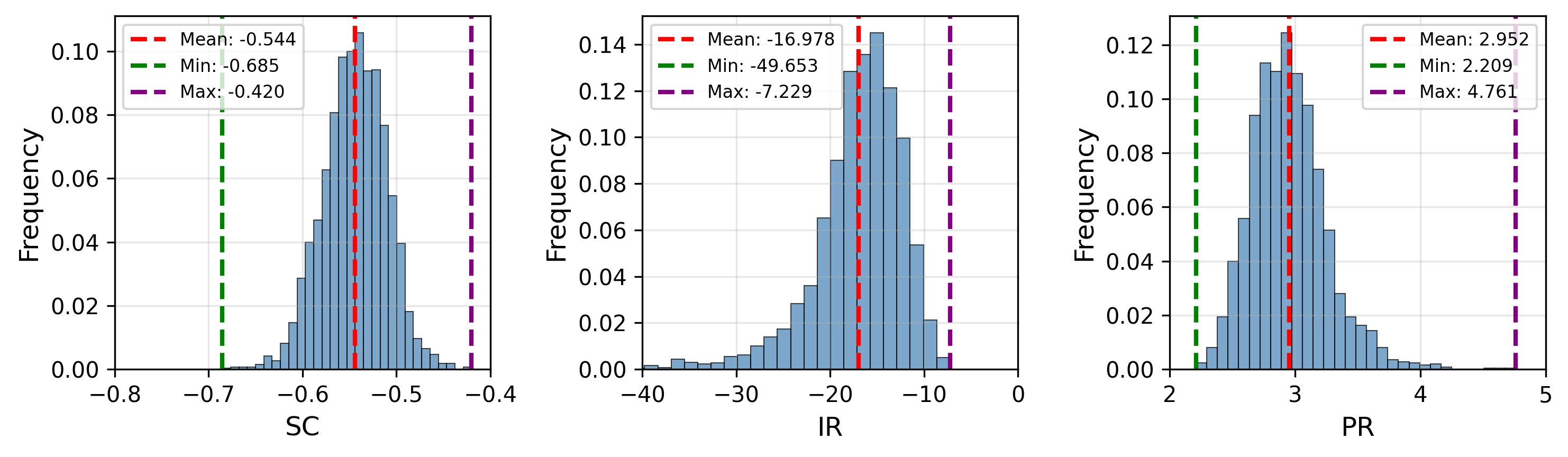}
\caption{\MBRS}
\label{fig:mbrs}
\end{subfigure}
\begin{subfigure}[t]{1.0\columnwidth}
\includegraphics[width=\linewidth]{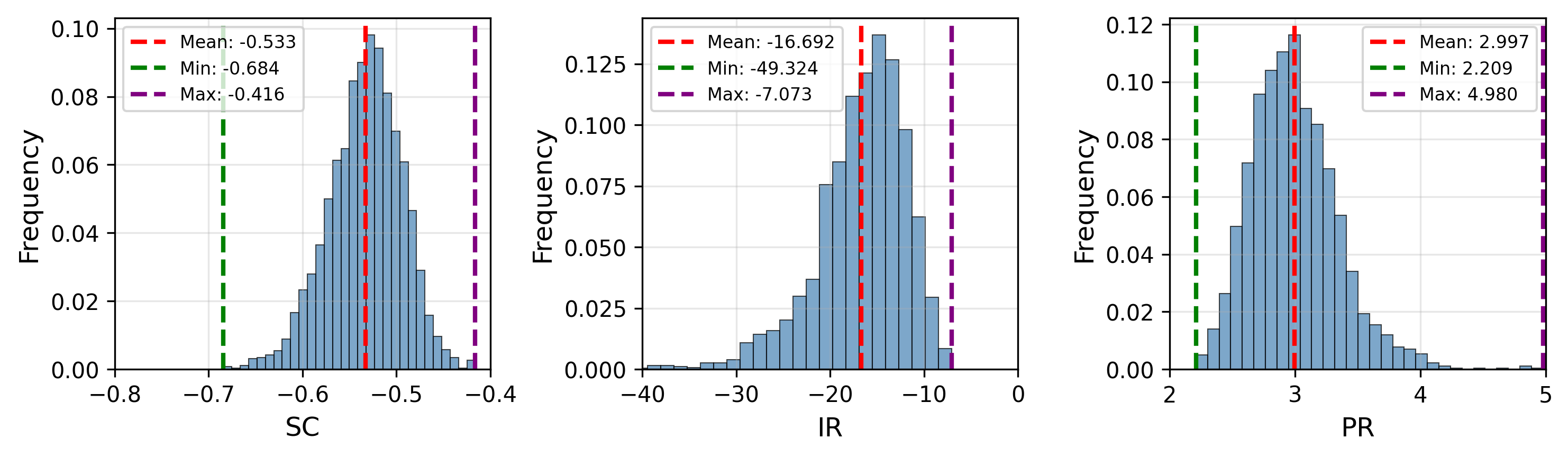}
\caption{\CIN}
\label{fig:cin}
\end{subfigure}

\begin{subfigure}[t]{1.0\columnwidth}
\includegraphics[width=\linewidth]{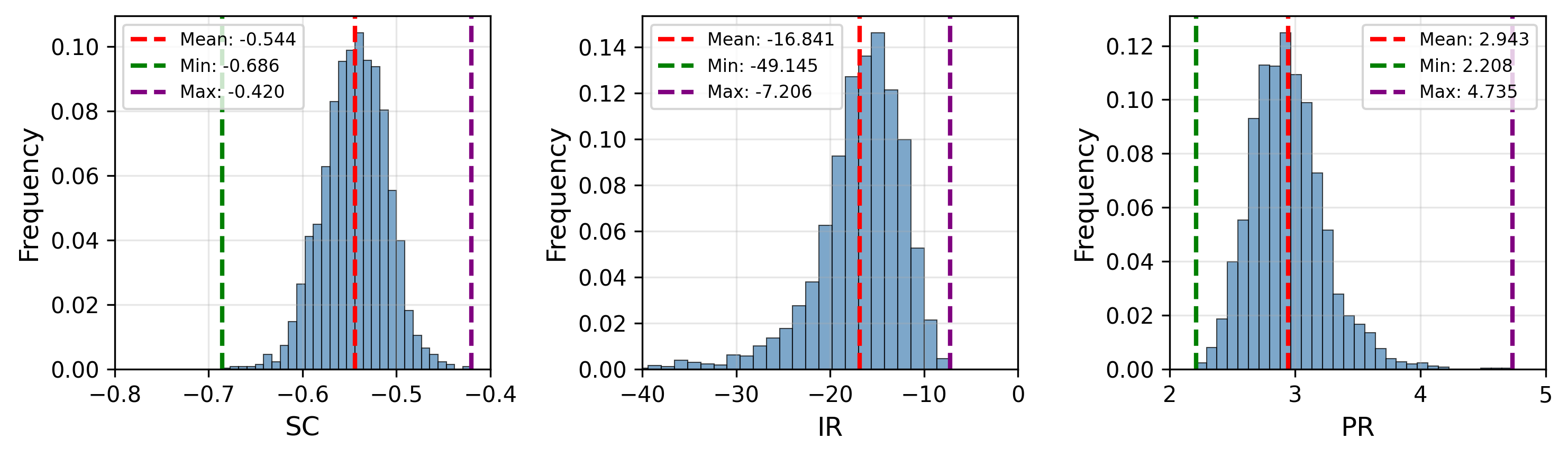}
\caption{\TrustMark}
\label{fig:trustmark}
\end{subfigure}
\begin{subfigure}[t]{1.0\columnwidth}
\includegraphics[width=\linewidth]{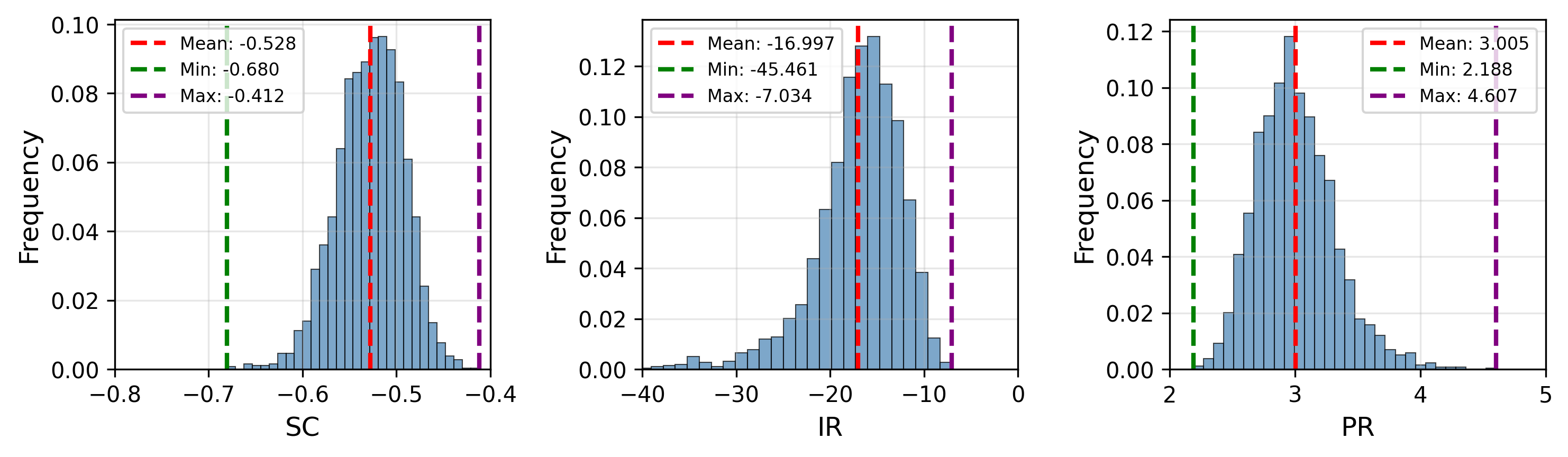}
\caption{\PIMoG}
\label{fig:pimog}
\end{subfigure}

\begin{subfigure}[t]{1.0\columnwidth}
\includegraphics[width=\linewidth]{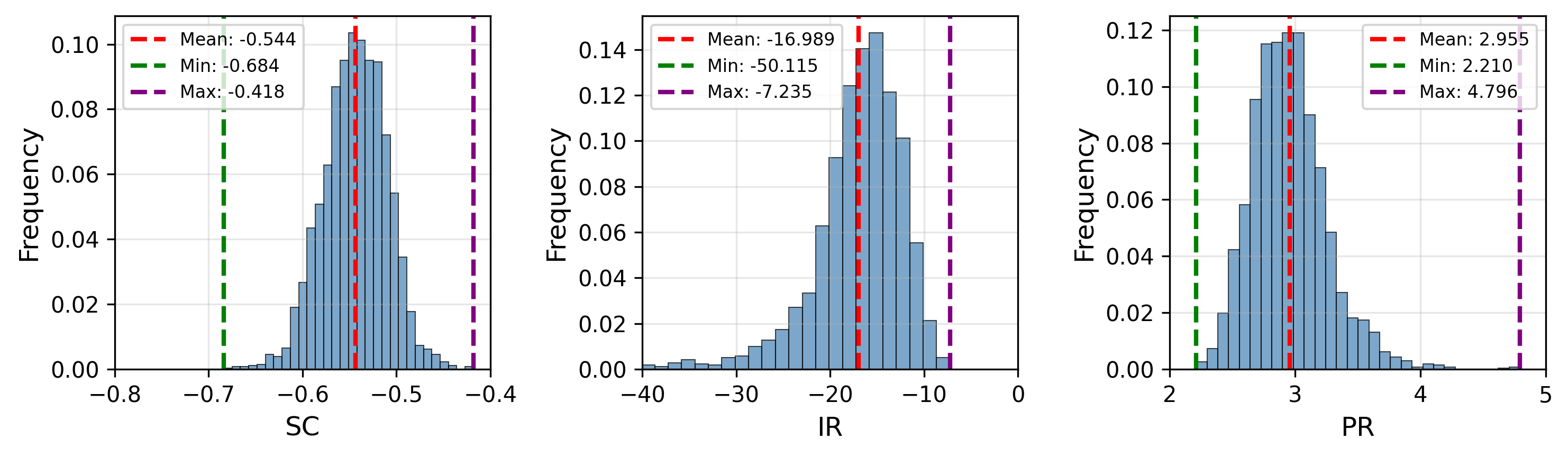}
\caption{\VINEB}
\label{fig:vine_b}
\end{subfigure}
\begin{subfigure}[t]{1.0\columnwidth}
\includegraphics[width=\linewidth]{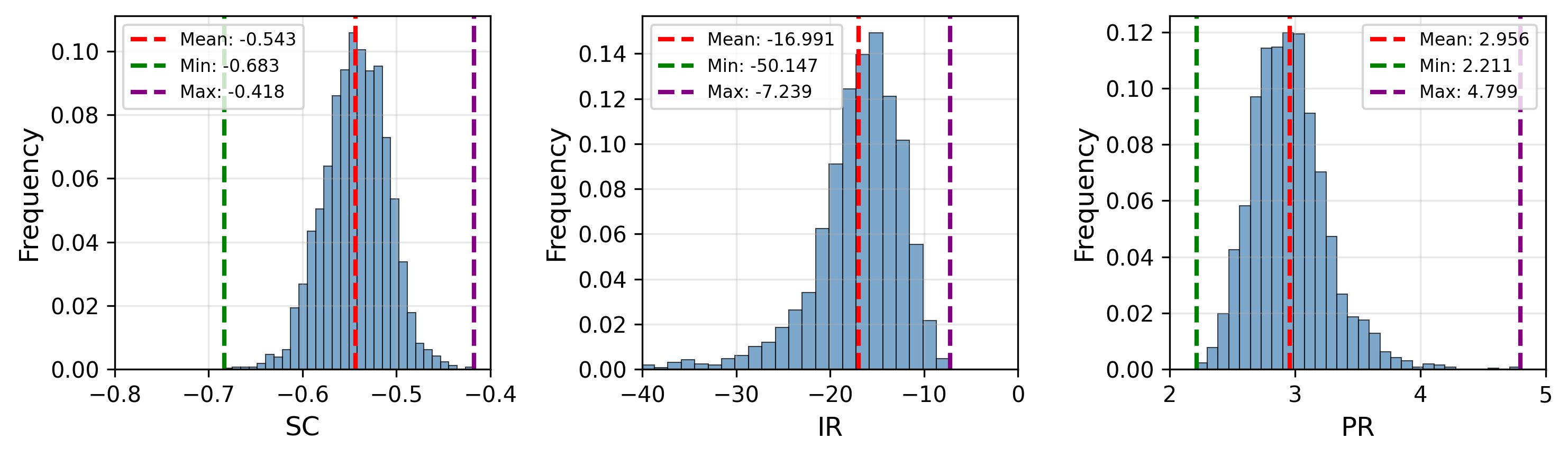}
\caption{\VINER}
\label{fig:vine}
\end{subfigure}

\begin{subfigure}[t]{1.0\columnwidth}
\includegraphics[width=\linewidth]{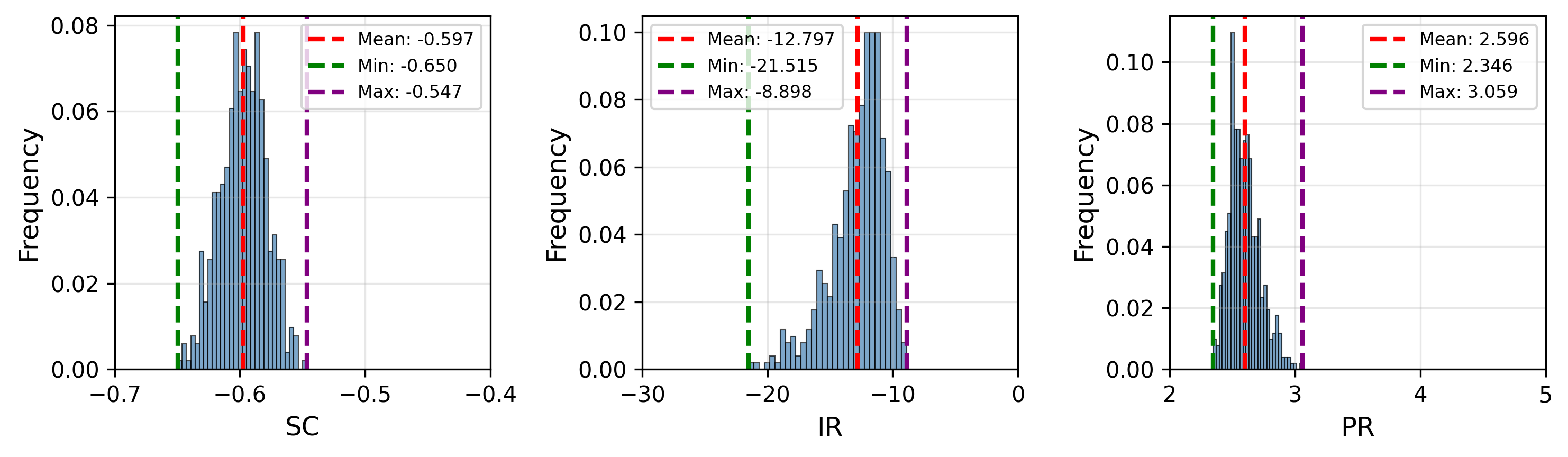}
\caption{\PTW}
\label{fig:ptw}
\end{subfigure}
\begin{subfigure}[t]{1.0\columnwidth}
\includegraphics[width=\linewidth]{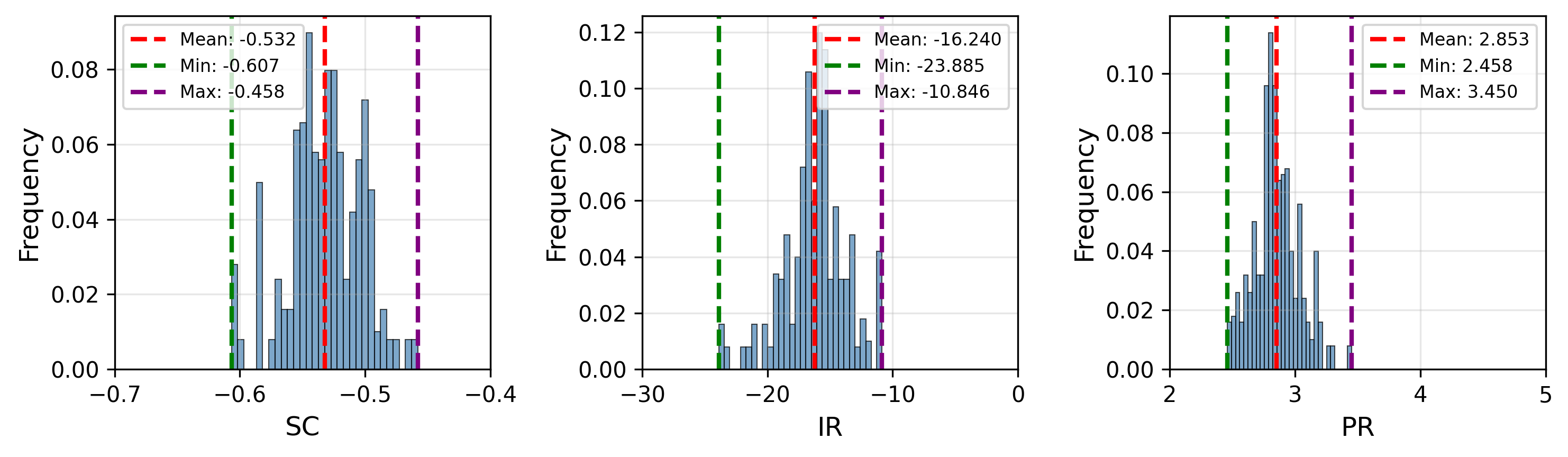}
\caption{\StableSignature}
\label{fig:stablesignature}
\end{subfigure}

\caption{Effects of \ICS on SC, IR, and PR for a set of 2,000 images from the \OpenImage dataset watermarked using each of the six post-processing schemes, \MBRS \cite{MBRS}, \CIN \cite{CIN}, \TrustMark \cite{TrustMark}, \PIMoG \cite{PIMoG}, \VINEB \cite{VINE}, and \VINER \cite{VINE}, and 500 watermarked images generated 
from each of two in-processing schemes, \PTW \cite{PTW} and \StableSignature \cite{StableSignature}.}
\label{fig:sc_ir_pr}
\vspace*{-2ex}
\end{figure*}

\emph{\underline{Dispersal Effect Validation.}} 
We empirically validate \ICS's impact on SC, with \(\tau = 0.02\)~\cite{CS-Signal-Processing, CS-Theory-Book, CS-Theory-Survey}), IR, and PR  using 2,000 images from the \OpenImage~\cite{OpenImage} dataset, each watermarked by six post-processing schemes—\MBRS~\cite{MBRS}, \CIN~\cite{CIN}, \TrustMark~\cite{TrustMark}, \PIMoG~\cite{PIMoG}, \VINEB~\cite{VINE}, and \VINER~\cite{VINE}—and 500 images generated by two in-processing schemes, \PTW~\cite{PTW} and \StableSignature~\cite{StableSignature}. 
As shown in \Cref{fig:sc_ir_pr}, across all eight watermarking schemes, \ICS consistently reduces coefficient density (SC), redistributes intensity (IR), and shifts positional distributions (PR) of non-zero latent components. 
These results empirically confirm the \emph{dispersal effect}—the foundation of \ComMark's ability to attenuate embedded watermark signals and hinder their detectability.

\subsubsection{Structural-Perceptual Loss (SPL)} To ensure reconstruction quality, we introduce a structural-perceptual loss \(\mathcal{L}_\text{SPL}\) that integrates two complementary image quality metrics: \SSIM \cite{SSIM-TIP15} and \LPIPS \cite{LPIPS}. SSIM imposes \emph{pixel-level structural consistency} while LPIPS enforces \emph{feature-level perceptual similarity} \cite{SSIM-TIP15, LPIPS}. This complementary constraint ensures that the reconstructed images preserve both the semantic structure and the human-perceived realism of the originals \cite{unmark, ICS-CVPR, ICS-TPAMI2024}. \(\mathcal{L}_\text{SPL}\) is defined as: 
\begin{equation}
\label{eq:dml_loss}
\mathcal{L}_\text{SPL} = \mathcal{L}_{\text{SSIM}} + \mathcal{L}_{\text{LPIPS}}
\end{equation}
where \(\mathcal{L}_\text{SSIM}\) and \(\mathcal{L}_\text{LPIPS}\) uses \SSIM and \LPIPS, respectively.

\SSIM (Structural Similarity Index Measure) \cite{SSIM-TIP15} evaluates the structural similarity between two images by comparing their \emph{pixel-level} attributes like luminance and contrast:
\begin{equation}
\label{eq:ssim}
\small
\text{SSIM}(x_m, \tilde{x}) = \frac{(2\mu_{x_m} \mu_{\tilde{x}} + C_1)(2\sigma_{{x_m}\tilde{x}} + C_2)}{(\mu_{x_m}^2 + \mu_{\tilde{x}}^2 + C_1)(\sigma_{x_m}^2 + \sigma_{\tilde{x}}^2 + C_2)}
\end{equation}
For watermarked image \(x_m\) and reconstructed image \(\tilde{x}\), \(\mu_{x_m}\) and \(\mu_{\tilde{x}}\) are their mean intensities, \(\sigma_{x_m}\) and \(\sigma_{\tilde{x}}\) their standard deviations, and \(\sigma_{{x_m}\tilde{x}}\) their covariance. Constants \(C_1\) and \(C_2\) ensure numerical stability to avoid division by zero \cite{SSIM-TIP15}.

\SSIM values typically range from 0 to 1, with higher values indicating greater structural similarity between images \cite{SSIM-TIP15}. To integrate this into our attack framework, we define the SSIM-based loss \(\mathcal{L}_{\text{SSIM}}\) as  follows:
\begin{equation}
\label{eq:ssim_loss}
\mathcal{L}_{\text{SSIM}} = 1 - \text{SSIM}(x_m, \tilde{x})
\end{equation}
\(\mathcal{L}_{\text{SSIM}}\) enforces pixel-level structural consistency, ensuring that the reconstructed image \(\tilde{x}\) preserves the critical structural patterns of the watermarked image \(x_m\).

In contrast, \LPIPS (Learned Perceptual Image Patch Similarity) \cite{LPIPS} assesses \emph{feature-level} perceptual similarity by measuring distances between deep representations extracted using pre-trained models such as AlexNet \cite{AlexNet}:
\begin{equation}
\label{eq:lpips}
\small
\text{LPIPS}(x_m, \tilde{x}) = \sum_{l} \frac{1}{H_l W_l C_l} \left\| \phi_l^M(x_m) - \phi_l^M(\tilde{x}) \right\|_2^2
\end{equation}
Here, $\phi_l^M$ represents the feature map of the \(l\)-th layer in a CNN model $M$, with \(W_l\), \(H_l\), and \(C_l\) indicating the width, height, and number of channels of that layer, respectively.

A lower \(\text{LPIPS}\) score indicates greater perceptual similarity between images~\cite{LPIPS}. Thus, we use \(\text{LPIPS}\) as a loss (\(\mathcal{L}_\text{LPIPS}\)) to quantify discrepancies between the original and reconstructed images:
\begin{equation}
\label{eq:lpips_loss}
\mathcal{L}_\text{LPIPS} = \text{LPIPS}(x_m, \tilde{x})
\end{equation}
Unlike \(\mathcal{L}_\text{SSIM}\), which evaluates pixel statistics for structural integrity, \(\mathcal{L}_\text{LPIPS}\) preserves deeper perceptual features, ensuring the perceptual coherence of the reconstructed image \(\tilde{x}\).

By combining both metrics in a unified loss, \ComMark preserves original image content at pixel and feature levels, maintaining fine-grained structural consistency and perceptual coherence between watermarked and reconstructed images.

\subsection{The Attack Framework}
\label{sec:commark}

We instantiate the above design principles through \ComMark, a query-free black-box attack framework leveraging \ICS\ sparsity to disrupt defensive watermarking schemes used for deepfake detection. 
\ComMark\ comprises two main modules: \emph{sparse encoding} and \emph{image reconstruction}.

\subsubsection{Sparse Encoding}
\label{subsec:sparse_encoding}

In \ComMark, a CNN-based sparse encoder \(\mathcal{T}_{\text{CNN}}\), guided by the sparse encoding loss \(\mathcal{L}_\text{SEL}\), transforms the watermarked image \(x_m\) into a sparse latent representation \(\mathcal{Z}\), enforcing the SC (\Cref{eq:SC}), IR (\Cref{eq:IR}), and PR (\Cref{eq:PR}) dispersal effects:
\begin{equation}
\label{eq:sparse_encoding}
  \mathcal{Z} = \mathcal{T}_{\text{CNN}}(x_m)
\end{equation} 
\(\mathcal{T}_{\text{CNN}}\) consists of four sequential blocks, each incorporating a convolutional layer to extract feature maps and an attention layer for further refinement. The attention layer employs adaptive average and max pooling \cite{CNN-Attention-CVPR2019, CNN-Attention-ICCV2017, CNN-Attention-TIP2018} to generate compact descriptors for channel-wise feature enhancement in \(\mathcal{Z}\). Specifically, each block \(b\), where \(1 \leqslant b \leqslant 4\), captures feature maps \(\mathcal{F}_b \in \mathbb{R}^{C_b \times H_b \times W_b}\) and aggregates spatial information through adaptive pooling, producing average and max descriptors \(\mathcal{F}_{\text{avg}}^{b} = \texttt{AdaptiveAvgPool}(\mathcal{F}_b)\) and \(\mathcal{F}_{\text{max}}^{b} = \texttt{AdaptiveMaxPool}(\mathcal{F}_b)\). These descriptors are processed through a shared fully connected network \(f\) and a sigmoid activation \(\mathcal{A}\), resulting in weight vectors \(w_{\text{avg}}^{b} = \mathcal{A}\bigl(f(\mathcal{F}_{\text{avg}}^{b})\bigr)\) and \(w_{\text{max}}^{b} = \mathcal{A}\bigl(f(\mathcal{F}_{\text{max}}^{b})\bigr)\). The final attention weights for each block \(b\) are calculated as \(w^{b} = w_{\text{avg}}^{b} + w_{\text{max}}^{b}\), applied to enhance or suppress the feature maps \(\mathcal{F}_b\) through element-wise multiplication \(\otimes\), yielding the weighted feature map \(\mathcal{F}_b' = \mathcal{F}_b \otimes w^{b}\) \cite{CBAM-ECCV2018}. This process is replicated across all blocks, culminating in the final latent representation \(\mathcal{Z} = \mathcal{F}_4'\). This attention-driven encoding optimizes the sparsification process to ensure that crucial image features are preserved while less pertinent ones are minimized \cite{CNN-Attention-CVPR2019, CNN-Attention-ICCV2017, CNN-Attention-TIP2018}.

A sparse encoding loss \(\mathcal{L}_\text{SEL}\)~\cite{ICS-CVPR, ICS-TIP, ICS-TPAMI}, defined in \Cref{eq:sparse_encoding_loss}, guides \SparseEncoder\ during training to encode the watermarked image \(x_m\) while enforcing sparsity in \(\mathcal{Z}\).

\subsubsection{Image Reconstruction}
\label{subsec:reconstruction}

Following sparse encoding, the main challenge in reconstruction is ensuring that the recovered image \(\tilde{x}\) from \(\mathcal{Z}\) remains visually convincing as a deepfake~\cite{unmark} while disrupting the embedded watermark (\Cref{sec:threat_model}). 
Traditional \ICS\ computes measurements \(y\) using a linear measurement matrix and an invertible sparsifying basis (\Cref{eq:compressive_sensing}); in contrast, \ComMark\ treats \(\mathcal{Z}\) as both the sparse representation and  measurement, removing the need for explicit matrices. 
A CNN-based reconstruction module \(\mathcal{R}_{\text{CNN}}\) then reconstructs the image as:
\begin{equation}
\label{eq:reconstruction}
    \tilde{x} = \mathcal{R}_{\text{CNN}}(\mathcal{Z})
\end{equation}
which is optimized with a structural–perceptual loss \(\mathcal{L}_\text{SPL}\) (\Cref{eq:dml_loss}) to preserve pixel-level structure and feature-level coherence. 
\(\mathcal{R}_{\text{CNN}}\) mirrors \(\mathcal{T}_{\text{CNN}}\), serving as its inverse to reconstruct the image from the sparse latent representation.

\subsubsection{\ComMark: An Attack Framework}
\label{subsec:commark_an_attack_framework}

As shown in \Cref{algorithm:commark_training_algorithm},
\ComMark co-trains its \(\mathcal{T}_\text{CNN}\) and \(\mathcal{R}_\text{CNN}\) modules using a tailored loss that combines the sparse encoding loss \(\mathcal{L}_\text{SEL}\) (\Cref{eq:sparse_encoding_loss}) to minimize the latent representation \(\mathcal{Z}\), and the structural-perceptual loss \(\mathcal{L}_\text{SPL}\) (\Cref{eq:dml_loss}) to ensure image integrity. The total loss \(\mathcal{L}\) is defined as:
\begin{equation}
\label{eq:total_loss}
    \mathcal{L} = \alpha \mathcal{L}_{\text{SEL}} + \beta \mathcal{L}_{\text{SPL}}
\end{equation}
where \(\alpha\) and \(\beta\) are weights that balance between sparsity and image integrity. The training parameters for \ComMark, including the number of training epochs \(N\) and the weights \(\alpha\) and \(\beta\), are optimized as discussed in \Cref{subsec:experiment_setup}.

\Cref{algorithm:commark} outlines \ComMark's procedure for attacking defensive image watermarking schemes. 
A \ComMark\ attack is considered successful if the watermark detector $\mathcal{W}$ fails to recover the embedded watermark \(m\) from the reconstructed image, that is, when \(\mathcal{W}(\tilde{x}) \neq m\).

\begin{algorithm}[t]
\LinesNumbered
\KwIn{Training image dataset $\mathcal{C}$,  epochs \(N\), and  weights \(\alpha\) and \(\beta\) from \Cref{eq:total_loss}.}

\KwOut{Trained parameters \(\Theta_{\mathcal{T}}\) and \(\Theta_{\mathcal{R}}\).}

\For{\(epoch = 1\) \KwTo \(N\)}{
    \ForEach{\(image \) \(x\) in \(\mathcal{C}\)}{

        \tcp{\small Sparse encoding}
        \(\mathcal{Z} \Leftarrow  \mathcal{T}_{\text{CNN}}(x)\);

        \tcp{\small Image reconstruction}
        \(\tilde{x} \Leftarrow \mathcal{R}_{\text{CNN}}((\mathcal{Z}) \);

        \(\mathcal{L} \Leftarrow \alpha \mathcal{L}_{\text{SEL}} + \beta \mathcal{L}_{\text{SPL}}\);

        \(\Theta_{\mathcal{T}}, \Theta_{\mathcal{R}} \Leftarrow \text{Minimize}\) \(\mathcal{L}\); 
    }
}

\caption{Training \ComMark.}
\label{algorithm:commark_training_algorithm}
\end{algorithm}

\begin{algorithm}[t]
\LinesNumbered
\KwIn{Watermarked image \(x_m\),  watermark detector \(\mathcal{W}\), watermark \(m\), and \SparseEncoder and  \Reconstructor (\Cref{algorithm:commark_training_algorithm}).}
\KwOut{Attacked image \(\tilde{x}\).}

\tcp{\small Extract  \(\mathcal{Z}\) from \(x_m\)}
\(\mathcal{Z} \Leftarrow  \mathcal{T}_{\text{CNN}}(x_m)\);

\tcp{\small Reconstruct the attacked image \(\tilde{x}\) from \(\mathcal{Z}\)}
\(\tilde{x} \Leftarrow  \mathcal{R}_{\text{CNN}}(\mathcal{Z})\);

\caption{\ComMark's attack framework.}
\label{algorithm:commark}
\end{algorithm}
\section{Evaluation}
\label{sec:evaluation}

We evaluate \ComMark's effectiveness against advanced defensive image watermarking schemes for deepfakes~\cite{MBRS, CIN, TrustMark, PIMoG, VINE, PTW, StableSignature}, showing that it outperforms existing attack methods and underscores the need for stronger watermark-based deepfake defenses. 
Our evaluation addresses four key research questions (RQs):

\begin{description}[leftmargin=0pt, itemsep=3pt]
\setlength{\parindent}{1em}
\setlength{\parskip}{0pt}

\item[\textbf{RQ1 Attack Capability.}] How does \ComMark perform compared to prior attacks on defensive watermarking?

\item[\textbf{RQ2 Image Integrity.}] Does \ComMark preserve image quality and fidelity through \ICS-based reconstruction with sparse latent representations?

\item[\textbf{RQ3 Computational Cost.}] Are \ComMark's computational demands practical for real-world attacks?

\item[\textbf{RQ4 Mitigation Robustness.}] How effective is \ComMark against mitigation strategies, including image super-resolution, sparse watermarking, and adversarial training?

\end{description}

\subsection{Experiment Setup}
\label{subsec:experiment_setup}

\begin{description}[leftmargin=0pt]
\setlength{\parindent}{1em}  
\setlength{\parskip}{0pt}
\item [Datasets.] We used two public datasets, OpenImage~\cite{OpenImage} and COCO~\cite{COCO}, selecting 50,000 and 40,000 images, respectively, to train \ComMark's sparse encoder (\SparseEncoder) and reconstruction module (\Reconstructor).

\smallskip
\item [Defensive Watermarking Schemes.] We assess \ComMark against eight representative watermarking schemes used for deepfake protection: six post-processing schemes (\MBRS \cite{MBRS}, \CIN \cite{CIN}, \TrustMark \cite{TrustMark}, \PIMoG \cite{PIMoG}, \VINEB \cite{VINE}, and \VINER \cite{VINE}) and two in-processing schemes (\PTW \cite{PTW} and \StableSignature \cite{StableSignature}).
\textit{\MBRS \footnote{\href{https://github.com/jzyustc/MBRS}{https://github.com/jzyustc/MBRS}}} employs a deep encoder–decoder architecture and enhances robustness by randomly applying random noise, simulated JPEG, or real JPEG during training, enabling it to effectively handle such distortions.
\textit{\CIN \footnote{\href{https://github.com/rmpku/CIN}{https://github.com/rmpku/CIN}}} improves watermark robustness against distortion while preserving invisibility. It adopts an invertible network with diffusion, extraction, fusion, and split modules for symmetric watermark embedding and extraction.
\textit{\TrustMark \footnote{\href{https://github.com/adobe/trustmark}{https://github.com/adobe/trustmark}}} balances watermark invisibility and recovery accuracy via adversarial learning and spatio-spectral losses \cite{TrustMark}. It includes a noise module during training to simulate distortion and improve robustness. Similar to \TrustMark, \textit{\PIMoG \footnote{\href{https://github.com/FangHanNUS/PIMoG-An-Effective-Screen-shooting-Noise-Layer-Simulation-for-Deep-Learning-Based-Watermarking-Netw}{https://github.com/FangHanNUS/PIMoG}}} enhances robustness by integrating a differentiable noise layer during training to model distortion noises.
\textit{\VINEB \footnote{\href{https://github.com/Shilin-LU/VINE}{https://github.com/Shilin-LU/VINE-B}}} improves the watermark encoder by using a condition adaptor to map images and watermarks into text sequences, which are processed by the text-to-image model SDXL-Turbo \cite{SDXL-Turbo} to generate watermarked images.
\textit{\VINER \footnote{\href{https://github.com/Shilin-LU/VINE}{https://github.com/Shilin-LU/VINE-R}}} builds on \VINEB by adding an adversarial learning module, similar to \TrustMark and \PIMoG, to enhance the watermark extractor’s ability to recover watermarks under diverse distortions.
\textit{\PTW \footnote{\href{https://github.com/nilslukas/gan-watermark}{https://github.com/nilslukas/PTW}}}, a leading in-processing watermarking scheme for StyleGAN models \cite{StyleGAN2, StyleGAN3}, uses pivotal tuning \cite{PTW} to freeze the original model’s parameters as a pivot while fine-tuning a secondary model with a watermark regularization term, enabling generation of watermarked images.
\textit{\StableSignature \footnote{\href{https://github.com/facebookresearch/stable_signature}{https://github.com/StableSignature}}} ensures all generated images embed a binary signature by fine-tuning the Latent Diffusion Model (LDM) decoder \cite{LDM} with a pre-trained watermark encoder, decoder, and extractor.

We used pre-trained watermarking schemes: six post-processing schemes trained on OpenImage~\cite{OpenImage} and COCO~\cite{COCO}, and two in-processing schemes, \StableSignature\ trained on LAION-5B~\cite{LAION-5B} and \PTW\ on AFHQv2~\cite{AFHQv2}. 
\Cref{tab:bit_length} (\appref{sec:appendix_watermark_length}) lists their watermark bit lengths. 
Following each scheme’s official protocol, we generated 2,000 watermarked images per post-processing scheme from OpenImage and COCO, \emph{ensuring no overlap with \ComMark's training data}. 
Image sizes were set to \(256 \times 256\) per default settings~\cite{VINE, MBRS, TrustMark, PIMoG}. 
For the in-processing schemes, we produced 500 watermarked images each at \(512 \times 512\); \StableSignature\ required Stable Diffusion prompts~\cite{Stable-Diffusion-Prompts}, while \PTW\ required none.

\smallskip
\item [Watermarking Attacks.] We compared \ComMark with three query-free, black-box attack methods: \DistortionAttack \cite{WAVES}, \RegenVAE \cite{WatermarkAttacker}, and \RegenDiffusion \cite{WatermarkAttacker, Regeneration-Attack-Diff1, Regeneration-Attack-Diff2}, as these represent the only practical options \cite{unmark}. Adversarial attacks \cite{WM-adversarial-CCS, WM-adversarial-ICLR, WM-adversarial-TCSVT} were excluded due to their unrealistic assumptions (\Cref{sec:watermarking_attacks}). For \DistortionAttack, we employed its publicly available implementation\footnote{\href{https://github.com/umd-huang-lab/WAVES/tree/main/distortions}{https://github.com/umd-huang-lab/WAVES/distortions}} and applied five distortion techniques: brightness, contrast, blurring, Gaussian noise, and compression, using attack strengths as specified in \cite{WAVES}. For the two deep-learning-based attacks, \RegenVAE and \RegenDiffusion, we used their implementation\footnote{\href{https://github.com/XuandongZhao/WatermarkAttacker}{https://github.com/XuandongZhao/WatermarkAttacker}}, employing the pre-trained \texttt{bmshj2018} model \cite{bmshj2018} for \RegenVAE and \texttt{Stable Diffusion v1-4} \cite{Stable-Diffusion-V1-4} for \RegenDiffusion, with a noise step of 60 for the latter per its default settings.

\smallskip
\item [Computing Platform.] All experiments ran on an NVIDIA RTX 4090 GPU with 24GB RAM, using CUDA 12.0.

\smallskip
\item [\ComMark Training Protocols.] \ComMark, comprising a sparse encoder (\SparseEncoder) and an image reconstruction module (\Reconstructor), 
as shown in \Cref{algorithm:commark_training_algorithm},
is trained on \OpenImage and \COCO datasets without access to watermarking schemes or detectors. Training employs the Adam optimizer with a 0.001 initial learning rate over 50 epochs. In \Reconstructor, the sparse and structural-perceptual loss weights in \Cref{eq:total_loss} are set to \(\alpha = 10\) and \(\beta = 0.1\), respectively. Co-training typically converges within 10 epochs for both datasets.

\smallskip
\item [Evaluation Metrics.] Following \cite{WatermarkAttacker, VINE, PTW}, we used two standard metrics: \emph{bit accuracy} (\BitAcc) and \emph{detection accuracy} (\DetectAcc), to evaluate attack effectiveness. \BitAcc quantifies the ratio of correctly recovered watermark bits for a watermarked image, while \DetectAcc, defined as TPR@0.1\%FPR, measures the true positive rate under a stringent false positive constraint. \BitAcc is averaged across all samples and does not necessarily correlate with \DetectAcc \cite{VINE, WatermarkAttacker, StableSignature}. Smaller values indicate better attack performance for both metrics, denoted as \BitAcc\(\downarrow\) and \DetectAcc\(\downarrow\) in our tables.

\end{description}

\begin{table*}[t]
\centering
\captionsetup{font=normalsize}
\caption{{Comparison of \ComMark with \DistortionAttack, \RegenVAE, and \RegenDiffusion attacks on six post-processing watermarking schemes—\MBRS, \CIN, \TrustMark, \PIMoG, \VINEB, and \VINER{}—using two metrics: \BitAcc and \DetectAcc (TPR@0.1\%FPR) (\Cref{subsec:experiment_setup}). The best value in each column is highlighted in \textbf{bold}, while the second best is \underline{underlined}.}
}
\small
\scalebox{0.87}{
\addtolength{\tabcolsep}{-1.45ex}
\begin{tabular}{lc|cc|cc|cc|cc|cc|cc}
\toprule
\multirow{2}{*}{\makecell{Dataset}} 
& \multirow{2}{*}{\makecell{Attack \\ Method}} 
& \multicolumn{2}{c|}{MBRS} 
& \multicolumn{2}{c|}{CIN} & \multicolumn{2}{c|}{TrustMark} & \multicolumn{2}{c|}{PIMoG} 
& \multicolumn{2}{c|}{VINE-B} & \multicolumn{2}{c}{VINE-R} \\
\cmidrule(lr){3-4} \cmidrule(lr){5-6} \cmidrule(lr){7-8} \cmidrule(lr){9-10} \cmidrule(lr){11-12} \cmidrule(lr){13-14}
& & \BitAcc\(\downarrow\) & \DetectAcc\(\downarrow\) & \BitAcc\(\downarrow\) & \DetectAcc\(\downarrow\) & \BitAcc\(\downarrow\) & \DetectAcc\(\downarrow\) & \BitAcc\(\downarrow\) & \DetectAcc\(\downarrow\) 
& \BitAcc\(\downarrow\) & \DetectAcc\(\downarrow\) & \BitAcc\(\downarrow\) & \DetectAcc\(\downarrow\)  \\
\midrule
\multirow{5}{*}{\makecell{OpenImage}} &
No Attack & 99.5\% & 100.0\% & 100.0\% & 100.0\% & 100.0\% & 100.0\% & 100.0\% & 100.0\% & 100.0\% & 100.0\% & 100.0\%  & 100.0\% \\
& Distortion & 84.6\% & 83.0\% & 89.2\% & 80.1\% & 75.7\% & 75.3\% & \underline{85.5\%} & \underline{73.2\%} & \underline{86.0\%} & \underline{73.5\%} & 87.8\% & 78.1\%  \\
& \RegenVAE & \underline{67.2\%} & \underline{73.7\%} & \textbf{56.3\%} & \textbf{6.7\%} & 82.3\% & 82.4\% & 99.6\% & 100.0\% & 97.5\% & 100.0\% & \underline{86.3\%} & \underline{76.3\%}  \\
& \RegenDiffusion
& 96.3\% & 99.4\% & 74.8\% & 51.7\% & \textbf{58.3\%} & \textbf{54.6\%} & 88.4\% & 76.8\% & 92.4\% & 99.3\% & 99.6\% & 100.0\%  \\
\cmidrule{2-14}
\rowcolor{red!10} & \textbf{\ComMark} & \textbf{63.6\%} & \textbf{23.3\%} & \underline{65.2\%} & \underline{21.0\%} & \underline{67.8\%} & \underline{67.9\%} & \textbf{71.5\%} & \textbf{25.1\%} & \textbf{62.1\%} & \textbf{24.0\%} & \textbf{66.5\%}  & \textbf{48.7\%} \\
\midrule
\multirow{5}{*}{\makecell{COCO}} &
No Attack & 99.6\% & 100.0\% & 100.0\% & 100.0\% & 100.0\% & 100.0\% & 100.0\% & 100.0\% & 100.0\% & 100.0\% & 100.0\% & 100.0\%  \\
& Distortion & 84.3\% & 85.0\% & 89.6\% & 80.2\% & 65.6\% & 65.0\% & \underline{85.0\%} & \underline{71.9\%} & \underline{85.8\%} & \underline{72.7\%} & \underline{84.7\%} & \underline{70.6\%}  \\
& \RegenVAE & \underline{67.9\%} & \underline{72.6\%} & \textbf{58.4\%} & \textbf{6.0\%} & 84.6\% & 84.7\% & 99.6\% & 100.0\% & 97.6\% & 100.0\% & 97.5\%  & 100.0\% \\
& \RegenDiffusion
& 98.6\% & 100.0\% & 72.5\% & 46.9\% & \underline{62.4\%} & \underline{61.8\%} & 90.2\% & 83.8\% & 91.6\% & 98.5\% & 99.4\% & 100.0\% \\
\cmidrule{2-14}
\rowcolor{red!10} & \textbf{\ComMark} & \textbf{60.3\%} & \textbf{22.1\%} & \underline{64.3\%} & \underline{18.6\%} & \textbf{57.6\%} & \textbf{57.7\%} & \textbf{72.4\%} & \textbf{26.3\%} & \textbf{63.1\%} & \textbf{24.6\%} & \textbf{66.4\%} & \textbf{49.6\%} \\
\bottomrule
\end{tabular}
}
\label{table:asr_main_table_post}
\vspace*{0ex}
\end{table*}

\begin{table}[t]
\centering
\captionsetup{font=normalsize}
\caption{Comparison of \ComMark with \DistortionAttack, \RegenVAE, and \RegenDiffusion attacks on \BitAcc and \DetectAcc (TPR@0.1\%FPR) (\Cref{subsec:experiment_setup})  for two in-processing watermarking schemes: \PTW and \StableSignature.}
\small
\scalebox{0.93}{
\addtolength{\tabcolsep}{-.3ex}
\begin{tabular}{l|cc|cc}
\toprule
\multirow{2}{*}{\makecell{Attack \\ Method}} 
& \multicolumn{2}{c|}{\PTW}
& \multicolumn{2}{c}{\StableSignature} \\
\cmidrule(lr){2-3} \cmidrule(lr){4-5}
& \BitAcc\(\downarrow\) & \DetectAcc\(\downarrow\) & \BitAcc\(\downarrow\) & \DetectAcc\(\downarrow\)   \\
\midrule
No Attack & 98.0\% & 100.0\% & 100.0\% & 100.0\%  \\
Distortion & 86.1\% & 82.4\% & 87.2\% & 81.2\% \\
\RegenVAE & 83.5\% & 79.8\% & 61.4\% & 46.3\%  \\
\RegenDiffusion & \underline{72.1\%} & \underline{36.5\%} & \underline{58.4\%} & \underline{34.8\%}  \\
\cmidrule{1-5}
\rowcolor{red!10} \textbf{\ComMark} & \textbf{68.3\%} & \textbf{26.4\%} & \textbf{55.9\%} & \textbf{24.6\%}  \\
\bottomrule
\end{tabular}
}
\label{table:asr_main_table_in}
\vspace*{-2ex}
\end{table}

\subsection{RQ1. \ComMark's Attack Capabilities}
\label{subsec:RQ1}

We present that \ComMark, with its default weight settings of \(\alpha = 10\) and \(\beta = 0.1\), significantly outperforms \DistortionAttack \cite{WAVES}, \RegenVAE \cite{WatermarkAttacker}, and \RegenDiffusion \cite{WatermarkAttacker, Regeneration-Attack-Diff1, Regeneration-Attack-Diff2}. Adversarial attacks \cite{WM-adversarial-CCS, WM-adversarial-ICLR, WM-adversarial-TCSVT} were excluded due to their unrealistic assumptions \cite{unmark} (\Cref{sec:watermarking_attacks}).

\Cref{table:asr_main_table_post} highlights \ComMark's superior attack performance against six post-processing schemes—\MBRS, \CIN, \TrustMark, \PIMoG, \VINEB, and \VINER{}—substantially surpassing \DistortionAttack, \RegenVAE, and \RegenDiffusion. On \OpenImage, \ComMark reduces \BitAcc by an average of 21.8\%, 15.8\%, and 19.3\%, and decreases \DetectAcc by an average of 54.4\%, 9.9\%, and 51.0\%, respectively. Similarly, on COCO, \ComMark achieves average reductions of 22.0\%, 17.5\%, and 23.6\% in \BitAcc, and 53.5\%, 15.2\%, and 56.5\% in \DetectAcc, respectively. \ComMark also demonstrates consistent effectiveness against the two in-processing schemes—\PTW and \StableSignature{}—as shown in \Cref{table:asr_main_table_in}, surpassing \DistortionAttack, \RegenVAE, and \RegenDiffusion. On average, \ComMark reduces \BitAcc by 28.3\%, 13.6\%, and 4.8\% and decreases \DetectAcc by 68.8\%, 56.8\%, and 28.5\% compared to \DistortionAttack, \RegenVAE, and \RegenDiffusion, respectively.

\DistortionAttack shows limited effectiveness across all watermarking schemes, as many \cite{VINE, PIMoG, MBRS, PTW, TrustMark, CIN} incorporate adversarial training modules to enhance robustness against common image distortions (\Cref{subsec:experiment_setup}).

\RegenVAE is most effective against \CIN, as its invertible VAE-based module maps watermarked images back to their original counterparts, making it particularly vulnerable to \RegenVAE. However, \RegenVAE proves less effective for other watermarking schemes, or nearly ineffective (e.g., for \PIMoG and \VINEB), because VAE-based regeneration preserves the integrity of image latent representations—the carriers for watermarks—unintentionally retaining embedded watermark along with critical image features.

\RegenDiffusion employs a diffusion denoising to remove watermarks, achieving strong \DetectAcc against in-processing schemes (\Cref{table:asr_main_table_in}) but struggling with most post-processing schemes like \MBRS, \VINEB, and \VINER (\Cref{table:asr_main_table_post}). In-processing schemes embed watermarks during image generation, inherently involving noise addition and denoising, which \RegenDiffusion exploits. In contrast, post-processing schemes embed watermarks after image generation, making them more resistant to denoising-based attacks. For OpenImage, an exception is \TrustMark, as it incorporates an additional denoising process to remove watermarks and recover unwatermarked images (e.g., for re-watermarking), rendering it especially susceptible to \RegenDiffusion.

\ComMark leverages \ICS theory \cite{ICS-CVPR, ICS-TIP, ICS-TPAMI} to sparsify the latent representations of watermarked images, reducing \BitAcc from 100.0\% to 64.7\% and \DetectAcc from 100.0\% to 32.9\% on average across the eight advanced defensive watermarking schemes \cite{MBRS, CIN, TrustMark, PIMoG, VINE, PTW, StableSignature}. In contrast, baseline attacks—\DistortionAttack, \RegenVAE, and \RegenDiffusion{}—achieve average \BitAcc reductions of 84.1\%, 80.7\%, and 82.5\%, and average \DetectAcc reductions of 76.6\%, 73.5\%, and 74.6\%, respectively.

These results highlight \ComMark's effectiveness in weakening the robustness of state-of-the-art watermarking schemes for deepfake detection. The sharp drop in \BitAcc indicates the inability to reliably extract embedded watermark bits from \ComMark-attacked deepfakes, while the decline in \DetectAcc reflects the failure to consistently detect and authenticate watermarks in such content.

\begin{figure*}[ht]
\centering
\begin{subfigure}{0.102\textwidth}
\centering
\includegraphics[width=1.8cm]{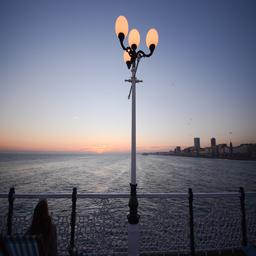}
\end{subfigure}
\hspace*{-1.ex}
\begin{subfigure}{0.55\textwidth}
\centering
\includegraphics[width=1.8cm]{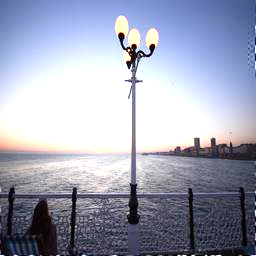}
\includegraphics[width=1.8cm]{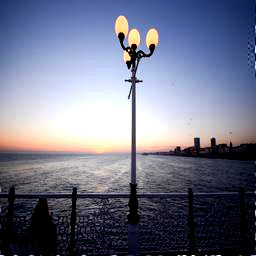}
\includegraphics[width=1.8cm]{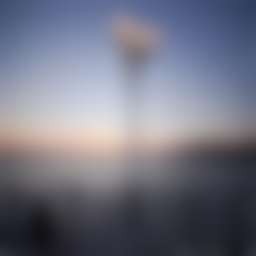}
\includegraphics[width=1.8cm]{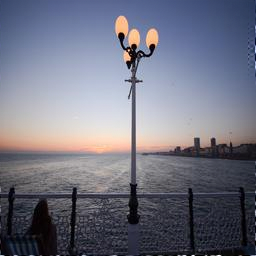}
\includegraphics[width=1.8cm]{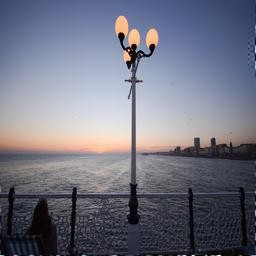}
\end{subfigure}
\hspace*{-1.ex}
\begin{subfigure}{0.109\textwidth}
\centering
\includegraphics[width=1.8cm]{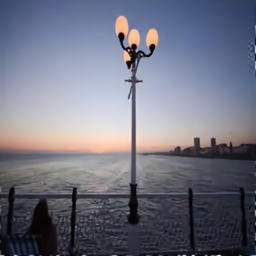}
\end{subfigure}
\hspace*{.0ex}
\begin{subfigure}{0.102\textwidth}
\centering
\includegraphics[width=1.8cm]{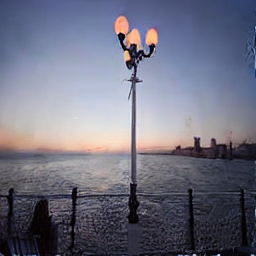}
\end{subfigure}
\hspace*{.0ex}
\begin{subfigure}{0.102\textwidth}
\centering
\includegraphics[width=1.8cm]{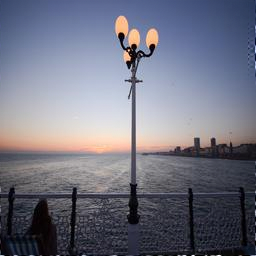}
\end{subfigure}

\begin{subfigure}{0.102\textwidth}
\centering
\includegraphics[width=1.8cm]{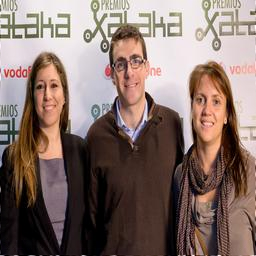}
\subcaption{WMs}
\end{subfigure}
\hspace*{-1.ex}
\begin{subfigure}{0.55\textwidth}
\centering
\includegraphics[width=1.8cm]{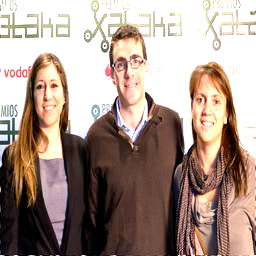}
\includegraphics[width=1.8cm]{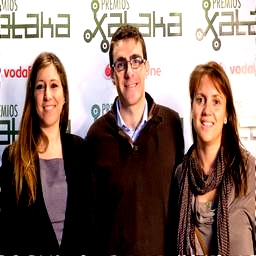}
\includegraphics[width=1.8cm]{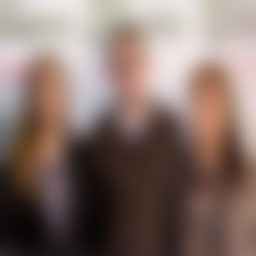}
\includegraphics[width=1.8cm]{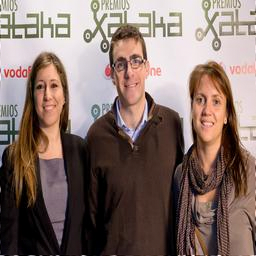}
\includegraphics[width=1.8cm]{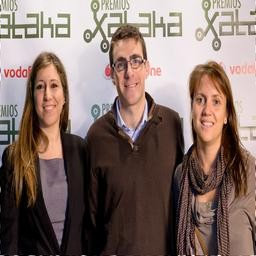}
\caption{\DistortionAttack}
\end{subfigure}
\hspace*{-1.ex}
\begin{subfigure}{0.109\textwidth}
\centering
\includegraphics[width=1.8cm]{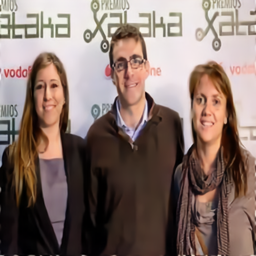}
\subcaption{\RegenVAE}
\end{subfigure}
\hspace*{.0ex}
\begin{subfigure}{0.102\textwidth}
\centering
\includegraphics[width=1.8cm]{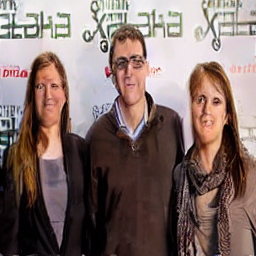}
\subcaption{\RegenDiffusion}
\end{subfigure}
\hspace*{.0ex}
\begin{subfigure}{0.102\textwidth}
\centering
\includegraphics[width=1.8cm]{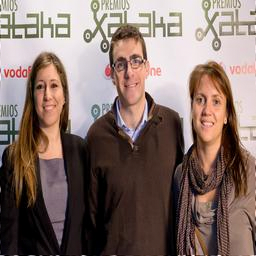}
\subcaption{\ComMark}
\end{subfigure}
\caption{{Illustration of \ComMark, \DistortionAttack (brightness, contrast, blurring, Gaussian noise, and compression in that order), \RegenVAE, and \RegenDiffusion on image integrity using two \VINER-generated watermarked images (WMs) from OpenImage.}}
\label{fig:regenvae_diffusion_commark}
\end{figure*}

\begin{table*}[ht]
\centering
\captionsetup{font=normalsize}
\caption{Comparison of \ComMark with \DistortionAttack, \RegenVAE, and \RegenDiffusion on post-attack image integrity  for six post-processing watermarking schemes, evaluated using four standard metrics: \PSNR, \SSIM, \FID, and \LPIPS.}
\small
\scalebox{0.66}{
\addtolength{\tabcolsep}{-1.45ex}
\begin{tabular}{lc|cccc|cccc|cccc|cccc|cccc|cccc}
\toprule
\multirow{3}{*}{\makecell{Dataset}} 
& \multirow{3}{*}{\makecell{Attack \\ Method}} 
& \multicolumn{4}{c|}{MBRS} 
& \multicolumn{4}{c|}{CIN} & \multicolumn{4}{c|}{TrustMark} & \multicolumn{4}{c|}{PIMoG} 
& \multicolumn{4}{c|}{VINE-B} & \multicolumn{4}{c}{VINE-R} \\
\cmidrule(lr){3-6} \cmidrule(lr){7-10} \cmidrule(lr){11-14} \cmidrule(lr){15-18} \cmidrule(lr){19-22} \cmidrule(lr){23-26} 
& & PSNR\(\uparrow\) & SSIM\(\uparrow\) & FID\(\downarrow\) & LPIPS\(\downarrow\) & PSNR\(\uparrow\) & SSIM\(\uparrow\) & FID\(\downarrow\) & LPIPS\(\downarrow\) & PSNR\(\uparrow\) & SSIM\(\uparrow\) & FID\(\downarrow\) & LPIPS\(\downarrow\) & PSNR\(\uparrow\) & SSIM\(\uparrow\) & FID\(\downarrow\) & LPIPS\(\downarrow\)
& PSNR\(\uparrow\) & SSIM\(\uparrow\) & FID\(\downarrow\) & LPIPS\(\downarrow\) & PSNR\(\uparrow\) & SSIM\(\uparrow\) & FID\(\downarrow\) & LPIPS\(\downarrow\)\\
\midrule
\multirow{4}{*}{\makecell{OpenImage}} 
& Distortion & \underline{29.19} & \underline{0.80} & 1692.83 & 0.16 & \underline{28.77} & 0.79 & 1840.32 & 0.18 & \underline{30.15} & 0.82 & 1547.96 & 0.16 & \textbf{28.08} & 0.79 & 1822.96 & 0.17 & \textbf{29.94} & \underline{0.81} & 1654.71 & 0.17 & \textbf{29.90} & \underline{0.82} & 1601.09 & 0.16 \\
& \RegenVAE
& 26.67 & 0.79 & 36.98 & \textbf{0.06} & \textbf{28.95} & \textbf{0.86} & \underline{43.54} & \underline{0.07} & \textbf{30.52} & \textbf{0.90} & 29.43 & \textbf{0.06} & \underline{26.96} & \underline{0.81} & 49.37 & \textbf{0.06} & \underline{28.67} & 0.80 & 33.86 & \textbf{0.06} & 29.33 & \textbf{0.88} & 30.64 & \textbf{0.06} \\
& \RegenDiffusion 
& 22.24 & 0.65 & \textbf{12.78} & \underline{0.13} & 20.61 & 0.58 & 45.43 & 0.14 & 22.27 & 0.65 & \textbf{11.87} & \underline{0.13} & 19.99 & 0.57 & \underline{45.28} & 0.15 & 22.06 & 0.65 & \textbf{13.16} & \underline{0.13} & 22.43 & 0.67 & \textbf{15.86} & \underline{0.10} \\
\cmidrule{2-26}
\rowcolor{red!10} & \textbf{\ComMark} & \textbf{29.63} & \textbf{0.88} & \underline{22.49} & \textbf{0.06} & 28.49 & \underline{0.83} & \textbf{30.95} & \textbf{0.06} & 28.41 & \underline{0.87} & \underline{23.59} & \textbf{0.06} & 26.32 & \textbf{0.83} & \textbf{37.48} & \underline{0.07} & 27.68 & \textbf{0.87} & \underline{23.77} & \textbf{0.06} & \underline{29.47} & \textbf{0.88} & \underline{23.81} & \textbf{0.06} \\
\midrule
\multirow{4}{*}{\makecell{COCO}} 
& Distortion & \underline{27.55} & 0.79 & 1767.71 & 0.16 & \underline{27.98} & 0.79 & 1880.79 & 0.17 & 27.76 & \underline{0.80} & 1812.16 & 0.16 & \underline{27.72} & 0.79 & 1860.57 & 0.17 & \textbf{27.52} & \underline{0.80} & 1807.62 & 0.17 & \underline{26.83} & \underline{0.79} & 1843.06 & 0.17 \\
& \RegenVAE
& 26.96 & \underline{0.82} & 40.63 & \textbf{0.06} & \textbf{29.04} & \textbf{0.85} & 48.47 & \textbf{0.05} & \textbf{29.41} & \textbf{0.83} & 35.43 & \textbf{0.06} & \textbf{30.15} & \underline{0.80} & \underline{44.87} & \textbf{0.06} & 21.72 & 0.76 & 37.95 & \underline{0.09} & 23.26 & 0.76 & 45.94 & \underline{0.10} \\
& \RegenDiffusion 
& 21.48 & 0.61 & \textbf{17.31} & \underline{0.14} & 22.53 & 0.65 & \underline{46.73} & 0.14 & 21.53 & 0.61 & \textbf{15.90} & 0.13 & 19.83 & 0.56 & 46.74 & 0.15 & 21.36 & 0.61 & \textbf{17.85} & 0.14 & 21.25 & 0.60 & \textbf{18.11} & 0.14 \\
\cmidrule{2-26}
 \rowcolor{red!10} & \textbf{\ComMark} & \textbf{28.42} & \textbf{0.86} & \underline{23.59} & \textbf{0.06} & 26.08 & \underline{0.80} & \textbf{38.73} & \underline{0.07} & \underline{29.27} & \textbf{0.83} & \underline{27.81} & \underline{0.07} & 26.50 & \textbf{0.82} & \textbf{40.28} & \underline{0.07} & \underline{26.42} & \textbf{0.83} & \underline{27.91} & \textbf{0.07} & \textbf{28.68} & \textbf{0.83} & \underline{27.89} & \textbf{0.07} \\
\bottomrule
\end{tabular}
}
\label{table:quality_main_table}
\end{table*}

\begin{table}[ht]
\centering
\captionsetup{font=normalsize}
\caption{{Comparison of \ComMark with \DistortionAttack, \RegenVAE, and \RegenDiffusion on post-attack image integrity against two in-processing schemes—with respect to the four standard metrics: \PSNR, \SSIM, \FID, and \LPIPS.}}
\small
\scalebox{0.85}{
\addtolength{\tabcolsep}{-1.35ex}
\begin{tabular}{l|cccc|cccc}
\toprule
\multirow{2}{*}{\makecell{Attack \\ Method}} 
& \multicolumn{4}{c|}{\PTW}
& \multicolumn{4}{c}{\StableSignature} \\
\cmidrule(lr){2-5} \cmidrule(lr){6-9}
& PSNR\(\uparrow\) & SSIM\(\uparrow\) & FID\(\downarrow\) & LPIPS\(\downarrow\) & PSNR\(\uparrow\) & SSIM\(\uparrow\) & FID\(\downarrow\) & LPIPS\(\downarrow\)   \\
\midrule
Distortion & 30.22 & \underline{0.87} & 1950.72 & 0.18 & \underline{29.71} & 0.83 & 1508.93 & 0.16 \\
\RegenVAE & 30.80 & \textbf{0.90} & \textbf{16.53} & \textbf{0.05} & \textbf{33.23} & \textbf{0.89} & 23.65 & \textbf{0.06}   \\
\RegenDiffusion & 21.56 & 0.60 & 22.35 & 0.15 & 21.73 & 0.63 & \textbf{20.62} & 0.14   \\
\cmidrule{2-9}
\rowcolor{red!10} \textbf{\ComMark} & \textbf{32.46} & \underline{0.87} & \underline{21.32} & \underline{0.06} & 29.43 & \underline{0.88} & \underline{23.35} & \textbf{0.06}   \\
\bottomrule
\end{tabular}
}
\label{table:quality_main_table_in}
\vspace*{-2ex}
\end{table}

\subsection{RQ2. \ComMark's Image Integrity}
\label{subsec:RQ2}

\Cref{table:quality_main_table,table:quality_main_table_in} compare \ComMark with three baseline attacks—\DistortionAttack, \RegenVAE, and \RegenDiffusion{}—on post-attack image integrity using four standard metrics: \PSNR, \SSIM, \FID and \LPIPS. While \RegenVAE achieves the highest overall image integrity, it does so at the expense of significantly reduced attack effectiveness (\Cref{table:asr_main_table_post,table:asr_main_table_in}), as discussed in \Cref{subsec:RQ1}. In contrast, \ComMark outperforms all three methods in attack effectiveness while maintaining comparable or superior image integrity to \RegenVAE.

Following \cite{WAVES, WatermarkAttacker, VINE}, we use \PSNR \cite{psnr}, \SSIM \cite{SSIM-TIP15}, \FID \cite{FID-CVPR20}, and \LPIPS \cite{LPIPS} to assess post-attack image integrity, with \PSNR and \SSIM measuring quality, \FID and \LPIPS reflecting fidelity. \PSNR quantifies pixel-level similarity (higher is better), \SSIM evaluates structural similarity (0 to 1, higher is better), \FID measures Fréchet distance of feature distribution (lower is better), and \LPIPS assess the similarity of learned deep features (0 to 1, lower is better).

We analyze the results in \Cref{table:quality_main_table,table:quality_main_table_in}, where \PSNR~\(\uparrow\), \SSIM~\(\uparrow\), \FID~\(\downarrow\), and \LPIPS~\(\downarrow\) indicate whether higher or lower values are preferred. 
\DistortionAttack, which applies brightness, contrast, blurring, Gaussian noise, and compression (\Cref{sec:watermarking_attacks}), severely degrades image fidelity, as shown in \Cref{fig:regenvae_diffusion_commark}
(main paper) and 
Figures~\ref{fig:regenvae_diffusion_commark_mbrs}--\ref{fig:regenvae_diffusion_commark_PTW} (\appref{sec:appendix_visual_examples_for_all_schmes}), yielding \FID\ values above 1,500 across all schemes. 
Such high FID scores reflect the severe fidelity loss caused by heavy distortions (e.g., blurring), which shift image distributions away from natural ones~\cite{FID-CVPR20}. 
On average, \DistortionAttack\ achieves a \PSNR\ of 28.66, \SSIM\ of 0.81, \FID\ of 1,750.10, and \LPIPS\ of 0.17 across the eight watermarking schemes.

While \RegenDiffusion achieves the best \FID, averaging 23.40 across all schemes and showing excellent preservation of feature distribution, it performs poorly in \PSNR and \SSIM, with averages of 21.65 and 0.62, respectively. This indicates significant visual degradation. Furthermore, its average \LPIPS of 0.14, higher than those of \RegenVAE and \ComMark, reflects notable changes in deep feature values. The denoising process in \RegenDiffusion compromises pixel-level details and structural information, leading to noticeable quality loss as evidenced in \Cref{fig:regenvae_diffusion_commark} (main paper) and Figures~\ref{fig:regenvae_diffusion_commark_mbrs}--\ref{fig:regenvae_diffusion_commark_PTW} (\appref{sec:appendix_visual_examples_for_all_schmes}). The diffusion model particularly struggles with human subjects \cite{StableSignature}, failing to preserve complex shapes and details, resulting in significant distortions.

\RegenVAE excels at preserving image integrity, achieving an average \PSNR of 29.74, \SSIM of 0.85, \FID of 33.74, and \LPIPS of 0.06. VAEs \cite{VAE} encode images into latent representations with probabilistic noise, capturing both critical image features and undesired watermark-related attributes. While this maintains high image integrity during reconstruction, it significantly compromises attack effectiveness, as evidenced in \Cref{table:asr_main_table_post,table:asr_main_table_in}.

In contrast, \ComMark achieves an average \PSNR of 28.45, \SSIM of 0.85, \FID of 28.07, and \LPIPS of 0.06, notably surpassing \DistortionAttack and \RegenDiffusion in image integrity by maintaining high quality and fidelity. Leveraging \ICS theory \cite{ICS-CVPR, ICS-TIP, ICS-TPAMI}, \ComMark sparsely encodes watermarked images into latent representations, preserving crucial image features while removing watermark-related features. While minor visual deviations are acceptable since these are deepfakes rather than authentic artworks, \ComMark nonetheless preserves natural appearance and perceptual quality. This results in comparable or superior image integrity metrics: \PSNR, \SSIM, \FID, and \LPIPS (\Cref{table:quality_main_table,table:quality_main_table_in}), with visual demonstrations in \Cref{fig:regenvae_diffusion_commark} (main paper) and Figures~\ref{fig:regenvae_diffusion_commark_mbrs} --\ref{fig:regenvae_diffusion_commark_PTW} (\appref{sec:appendix_visual_examples_for_all_schmes}), alongside significantly better attack performance (\Cref{table:asr_main_table_post,table:asr_main_table_in}).

\begin{table}[t]
\centering
\captionsetup{font=normalsize}
\caption{Comparison of attack times per image between \ComMark and \DistortionAttack, \RegenVAE, and \RegenDiffusion, averaged over 100 samples per image size.}
\scalebox{0.95}{
\begin{tabular}{cccc>{\columncolor{red!10}}c}
\toprule
Image Size & Distortion & \RegenVAE & \RegenDiffusion & \ComMark \\
\cmidrule(lr){1-5}
\multirow{1}{*}{$256\times 256$} & \textbf{0.03s} & 0.75s & 2.73s & \underline{0.56s} \\
\cmidrule(lr){1-5}
\multirow{1}{*}{$512\times 512$} & \textbf{0.11s} & 0.79s & 3.30s & \underline{0.58s} \\
\bottomrule
\end{tabular}
}
\label{table:comparison_time}
\vspace*{0ex}
\end{table}

\subsection{RQ3. \ComMark's Computational Costs}  
\label{subsec:RQ3}

Computational costs are crucial for assessing the practicality of an attack, as excessive time or memory usage limits applicability. We compare \ComMark with \DistortionAttack, \RegenVAE, and \RegenDiffusion in terms of average attack time per watermarked image and memory consumption. Results, averaged over 100 samples for each image size ($256 \times 256$ for post-processing and $512 \times 512$ for in-processing schemes, per \Cref{subsec:experiment_setup}), are reported for both configurations.

\Cref{table:comparison_time,table:comparison_memory} summarize our findings. \DistortionAttack \cite{WAVES} is fast and memory-efficient due to its reliance on simple image editing instead of deep learning. We therefore focus on the three deep-learning-based methods: \ComMark, \RegenVAE \cite{WatermarkAttacker}, and \RegenDiffusion \cite{WatermarkAttacker, Regeneration-Attack-Diff1, Regeneration-Attack-Diff2}. Among them, \ComMark is the most efficient. For $256\times 256$ images, it reduces memory usage by 2.4\(\times\) and 14.9\(\times\), and improves speed by 1.3\(\times\) and 4.9\(\times\), compared to \RegenVAE and \RegenDiffusion. For $512\times 512$ images, the gains are similar: 2.2\(\times\) and 13.5\(\times\) in memory, and 1.4\(\times\) and 5.7\(\times\) in speed.

\ComMark's efficiency stems from the sparsity concept in \ICS, enabling it to use only four convolutional layers in both its sparse encoding and reconstruction modules. By focusing on the essential tasks of watermark disruption and image reconstruction, \ComMark avoids unnecessary operations found in \RegenVAE and \RegenDiffusion. \RegenVAE relies on complex variational networks with downsampling, upsampling, and probabilistic modelling, increasing computational and memory demands. \RegenDiffusion uses iterative denoising processes, typical of diffusion models, leading to significantly higher runtimes and memory use. This makes \ComMark a more practical and scalable attack method.

\begin{table}[t]
\centering
\captionsetup{font=normalsize}
\caption{Comparison of memory usage per image between \ComMark and \DistortionAttack, \RegenVAE, and \RegenDiffusion, averaged over 100 samples per image size.}
\scalebox{0.95}{
\begin{tabular}{cccc>{\columncolor{red!10}}c}
\toprule
Image Size & Distortion & \RegenVAE & \RegenDiffusion & \ComMark \\
\cmidrule(lr){1-5}
\multirow{1}{*}{$256\times 256$} & \textbf{0.57 MB} & 396.87 MB & 2469.34 MB & \underline{165.75 MB} \\
\cmidrule(lr){1-5}
\multirow{1}{*}{$512\times 512$} & \textbf{0.90 MB} & 398.38 MB & 2472.34 MB & \underline{183.68 MB} \\
\bottomrule
\end{tabular}
}
\label{table:comparison_memory}
\vspace*{-2ex}
\end{table}

\begin{table*}[t]
\centering
\captionsetup{font=normalsize}
\caption{\ComMark's attack performance with \ISR mitigation (\ComMark-\ISR) across four post-processing watermarking schemes—\CIN, \PIMoG, \VINEB, and \VINER{}—evaluated using \BitAcc and \DetectAcc (TPR@0.1\%FPR) (\Cref{subsec:experiment_setup}).}
\small
\scalebox{.88}{
\addtolength{\tabcolsep}{-.ex}
\begin{tabular}{ll|cc|cc|cc|cc}
\toprule
\multirow{2}{*}{\makecell{Dataset}} 
& \multirow{2}{*}{\makecell{Attack Method}} 
& \multicolumn{2}{c|}{CIN}
& \multicolumn{2}{c|}{PIMoG} 
& \multicolumn{2}{c|}{VINE-B} 
& \multicolumn{2}{c}{VINE-R} \\
\cmidrule(lr){3-4} \cmidrule(lr){5-6} \cmidrule(lr){7-8} \cmidrule(lr){9-10}
& & \BitAcc\(\downarrow\) & \DetectAcc\(\downarrow\) 
& \BitAcc\(\downarrow\) & \DetectAcc\(\downarrow\) 
& \BitAcc\(\downarrow\) & \DetectAcc\(\downarrow\) 
& \BitAcc\(\downarrow\) & \DetectAcc\(\downarrow\) \\
\midrule
\multirow{3}{*}{\makecell{OpenImage}} &
No Attack & 100.0\% & 100.0\% & 100.0\% & 100.0\% & 100.0\% & 100.0\% & 100.0\%  & 100.0\% \\
& \ComMark & \underline{65.2\%} & \underline{21.0\%} & \underline{71.5\%} & \underline{25.1\%} & \underline{62.1\%} & \underline{24.0\%} & \underline{66.5\%}  & \underline{48.7\%} \\
\cmidrule{2-10}
\rowcolor{red!10} & \ComMark-ISR & \textbf{45.5\%} & \textbf{0.5\%} & \textbf{62.9\%} & \textbf{1.4\%} & \textbf{61.1\%} & \textbf{11.3\%} & \textbf{64.7\%}  & \textbf{34.9\%} \\
\midrule
\multirow{3}{*}{\makecell{COCO}} &
No Attack & 100.0\% & 100.0\% & 100.0\% & 100.0\% & 100.0\% & 100.0\% & 100.0\% & 100.0\% \\
& \ComMark & \underline{64.3\%} & \underline{18.6\%} & \underline{72.4\%} & \underline{26.3\%} & \underline{63.1\%} & \underline{24.6\%} & \underline{66.4\%}  & \underline{49.6\%} \\
\cmidrule{2-10}
\rowcolor{red!10} & \ComMark-ISR & \textbf{44.9\%} & \textbf{0.2\%} & \textbf{62.9\%} & \textbf{1.7\%} & \textbf{61.6\%} & \textbf{14.4\%} & \textbf{64.9\%}  & \textbf{35.6\%} \\

\bottomrule
\end{tabular}
}
\label{table:asr_super_resolution_post}
\vspace*{-1ex}
\end{table*}

\subsection{RQ4. Mitigations Against \ComMark}
\label{subsec:RQ5}

Effective mitigations should enhance watermark robustness, allowing providers to reliably use defensive watermarking for deepfake detection (\Cref{sec:threat_model}). 
We evaluate three strategies: two new methods—image super-resolution and sparse watermarking—and traditional adversarial training~\cite{AT-1, AT-2, AT-3}. 
However, all three prove largely ineffective against \ComMark, underscoring the urgent need to develop stronger and more resilient watermarking defenses.

\begin{description}[leftmargin=0pt]
\setlength{\parindent}{1em}  
\setlength{\parskip}{0pt}

\item[Image Super-Resolution (\ISR).] \ISR enhances fine details by reconstructing high-resolution images from low-resolution inputs \cite{Image-SR-ICCV2023, Image-SR-TPAMI2020, Image-SR-TPAMI2022}. We explore \ISR as a mitigation strategy against \ComMark, marking, to the best of our knowledge, the first application of \ISR to counter watermarking attacks. We utilize the enhanced super-resolution generative adversarial network (\ESRGAN) \cite{ESRGAN} with its pre-trained \texttt{RRDBNet} model\footnote{\href{https://github.com/xinntao/ESRGAN}{https://github.com/xinntao/ESRGAN}}, configured with default settings, including 23 residual blocks and a scaling factor of 4.

\emph{We evaluated \ComMark's performance under image super-resolution (\ComMark-\ISR) on four post-processing schemes—\CIN, \PIMoG, \VINEB, and \VINER{}—each supporting arbitrary resolutions.} 
For 2,000 images from both \OpenImage\ and \COCO\ datasets (\(256 \times 256\), \Cref{subsec:experiment_setup}), \ISR\ was applied after \ComMark's attack to upscale the images to \(1024 \times 1024\) (scaling factor 4). 
This setup assesses whether \ISR\ reduces \ComMark-\ISR's attack effectiveness compared to \ComMark.

Interestingly, instead of mitigating \ComMark, \ISR amplifies its attack effectiveness. As shown in \Cref{table:asr_super_resolution_post}, across the four watermarking schemes—\CIN, \PIMoG, \VINEB, and \VINER{}—\ComMark reduces \BitAcc and \DetectAcc from $100.0\%$ (without attacks) to $66.4\%$ and $29.7\%$, on average. However, with \ComMark-ISR, these metrics drop further to $58.6\%$ and $12.1\%$,  thereby exacerbating the attack's impact.  
While \ISR restores fine details and enhances image integrity \cite{Image-SR-ICCV2023, Image-SR-TPAMI2020, Image-SR-TPAMI2022}, as shown in \Cref{fig:commark_isr}  (\appref{sec:appendix_isr}), its application to \ComMark-attacked images fails to mitigate the attack and instead makes the embedded watermarks even harder to recover. Furthermore, \ISR disperses residual watermark traces by introducing new details and altering pixel structures, complicating extraction by watermark detectors.

\smallskip
\item[Sparse Watermarking (SW).] Since \ComMark exploits sparsifying image latent representations that carry watermarks in encoder-decoder-based schemes, enhancing such schemes by incorporating a sparse encoding loss into the training of the encoder \(\mathcal{E}\), decoder \(\mathcal{D}\), and detector \(\mathcal{W}\) (\Cref{subsec:invisible_image_watermarking}), 
may potentially mitigate \ComMark.

When embedding a watermark \(m\) into an image \(x\) to produce the watermarked output \(x_m = \mathcal{D}(\mathcal{E}(x, m))\) (\Cref{subsec:invisible_image_watermarking}), existing schemes typically employ an integrity loss \(\mathcal{L}_{\text{integrity}}\) (e.g., perceptual loss) to ensure similarity between \(x\) and \(x_m\); a watermark extraction loss \(\mathcal{L}_{\text{extraction}}\) (e.g., BCE \cite{VINE} or distance-based loss \cite{TreeRing}) to align the embedded and extracted watermarks \(m\) and \(m'\); and a GAN loss \(\mathcal{L}_\text{GAN}\) to improve robustness against noise via adversarial learning \cite{VINE, TrustMark, PIMoG, MBRS, StableSignature}. We extend this formulation with our sparse encoding loss \(\mathcal{L}_{\text{SEL}}\) (\Cref{eq:sparse_encoding_loss})—applied to the latent representation \(\mathcal{E}(x, m)\) to enforce sparsity. The total loss for a new sparse watermarking (SW) scheme becomes:
\begin{equation}
    \small
    \label{eq:sparse_watermarking_loss}
    \hspace*{-1.5ex}
    \begin{aligned}
        \mathcal{L}_\text{ALL} = &  \ \lambda_\text{integrity}\mathcal{L}_{\text{integrity}}(x, x_m) 
         + \lambda_\text{extraction}\mathcal{L}_{\text{extraction}}(m, m') \\
        & + \lambda_\text{GAN}\mathcal{L}_\text{GAN}(x, x_m) 
         + {\lambda_\text{sparse}\mathcal{L}_{\text{SEL}}(\mathcal{E}(x, m))}
    \end{aligned}
\end{equation}
The composite loss \(\mathcal{L}_\text{ALL}\) balances image integrity, watermark extraction accuracy, and latent sparsity. However, training SW with \(\mathcal{L}_\text{ALL}\) is challenging, as \(\mathcal{L}_\text{SEL}\) aims to reduce redundancy, while \(\mathcal{L}_\text{extraction}\) and \(\mathcal{L}_\text{integrity}\) strive to retain watermark information and image integrity, limiting latent representations' effective capacity.

To evaluate \SW's effectiveness, we extended the post-processing watermarking scheme \PIMoG \cite{PIMoG} into a sparse variant, \PIMoG-\SW, using the composite loss \(\mathcal{L}_\text{ALL}\). Specifically, \(\mathcal{L}_\text{SEL}\) is from \Cref{eq:sparse_encoding_loss}; \(\mathcal{L}_\text{integrity} = \mathcal{L}_\text{perceptual} = \frac{1}{HWC} \left\| x_m - x \right\|_2^2\); and \(\mathcal{L}_\text{extraction} = \mathcal{L}_\text{BCE}\), where \(\mathcal{L}_\text{BCE}\) measures bitwise watermark similarity \cite{VINE, MBRS, PIMoG}. We also apply the standard adversarial loss \cite{VINE, CIN, MBRS}:
\begin{equation}
\small
\mathcal{L}_{\text{GAN}} = \mathbb{E}_{x} \left[ \log D_{\text{disc}}(x) \right] + \mathbb{E}_{x_m} \left[ \log \left( 1 - D_{\text{disc}}(x_m) \right) \right]
\end{equation}
where the discriminator \(D_{\text{disc}}\) learns to distinguish \(x\) from \(x_m\). This setup promotes robustness by pushing \(x_m\) closer to \(x\) in appearance while preserving the embedded watermark.

When training \PIMoG-\SW, we adhered to \PIMoG's default configuration of 100 training epochs. The weight hyperparameters in \(\mathcal{L}_\text{ALL}\) were set to \(\lambda_\text{integrity} = 1\), \(\lambda_\text{extraction} = 3\), \(\lambda_\text{GAN} = 0.001\), and \(\lambda_\text{sparse} = 10\) to balance the contributions for the four different loss components. Training was conducted using the same 50,000 images selected from the \OpenImage dataset, as detailed in \Cref{subsec:experiment_setup}.

We evaluated \PIMoG-\SW with and without \ComMark attacks, using the same 2,000 images as for \PIMoG (\Cref{subsec:RQ1}). Without attacks, \PIMoG-\SW achieves 80.7\% \BitAcc and 89.6\% \DetectAcc on average, compared to \PIMoG's 100.0\% for both metrics (\Cref{table:asr_main_table_post}), illustrating the challenges of training a sparse watermarking scheme with \(\mathcal{L}_\text{ALL}\). Under \ComMark, \PIMoG-\SW achieves 73.4\% \BitAcc and 35.8\% \DetectAcc on average, with \BitAcc and \DetectAcc higher than \PIMoG's 71.5\% and 25.1\%. Despite the inclusion of \(\mathcal{L}_\text{SEL}\), \PIMoG-\SW's overall performance remains subpar, allowing \ComMark to consistently outperform three state-of-the-art attacks (\Cref{table:asr_main_table_post}). While \(\mathcal{L}_\text{SEL}\) reduces redundancy, balancing image integrity (\(\mathcal{L}_\text{integrity}\)) and watermark retention (\(\mathcal{L}_\text{extraction}\)) forces a watermark to rely on less significant latent components. In contrast, \ComMark prioritizes essential image features, achieving an average \PSNR of 26.73, \SSIM of 0.83, \FID of 36.94, and \LPIPS of 0.07 while effectively eliminating redundant latent components (\Cref{eq:total_loss}) to disrupt sparsified watermarks.

\smallskip
\item[Adversarial Training (\AT).] \AT \cite{AT-1, AT-2, AT-3} is a common mitigation strategy that improves watermark detector robustness by retraining on adversarial examples. However, it is resource-intensive, requiring large-scale adversarial data generation and re-training \cite{unmark}. To evaluate its effectiveness, we fine-tuned \StableSignature's detector with 500 \ComMark-attacked images, following its guideline, and tested on 100 additional samples. \BitAcc rose from 55.9\% to 59.4\%, and \DetectAcc from 24.6\% to 32.3\%, but these gains remain limited. \ComMark still significantly outperforms prior attacks against \StableSignature (\Cref{table:asr_main_table_in}). \AT's limited impact stems from its assumption that attacked samples retain the watermark \cite{unmark}, which does not hold for \ComMark, and from its reliance on access to the watermark detector, reducing real-world applicability.

\end{description}

\begin{figure*}[t]
\centering
\begin{subfigure}{0.117\textwidth}
\centering
\includegraphics[width=2.26cm]{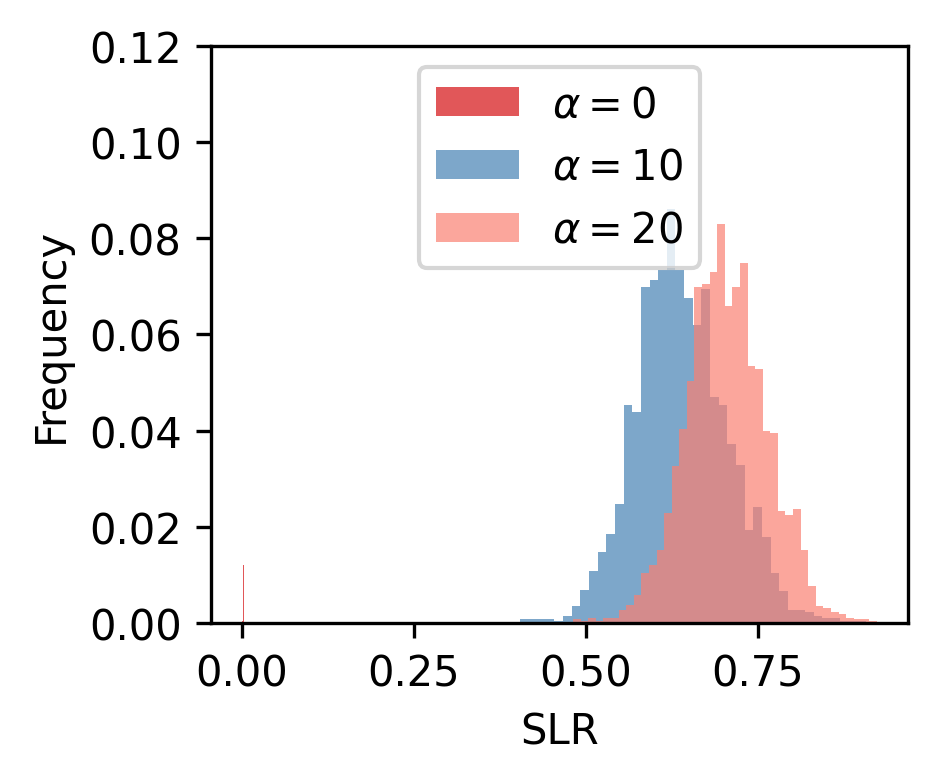}
\subcaption{\MBRS}
\end{subfigure}
\begin{subfigure}{0.117\textwidth}
\centering
\includegraphics[width=2.26cm]{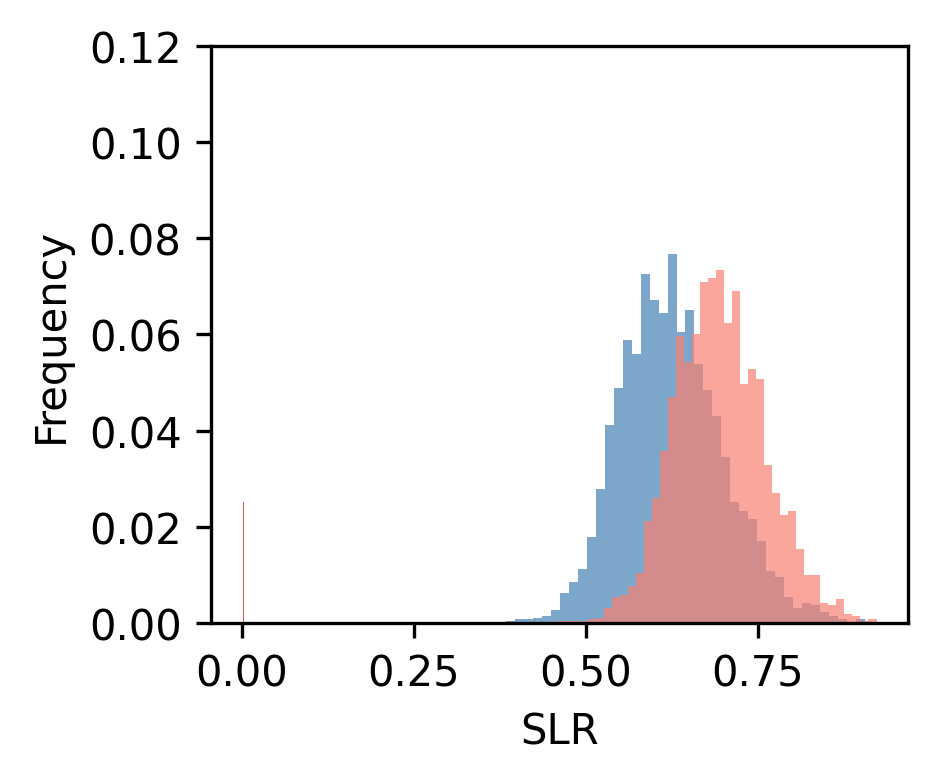}
\subcaption{\CIN}
\end{subfigure}
\begin{subfigure}{0.117\textwidth}
\centering
\includegraphics[width=2.26cm]{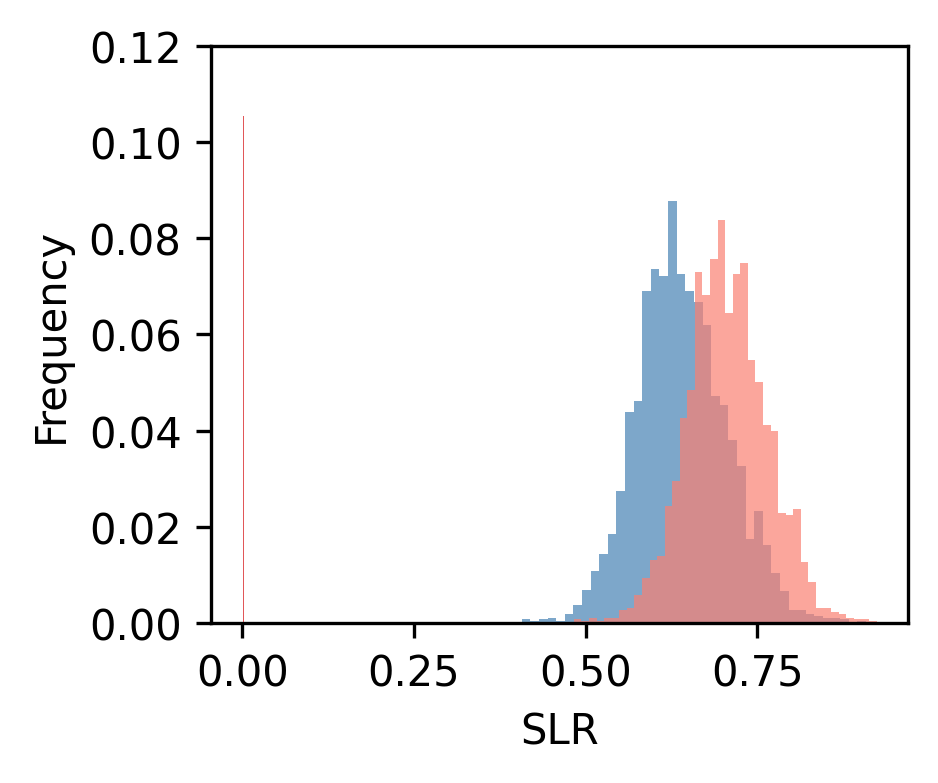}
\subcaption{\TrustMark}
\end{subfigure}
\begin{subfigure}{0.117\textwidth}
\centering
\includegraphics[width=2.26cm]{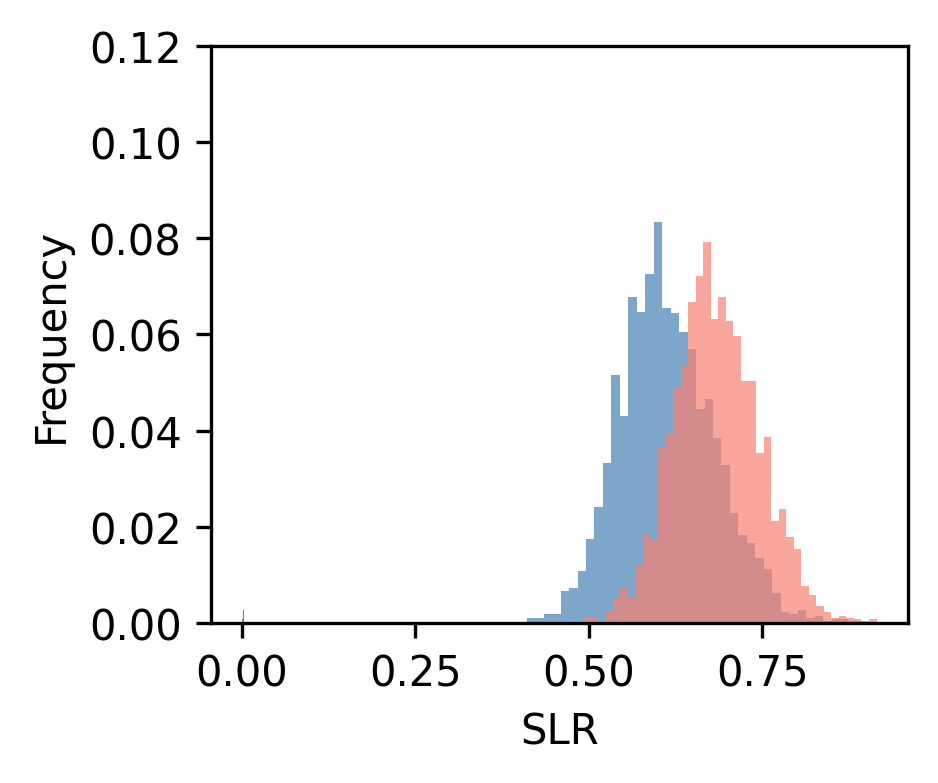}
\subcaption{\PIMoG}
\end{subfigure}
\begin{subfigure}{0.117\textwidth}
\centering
\includegraphics[width=2.26cm]{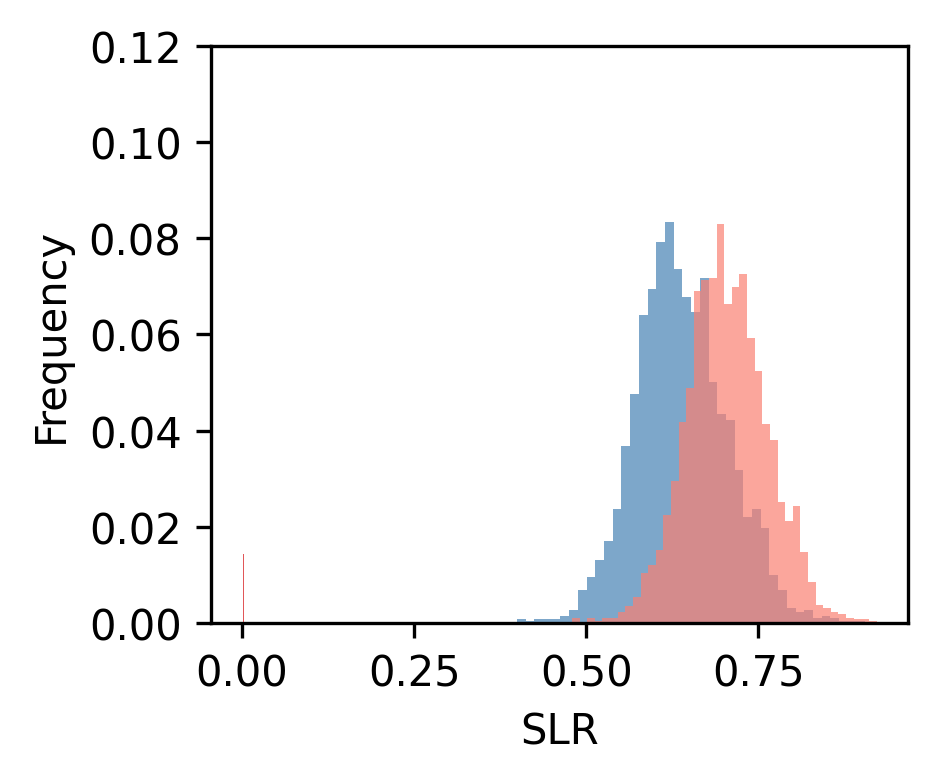}
\subcaption{\VINEB}
\end{subfigure}
\begin{subfigure}{0.117\textwidth}
\centering
\includegraphics[width=2.26cm]{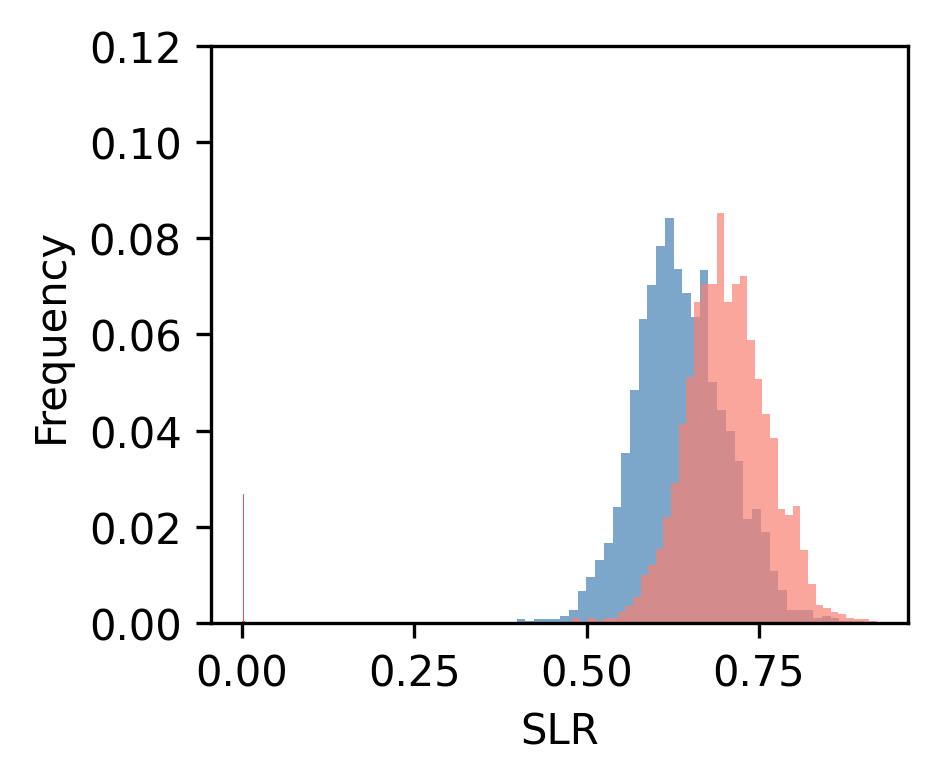}
\subcaption{\VINER}
\end{subfigure}
\begin{subfigure}{0.117\textwidth}
\centering
\includegraphics[width=2.26cm]{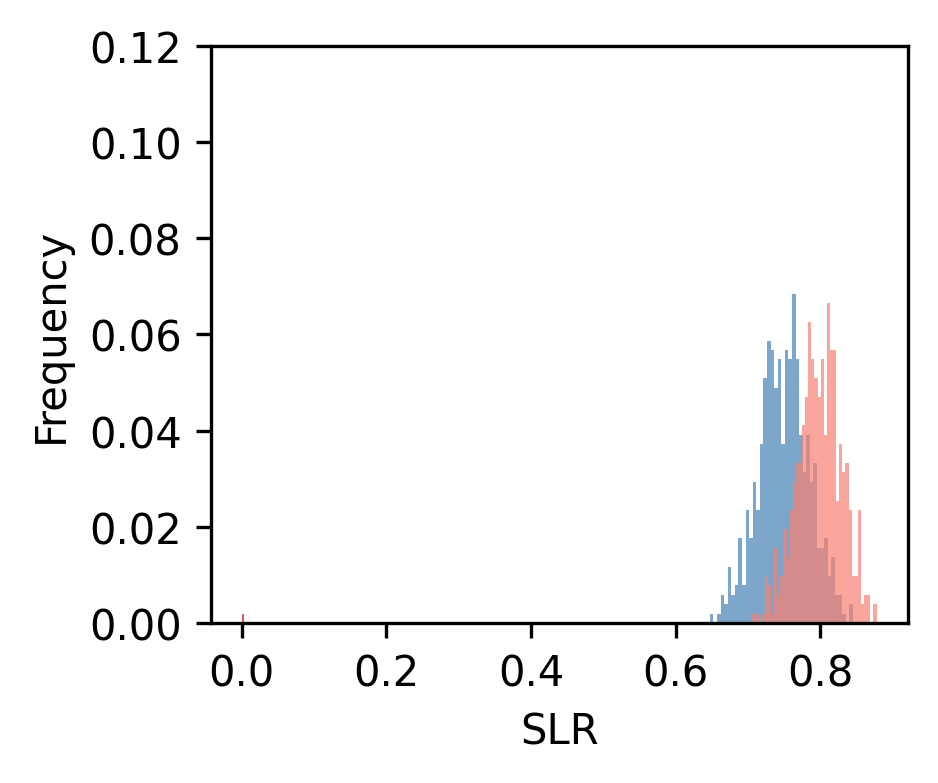}
\subcaption{\PTW}
\end{subfigure}
\begin{subfigure}{0.117\textwidth}
\centering
\includegraphics[width=2.26cm]{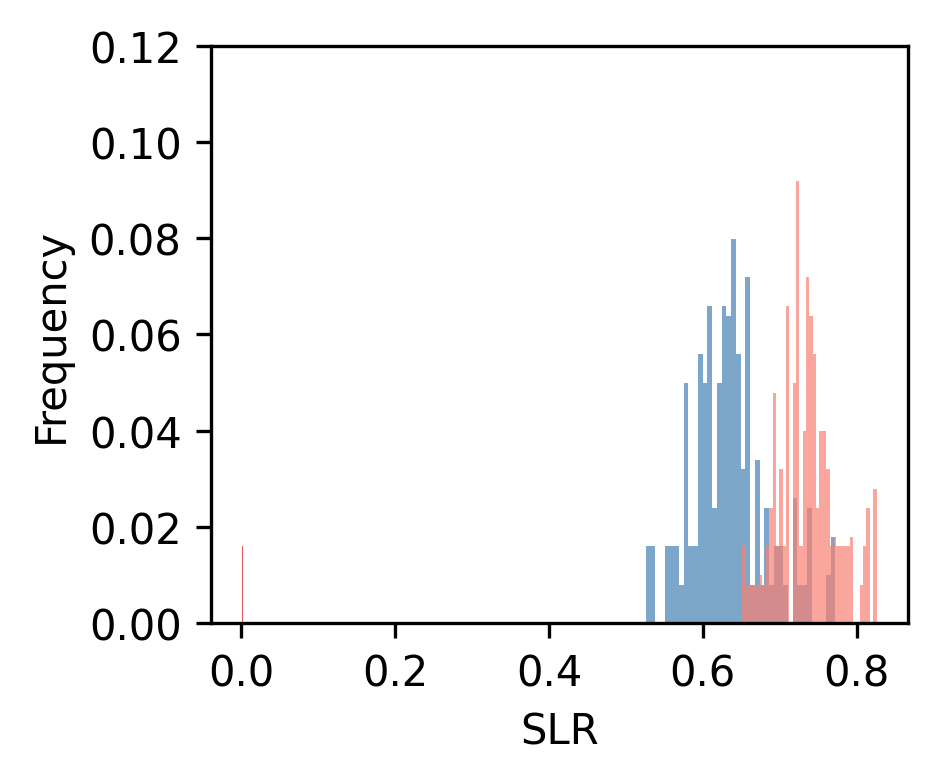}
\subcaption{SS}
\end{subfigure}
\caption{{
{Percentage distributions of watermarked images (y-axis: Frequency) across latent representation sparsity levels (x-axis: SLR) (\Cref{eq:sparsity_level}) for \ComMark under three \(\alpha\) settings: \(\alpha=0\) (no sparsity), \(\alpha=10\) (moderate sparsity), and \(\alpha=20\) (moderately high sparsity). Results include watermarked images from the eight watermarking schemes (\Cref{subsec:experiment_setup}).}
\label{fig:latent_sparsity_post}}}
\vspace*{-2ex}
\end{figure*}

\begin{figure*}[ht]
\centering
\begin{subfigure}{0.117\textwidth}
\centering
\includegraphics[width=2.31cm]{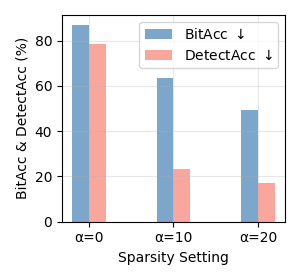}
\subcaption{\MBRS}
\end{subfigure}
\begin{subfigure}{0.117\textwidth}
\centering
\includegraphics[width=2.31cm]{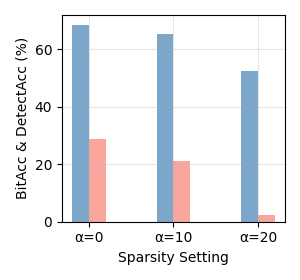}
\subcaption{\CIN}
\end{subfigure}
\begin{subfigure}{0.117\textwidth}
\centering
\includegraphics[width=2.31cm]{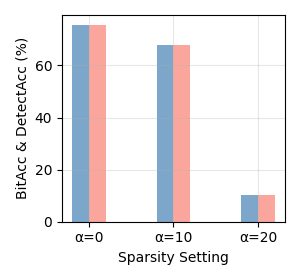}
\subcaption{\TrustMark}
\end{subfigure}
\begin{subfigure}{0.117\textwidth}
\centering
\includegraphics[width=2.31cm]{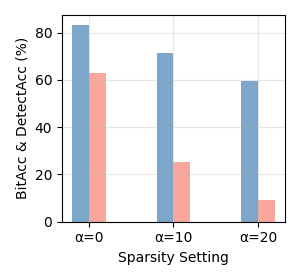}
\subcaption{\PIMoG}
\end{subfigure}
\begin{subfigure}{0.117\textwidth}
\centering
\includegraphics[width=2.31cm]{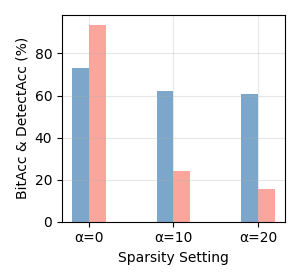}
\subcaption{\VINEB}
\end{subfigure}
\begin{subfigure}{0.117\textwidth}
\centering
\includegraphics[width=2.31cm]{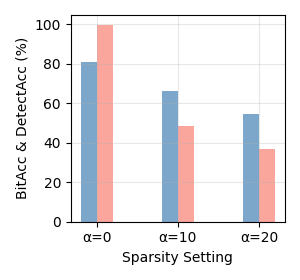}
\subcaption{\VINER}
\end{subfigure}
\begin{subfigure}{0.117\textwidth}
\centering
\includegraphics[width=2.31cm]{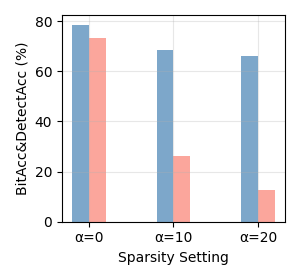}
\subcaption{\PTW}
\end{subfigure}
\begin{subfigure}{0.117\textwidth}
\centering
\includegraphics[width=2.31cm]{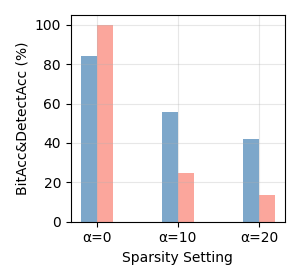}
\subcaption{SS}
\end{subfigure}
\caption{{
{Impact of latent representation sparsity levels on \ComMark's attack effectiveness across the eight watermarking schemes under three \(\alpha\) settings: \(\alpha=0\) (no sparsity), \(\alpha=10\) (moderate sparsity), and \(\alpha=20\) (moderately high sparsity).}
}}
\label{fig:asr_ablation_table_post}
\end{figure*}

\subsection{Ablation Study}
\label{subsec:RQ4}

\smallskip
\noindent
\textbf{Impact of Sparsity.}
Sparsity is central to \ComMark for effective signal compression and recovery. 
During the \ComMark optimization (\Cref{eq:sparse_encoding_loss}), excessive sparsity in latent representations \(\mathcal{Z}\) can cause information loss, while insufficient sparsity preserves redundancy and reduces efficiency.
Following \Cref{eq:SC}, we define the sparsity level of latent representations (SLR) as follows:
\begin{equation}
\label{eq:sparsity_level}
    \text{SLR} = \frac{|\{i \mid |\mathcal{Z}_i| < \tau\}|}{N_{\mathcal{Z}}}
\end{equation}
where \(N_{\mathcal{Z}}\) is the number of elements in \(\mathcal{Z}\) and \(\tau\) is the threshold. 
SLR measures the proportion of elements with magnitudes below \(\tau\), directly indicating enforced sparsity.

In \ComMark, sparsity is enforced through the sparse loss component (\Cref{eq:sparse_encoding_loss}) of the total loss (\Cref{eq:total_loss}), with weight \(\alpha\) controlling sparsity strength and influencing the SLR (\Cref{eq:sparsity_level}). 
To assess how varying sparsity affects watermark disruption, we evaluate three \(\alpha\) values—0 (none), 10 (moderate), and 20 (high)—while keeping other settings fixed. 
\ComMark\ is trained separately for each setting on   \OpenImage, following the protocol in \Cref{subsec:experiment_setup}, and evaluated against all eight watermarking schemes.

\Cref{fig:latent_sparsity_post} illustrates the smoothed histograms of sparsity level distributions (x-axis: SLR) for watermarked images, showing frequency percentages (y-axis) under three \(\alpha\) settings. 
We adopt \(\tau = 0.02\)~\cite{CS-Signal-Processing, CS-Theory-Book, CS-Theory-Survey} to measure sparsity, noting that \(\tau\) serves solely as the threshold for computing SLR and does not influence sparsity enforcement. 
When \(\alpha = 0\), SLR values (i.e., sparsity levels) concentrate near zero, producing nearly flat, visually indistinguishable curves that indicate minimal enforced sparsity.
As \(\alpha\) increases to 10 and 20, the SLR distributions shift rightward, reflecting stronger sparsity enforcement in the latent representations. 
This trend accords with \ICS\ theory, which removes redundancy while retaining essential reconstruction information, validating the role of the sparse encoding loss \(\mathcal{L}_{\text{SEL}}\) in \Cref{eq:total_loss} during \ComMark’s training optimization.

\Cref{fig:asr_ablation_table_post} shows that higher sparsity consistently enhances \ComMark's attack effectiveness across  eight watermarking schemes, as stronger sparsification removes more watermark-related information. However, as shown in \Cref{table:quality_ablation_table_post,table:quality_ablation_table_in} (\appref{sec:appendix_impact_of_sparsity}), excessive sparsity lowers \PSNR and \SSIM while raising \FID and \LPIPS, with a sharp FID increase at high sparsity indicating image degradation. To balance attack success and image quality (\Cref{sec:threat_model}), we empirically set \(\alpha = 10\), achieving the best trade-off across schemes~\cite{MBRS, CIN, TrustMark, PIMoG, VINE, StableSignature, PTW}.

\begin{table*}[t]
\centering
\captionsetup{font=normalsize}
\caption{{Impact of excluding \(\mathcal{L}_\text{LPIPS}\) (\ComMark-SSIM) or \(\mathcal{L}_\text{SSIM}\) (\ComMark-LPIPS) from \ComMark on post-attack image integrity for the six post-processing schemes using four standard metrics: \PSNR, \SSIM, \FID, and \LPIPS.}}
\small
\scalebox{0.71}{
\addtolength{\tabcolsep}{-1.35ex}
\begin{tabular}{l|cccc|cccc|cccc|cccc|cccc|cccc}
\toprule

\multirow{3}{*}{\makecell{Settings}} 
& \multicolumn{4}{c|}{MBRS} 
& \multicolumn{4}{c|}{CIN} & \multicolumn{4}{c|}{TrustMark} & \multicolumn{4}{c|}{PIMoG} 
& \multicolumn{4}{c|}{VINE-B} & \multicolumn{4}{c}{VINE-R} \\
\cmidrule(lr){2-5} \cmidrule(lr){6-9} \cmidrule(lr){10-13} \cmidrule(lr){14-17} \cmidrule(lr){18-21} \cmidrule(lr){22-25} 
& PSNR\(\uparrow\) & SSIM\(\uparrow\) & FID\(\downarrow\) & LPIPS\(\downarrow\) & PSNR\(\uparrow\) & SSIM\(\uparrow\) & FID\(\downarrow\) & LPIPS\(\downarrow\) & PSNR\(\uparrow\) & SSIM\(\uparrow\) & FID\(\downarrow\) & LPIPS\(\downarrow\) & PSNR\(\uparrow\) & SSIM\(\uparrow\) & FID\(\downarrow\) & LPIPS\(\downarrow\)
& PSNR\(\uparrow\) & SSIM\(\uparrow\) & FID\(\downarrow\) & LPIPS\(\downarrow\) & PSNR\(\uparrow\) & SSIM\(\uparrow\) & FID\(\downarrow\) & LPIPS\(\downarrow\) \\
\midrule

\rowcolor{red!10} \textbf{\ComMark} 
& 29.63 
& 0.88
& 22.49 
& 0.06
& 28.49 
& 0.83
& 30.95 
& 0.06
& 28.41 
& 0.87
& 23.59 
& 0.06
& 26.32 
& 0.83
& 37.48 
& 0.07
& 27.68 
& 0.87
& 23.77 
& 0.06
& 29.47 
& 0.88
& 23.81 
& 0.06 \\

\midrule
\textbf{\ComMark-SSIM} 
& 24.00 
& 0.80 
& 49.87 
& 0.23 
& 22.58 
& 0.80 
& 44.93 
& 0.21 
& 24.09 
& 0.84 
& 45.72 
& 0.20 
& 22.43 
& 0.79 
& 49.87 
& 0.23 
& 23.82 
& 0.83 
& 45.74 
& 0.21 
& 23.92 
& 0.83 
& 45.67 
& 0.21 \\

\textbf{\ComMark-LPIPS} 
& 22.26 
& 0.68 
& 32.73 
& 0.09 
& 21.69 
& 0.68 
& 40.41 
& 0.09 
& 22.36 
& 0.69 
& 32.87 
& 0.09 
& 21.34 
& 0.67 
& 44.14 
& 0.09 
& 22.16 
& 0.68 
& 32.91 
& 0.08 
& 22.16 
& 0.69 
& 32.80 
& 0.09 \\

\bottomrule
\end{tabular}
}
\label{table:quality_ablation_ssim_lpips_table}
\end{table*}

\begin{table*}[ht]
\centering
\captionsetup{font=normalsize}
\caption{{Impact of excluding \(\mathcal{L}_\text{LPIPS}\) (\ComMark-SSIM) or \(\mathcal{L}_\text{SSIM}\) (\ComMark-LPIPS) from \ComMark on attack performance for the six post-processing schemes using \BitAcc and \DetectAcc (TPR@0.1\%FPR) (\Cref{subsec:experiment_setup}).} 
}
\small
\scalebox{0.76}{
\addtolength{\tabcolsep}{-.4ex}
\begin{tabular}{l|cc|cc|cc|cc|cc|cc}
\toprule
\multirow{2}{*}{\makecell{Settings}} 
& \multicolumn{2}{c|}{MBRS} 
& \multicolumn{2}{c|}{CIN} & \multicolumn{2}{c|}{TrustMark} & \multicolumn{2}{c|}{PIMoG} 
& \multicolumn{2}{c|}{VINE-B} & \multicolumn{2}{c}{VINE-R} \\
\cmidrule(lr){2-3} \cmidrule(lr){4-5} \cmidrule(lr){6-7} \cmidrule(lr){8-9} \cmidrule(lr){10-11} \cmidrule(lr){12-13}
& \BitAcc\(\downarrow\) & \DetectAcc\(\downarrow\) & \BitAcc\(\downarrow\) & \DetectAcc\(\downarrow\) & \BitAcc\(\downarrow\) & \DetectAcc\(\downarrow\) & \BitAcc\(\downarrow\) & \DetectAcc\(\downarrow\) 
& \BitAcc\(\downarrow\) & \DetectAcc\(\downarrow\) & \BitAcc\(\downarrow\) & \DetectAcc\(\downarrow\)  \\

\midrule
\rowcolor{red!10} \textbf{\ComMark} & 63.6\% & 23.3\% & 65.2\% & 21.0\% & 67.8\% & 67.9\% & 71.5\% & 25.1\% & 62.1\% & 24.0\% & 66.5\%  & 48.7\% \\

\midrule
\textbf{\ComMark-SSIM} & 62.3\% & 19.8\% & 60.0\% & 16.9\% & 70.9\% & 71.1\% & 50.0\% & 0.0\% & 62.0\% & 19.9\% & 65.4\%  & 39.1\% \\

\textbf{\ComMark-LPIPS} & 61.9\% & 18.5\% & 55.5\% & 5.4\% & 54.4\% & 54.5\% & 49.3\% & 0.0\% & 61.9\% & 17.2\% & 64.8\%  & 33.0\% \\
\bottomrule
\end{tabular}
}
\label{table:asr_ablation_ssim_lpips_table}
\end{table*}

\smallskip
\noindent
\textbf{Impact of \(\mathcal{L}_\text{SSIM}\) and \(\mathcal{L}_\text{LPIPS}\).} 
\(\mathcal{L}_\text{SSIM}\) and \(\mathcal{L}_\text{LPIPS}\) enforce pixel-level similarity and feature-level alignment, respectively (\Cref{subsec:reconstruction}). Together, they enable \ComMark to maintain structural consistency and perceptual coherence in reconstructed images. To assess their complementarity, we examine two variants: \ComMark-SSIM, which removes \(\mathcal{L}_\text{LPIPS}\) from \(\mathcal{L}_\text{SPL}\) (\Cref{eq:dml_loss}), and \ComMark-LPIPS, which removes \(\mathcal{L}_\text{SSIM}\). Both variants are trained independently on the \OpenImage dataset and evaluated across eight watermarking schemes, with all other settings identical to \ComMark (\Cref{subsec:experiment_setup}).

\Cref{table:quality_ablation_ssim_lpips_table} (main paper) and \Cref{table:quality_ablation_ssim_lpips_table_in} (\appref{sec:appendix_ssim_lpips}) show that both \ComMark-SSIM and \ComMark-LPIPS underperform \ComMark in preserving image integrity, with lower \PSNR\ and \SSIM\ and higher \FID\ and \LPIPS.
As shown in \Cref{fig:ablation_ssim_lpips}, \ComMark-SSIM produces smoother but color-shifted reconstructions, whereas \ComMark-LPIPS achieves better perceptual alignment but introduces texture distortions. 
These results confirm the complementary roles of structural and perceptual constraints in maintaining image fidelity. 
However, \Cref{table:asr_ablation_ssim_lpips_table} (main paper) and \Cref{table:asr_ablation_ssim_lpips_table_in} (\appref{sec:appendix_ssim_lpips}) show stronger attack performance for both variants, indicating that relaxing reconstruction constraints increases feature dispersion and amplifies the dispersal effect (\Cref{sec:design_principle}), thereby enhancing watermark disruption.

These results highlight that combining \(\mathcal{L}_\text{SSIM}\) and \(\mathcal{L}_\text{LPIPS}\) is essential for preserving image integrity while maintaining the dispersal-driven disruption of watermark signals.

\section{Related Work}
\label{sec:related_work}

\begin{description}[leftmargin=0pt]
\setlength{\parindent}{1em}  
\setlength{\parskip}{0pt}

\item[Watermarking Schemes.] The rapid development of deep generative techniques \cite{LDM, GPT, art_creation} has enabled the creation of highly realistic images (deepfakes). While these advances benefit various applications, they also amplify risks of misinformation \cite{copyright-cvpr, copyright-mm, deepfake-cvpr}.
To prevent misuse of deepfakes, watermarking methods are broadly employed and categorized into post-processing schemes \cite{DCT, DTW-DCT, MBRS, CIN, TrustMark, PIMoG, VINE, StegaStamp}, which embed watermarks after image generation, and in-processing schemes \cite{TreeRing, PTW, StableSignature}, which integrate watermarking into the generation process.

\begin{figure}[t]
\centering
\begin{subfigure}{0.11\textwidth}
\centering
\includegraphics[width=1.8cm]{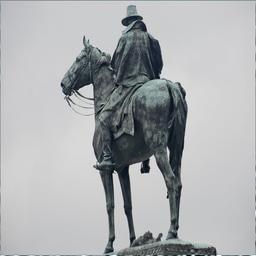}
\end{subfigure}
\begin{subfigure}{0.11\textwidth}
\centering
\includegraphics[width=1.8cm]{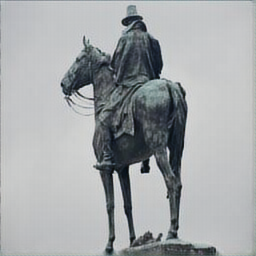}
\end{subfigure}
\begin{subfigure}{0.118\textwidth}
\centering
\includegraphics[width=1.8cm]{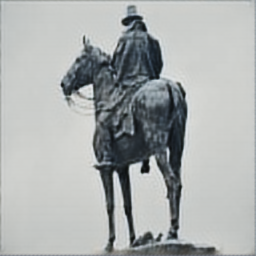}
\end{subfigure}
\begin{subfigure}{0.123\textwidth}
\centering
\includegraphics[width=1.8cm]{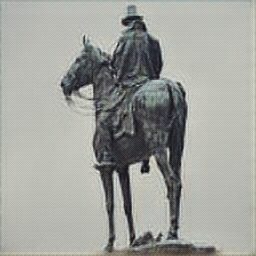}

\end{subfigure}
\centering
\begin{subfigure}{0.11\textwidth}
\centering
\includegraphics[width=1.8cm]{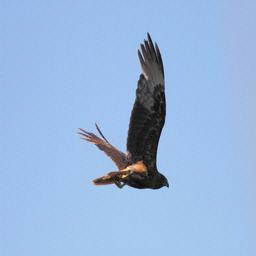}
\subcaption{\scriptsize WMs}
\end{subfigure}
\begin{subfigure}{0.11\textwidth}
\centering
\includegraphics[width=1.8cm]{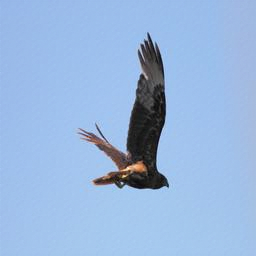}
\subcaption{\scriptsize \ComMark}
\end{subfigure}
\begin{subfigure}{0.118\textwidth}
\centering
\includegraphics[width=1.8cm]{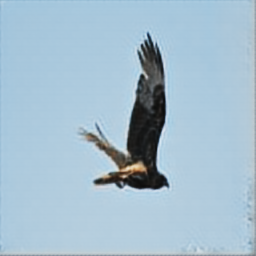}
\subcaption{\scriptsize \ComMark-SSIM}
\end{subfigure}
\begin{subfigure}{0.123\textwidth}
\centering
\includegraphics[width=1.8cm]{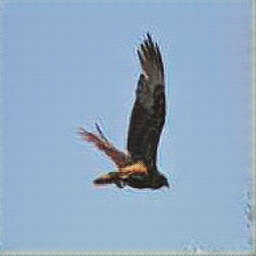}
\subcaption{\scriptsize \ComMark-LPIPS}
\end{subfigure}
\caption{{
{Comparison of \ComMark, \ComMark-SSIM, and \ComMark-LPIPS on image integrity using two
\VINER-generated watermarked images (WMs) from OpenImage.}
\label{fig:ablation_ssim_lpips}}}
\end{figure}

For post-processing watermarking, early approaches relied on frequency-domain transformations like DWT, DCT \cite{DCT}, and DWTDCT \cite{DTW-DCT}, embedding watermarks by modifying frequency bands. However, these methods lacked robustness and were easily attacked \cite{VINE, WAVES}. The advent of deep learning introduced encoder-decoder-based watermarking, which uses neural networks to enhance robustness against distortions. For instance, StegaStamp \cite{StegaStamp} encodes and decodes watermarks resilient to real-world distortions, including printing, photography, and environmental variations, though it often compromises image integrity \cite{StegaStamp, WAVES, VINE}. TrustMark \cite{TrustMark} employs spatio-spectral losses and resolution scaling to balance image integrity and watermark recovery. MBRS \cite{MBRS} uses adversarial training with random noise, simulated JPEG, and real JPEG distortions to improve robustness. CIN \cite{CIN} applies an invertible network with fusion-split modules for compression resistance. PIMoG \cite{PIMoG} incorporates a differentiable noise layer to address screen-shooting distortions. VINE-B \cite{VINE} encodes watermarks as text sequences using a condition adaptor and a text-to-image model, while VINE-R adds noise layers to further mitigate distortions.

For in-processing watermarking, \StableSignature \cite{StableSignature} embeds watermarks during diffusion generation by fine-tuning the decoder to insert a binary signature, ensuring all generated images are watermarked. A pre-trained watermark detector verifies watermark presence. \PTW \cite{PTW} targets GAN-based models and uses pivotal tuning to preserve the functionality of original generators by freezing parameters while fine-tuning secondary generators to embed watermarks with a regularization term. Unlike \StableSignature and \PTW, TreeRing \cite{TreeRing} fingerprints diffusion model outputs by embedding patterns into the initial noise vector during sampling. However, TreeRing significantly alters images due to these modifications, limiting its broader applicability \cite{WatermarkAttacker, WAVES, VINE}.

\smallskip
\item[Watermarking Attacks.] As detailed in \Cref{sec:watermarking_attacks}, there are three primary attack types: distortion, regeneration, and adversarial attacks. Distortion attacks, such as cropping, resizing, and noise addition, damage watermarks but often degrade image integrity, limiting practicality \cite{WAVES, PTW, StableSignature}. Advanced methods like Unmark \cite{unmark} manipulate spectral amplitudes but still significantly affect visual quality \cite{sok_aigc, unmark}. Regeneration attacks \cite{WatermarkAttacker, Regeneration-Attack-Diff1, Regeneration-Attack-Diff2} perturb images into noise and regenerate them using VAEs \cite{VAE} or diffusion models \cite{Diffusion-Model}, aiming to remove watermarks while preserving integrity. However, they struggle to disrupt watermarks embedded in latent representations \cite{VINE, WAVES, unmark}. Adversarial attacks \cite{WM-adversarial-CCS, WM-adversarial-ICLR, WM-adversarial-TCSVT} alter encoded features to deceive detectors but require access to watermarking pipelines and detectors, making them impractical in real-world settings \cite{WAVES, unmark, Hu-Transfer}. Even with surrogate models, transferability challenges reduce their effectiveness \cite{WAVES, unmark, Hu-Transfer}.

\end{description}
\section{Conclusion}  
\label{sec:conclusion}  

We present \ComMark, a query-free black-box attack framework that exposes the fragility of current watermarking-based deepfake defenses. Built upon Image Compressive Sensing (\ICS) theory, \ComMark exploits latent-space vulnerabilities in encoder–decoder watermarking models by enforcing sparsity to suppress watermark-bearing components while preserving perceptual and structural realism. Across eight state-of-the-art schemes, \ComMark significantly reduces watermark detection accuracy without compromising visual plausibility, demonstrating that existing deepfake watermarking mechanisms remain highly vulnerable to latent-space perturbations. These results call for a fundamental rethinking of defensive watermark design in generative AI. Future research should explore robustness through multi-level latent feature encoding, adaptive redundancy distribution, and watermark integration strategies that jointly optimize fidelity, security, and detectability in deepfake generation pipelines.

\section{Ethics Considerations}  
\label{sec:open_science}  

\ComMark is an attack framework designed to uncover vulnerabilities in defensive image watermarking schemes, particularly those based on deep encoder-decoder architectures. All experiments were conducted using publicly available, non-commercial watermarking schemes and datasets. In this research, the primary objective is to identify security risks in current methods and promote the development of more robust watermarking techniques for generative AI, without exploiting these vulnerabilities for malicious purposes. Additionally, mitigation strategies against \ComMark were thoroughly evaluated but proved 
ineffective.

\bibliographystyle{IEEEtran}
\bibliography{ComMark/ref}

\appendices

\begin{table*}[t]
\centering
\captionsetup{font=normalsize}
\caption{Impact of sparsity levels on \ComMark's  post-attack image integrity across six post-processing watermarking schemes under three \(\alpha\) settings: \(\alpha=0\) (no sparsity), \(\alpha=10\) (moderate sparsity), and \(\alpha=20\) (moderately high sparsity).}
\small
\scalebox{0.68}{
\addtolength{\tabcolsep}{-1.2ex}
\begin{tabular}{l|cccc|cccc|cccc|cccc|cccc|cccc}
\toprule
\multirow{3}{*}{\makecell{Sparsity \\ Setting}} 
& \multicolumn{4}{c|}{MBRS} 
& \multicolumn{4}{c|}{CIN} & \multicolumn{4}{c|}{TrustMark} & \multicolumn{4}{c|}{PIMoG} 
& \multicolumn{4}{c|}{VINE-B} & \multicolumn{4}{c}{VINE-R} \\
\cmidrule(lr){2-5} \cmidrule(lr){6-9} \cmidrule(lr){10-13} \cmidrule(lr){14-17} \cmidrule(lr){18-21} \cmidrule(lr){22-25} 
& PSNR\(\uparrow\) & SSIM\(\uparrow\) & FID\(\downarrow\) & LPIPS\(\downarrow\) & PSNR\(\uparrow\) & SSIM\(\uparrow\) & FID\(\downarrow\) & LPIPS\(\downarrow\) & PSNR\(\uparrow\) & SSIM\(\uparrow\) & FID\(\downarrow\) & LPIPS\(\downarrow\) & PSNR\(\uparrow\) & SSIM\(\uparrow\) & FID\(\downarrow\) & LPIPS\(\downarrow\)
& PSNR\(\uparrow\) & SSIM\(\uparrow\) & FID\(\downarrow\) & LPIPS\(\downarrow\) & PSNR\(\uparrow\) & SSIM\(\uparrow\) & FID\(\downarrow\) & LPIPS\(\downarrow\) \\
\midrule

\(\alpha=0\) & 30.06 & 0.92 & 22.93 & 0.04 & 29.53 & 0.88 & 34.54 & 0.05 & 29.76 & 0.92 & 22.78 & 0.04 & 28.31 & 0.85 & 35.71 & 0.06 & 29.47 & 0.90 & 23.70 & 0.04 & 31.15 & 0.93 & 22.16 & 0.04 \\

\(\alpha=10\)
& 29.63 & 0.88 & 22.49 & 0.06 & 28.49 & 0.83 & 30.95 & 0.06 & 28.41 & 0.87 & 23.59 & 0.06 & 26.32 & 0.83 & 37.48 & 0.07 & 27.68 & 0.87 & 23.77 & 0.06 & 29.47 & 0.88 & 23.81 & 0.06 \\
\(\alpha=20\) & 25.46 & 0.80 & 115.43 & 0.13 & 24.97 & 0.79 & 156.23 & 0.14 & 25.53 & 0.80 & 138.21 & 0.13 & 25.31 & 0.80 & 139.34 & 0.13 & 25.71 & 0.81 & 133.67 & 0.13 & 24.52 & 0.81 & 134.62 & 0.13\\
\bottomrule
\end{tabular}
}
\label{table:quality_ablation_table_post}
\vspace*{-2ex}
\end{table*}

\section{Watermarking Schemes' Bit Lengths}
\label{sec:appendix_watermark_length}

{\Cref{tab:bit_length} is referenced in \Cref{subsec:experiment_setup} to provide the watermark lengths used for the eight watermarking schemes.}

\begin{table}[htbp]
\centering
\captionsetup{font=normalsize}
\caption{Watermark bit lengths in eight schemes.}
\small
\scalebox{0.9}{
\begin{tabular}{c|c|c|c}
\toprule
\hline
\MBRS \cite{MBRS} & \CIN \cite{CIN} & \TrustMark \cite{TrustMark} & \PIMoG \cite{PIMoG} \\ \hline
30           & 30          & 100               & 30           \\ \hline
\VINEB \cite{VINE} & \VINER \cite{VINE} & \StableSignature \cite{StableSignature} & \PTW \cite{PTW} \\ \hline
100             & 100            & 48                     & 40        \\ \hline
\bottomrule
\end{tabular}
}
\label{tab:bit_length}
\vspace*{-2ex}
\end{table}

\section{Impact of Sparsity}
\label{sec:appendix_impact_of_sparsity}

{\Cref{table:quality_ablation_table_post,table:quality_ablation_table_in} are referenced in \Cref{subsec:RQ4} as part of analysis for the impact of sparsity.}

\begin{table}[h]
\centering
\captionsetup{font=normalsize}
\caption{Impact of sparsity levels on \ComMark's post-attack image integrity across two in-processing schemes.}
\small
\scalebox{0.76}{
\addtolength{\tabcolsep}{-.5ex}
\begin{tabular}{l|cccc|cccc}
\toprule
\multirow{2}{*}{\makecell{Sparsity \\ Setting}} 
& \multicolumn{4}{c|}{\PTW}
& \multicolumn{4}{c}{\StableSignature} \\
\cmidrule(lr){2-5} \cmidrule(lr){6-9}
& PSNR\(\uparrow\) & SSIM\(\uparrow\) & FID\(\downarrow\) & LPIPS\(\downarrow\) & PSNR\(\uparrow\) & SSIM\(\uparrow\) & FID\(\downarrow\) & LPIPS\(\downarrow\) \\
\midrule
\(\alpha=0\) & 35.49 & 0.94 & 7.78 & 0.03 & 32.25 & 0.93 & 12.40 & 0.03  \\
\(\alpha=10\) & 32.46 & 0.87 & 21.32 & 0.06 & 29.43 & 0.88 & 23.35 & 0.06 \\
\(\alpha=20\) & 28.12 & 0.83 & 105.67 & 0.11 & 27.48 & 0.82 & 95.16 & 0.10 \\
\bottomrule
\end{tabular}
}
\label{table:quality_ablation_table_in}
\vspace*{-2ex}
\end{table}

\section{Visual Comparison of Image Integrity for Image Super-Resolution (\ISR)}
\label{sec:appendix_isr}

\begin{figure}[htbp]
\centering
\begin{subfigure}[b]{0.12\textwidth}
\centering
\includegraphics[width=\textwidth]{ComMark/images/commark/000ac34008b0ba4c_wm.png}
\caption{\ComMark}
\end{subfigure}
\begin{subfigure}[b]{0.34\textwidth}
\centering
\includegraphics[width=\textwidth]{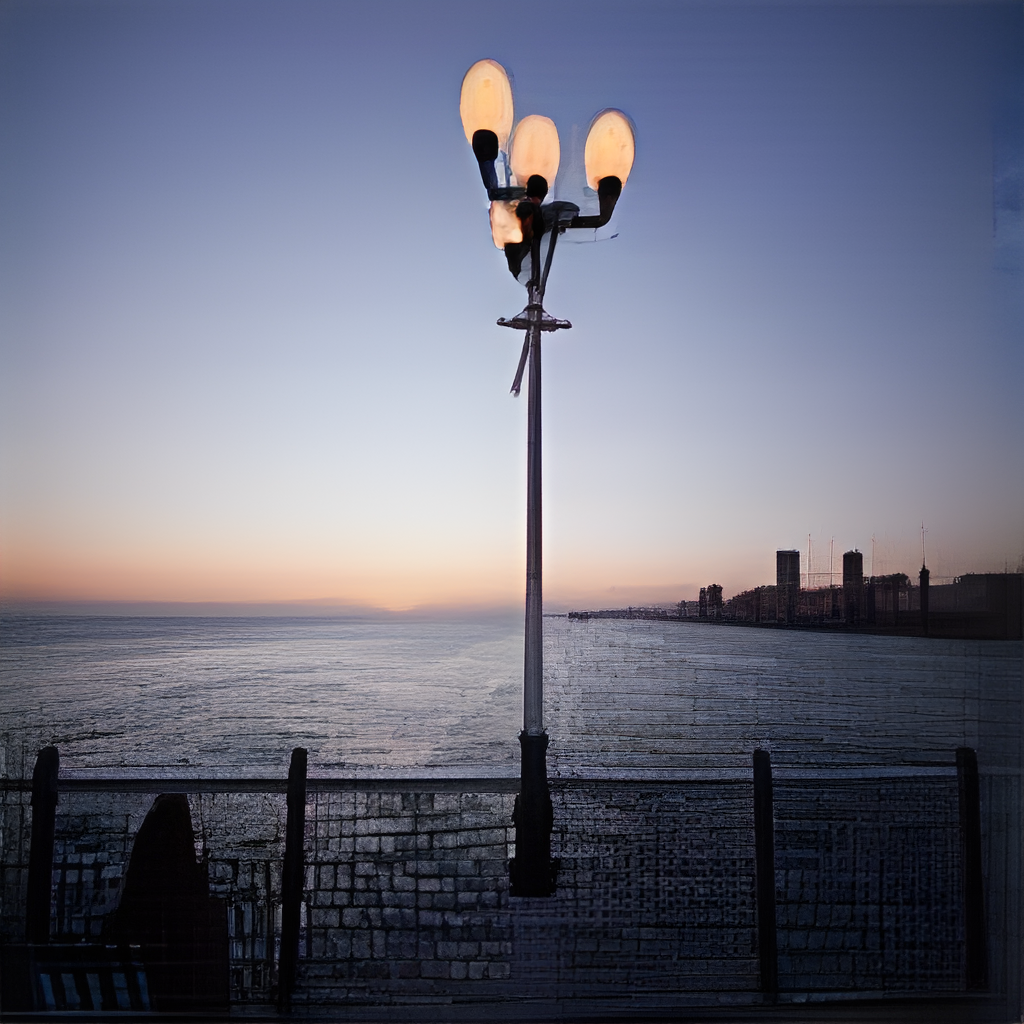}
\caption{\ComMark-ISR}
\end{subfigure}
\caption{Visual comparison of image integrity between \ComMark and \ComMark-ISR using an OpenImage sample watermarked by \VINER. (a) Image ($256 \times 256$) attacked by \ComMark. (b) The same image from (a) reconstructed to $1024 \times 1024$ using \ISR (\ComMark-ISR), allowing the reader to assess integrity differences visually.
}
\label{fig:commark_isr}
\vspace*{-.5ex}
\end{figure}

\Cref{fig:commark_isr} is referenced in \Cref{subsec:RQ5} as part of the analysis for RQ4.

\section{Impact of \(\mathcal{L}_\text{SSIM}\) and \(\mathcal{L}_\text{LPIPS}\)}
\label{sec:appendix_ssim_lpips}

{\Cref{table:quality_ablation_ssim_lpips_table_in,table:asr_ablation_ssim_lpips_table_in} are referenced in \Cref{subsec:RQ4} as part of analysis for the impact of \(\mathcal{L}_\text{SSIM}\) and \(\mathcal{L}_\text{LPIPS}\).}

\begin{table}[htbp]
\centering
\captionsetup{font=normalsize}
\caption{Impact of excluding \(\mathcal{L}_\text{LPIPS}\) (\ComMark-SSIM) or \(\mathcal{L}_\text{SSIM}\) (\ComMark-LPIPS) from \ComMark on post-attack image integrity for the two in-processing schemes using four standard metrics: \PSNR, \SSIM, \FID, and \LPIPS.}
\small
\scalebox{0.78}{
\addtolength{\tabcolsep}{-1.1ex}
\begin{tabular}{l|cccc|cccc}
\toprule

\multirow{3}{*}{\makecell{Settings}} 
& \multicolumn{4}{c|}{\PTW} 
& \multicolumn{4}{c}{\StableSignature} \\
\cmidrule(lr){2-5} \cmidrule(lr){6-9}
& PSNR\(\uparrow\) & SSIM\(\uparrow\) & FID\(\downarrow\) & LPIPS\(\downarrow\) & PSNR\(\uparrow\) & SSIM\(\uparrow\) & FID\(\downarrow\) & LPIPS\(\downarrow\)  \\
\midrule

\rowcolor{red!10} \textbf{\ComMark} 
& 32.46
& 0.87 
& 21.32 
& 0.06 
& 29.43 
& 0.88 
& 23.35 
& 0.06
\\

\midrule
\textbf{\ComMark-SSIM} 
& 26.54 
& 0.88 
& 32.65 
& 0.10 
& 23.97 
& 0.87 
& 38.38 
& 0.15 
\\

\textbf{\ComMark-LPIPS} 
& 26.32
& 0.73 
& 20.33 
& 0.04 
& 23.12 
& 0.79 
& 20.49 
& 0.07 
\\

\bottomrule
\end{tabular}
}
\label{table:quality_ablation_ssim_lpips_table_in}
\vspace*{-3ex}
\end{table}
\begin{table}[htbp]
\centering
\captionsetup{font=normalsize}
\caption{Impact of excluding \(\mathcal{L}_\text{LPIPS}\) (\ComMark-SSIM) or \(\mathcal{L}_\text{SSIM}\) (\ComMark-LPIPS) from \ComMark on attack performance for the two in-processing schemes using \BitAcc and \DetectAcc (TPR@0.1\%FPR) (\Cref{subsec:experiment_setup}). 
}
\small
\scalebox{0.8}{
\addtolength{\tabcolsep}{-.0ex}
\begin{tabular}{l|cc|cc}
\toprule
\multirow{2}{*}{\makecell{Settings}} 
& \multicolumn{2}{c|}{\PTW} 
& \multicolumn{2}{c}{\StableSignature} \\
\cmidrule(lr){2-3} \cmidrule(lr){4-5}
& \BitAcc\(\downarrow\) & \DetectAcc\(\downarrow\) 
& \BitAcc\(\downarrow\) & \DetectAcc\(\downarrow\)  \\

\midrule
\rowcolor{red!10} \textbf{\ComMark} & 68.3\% & 26.4\% & 55.9\% & 24.6\%  \\

\midrule
\textbf{\ComMark-SSIM} & 64.8\% & 21.3\% & 51.8\% & 23.0\% \\

\textbf{\ComMark-LPIPS} & 60.3\% & 19.7\% & 49.3\% & 20.4\%  \\
\bottomrule
\end{tabular}
}
\label{table:asr_ablation_ssim_lpips_table_in}
\end{table}

\section{Visual Comparison of Image Integrity across four Attack Methods on Watermarked Images}
\label{sec:appendix_visual_examples_for_all_schmes}

\begin{figure*}[htbp]

\begin{subfigure}{0.107\textwidth}
\centering
\includegraphics[width=1.8cm]{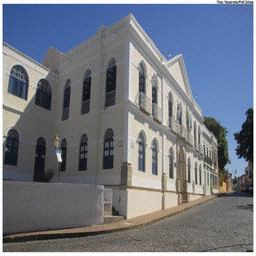}
\end{subfigure}
\hspace*{-1.ex}
\begin{subfigure}{0.55\textwidth}
\centering
\includegraphics[width=1.8cm]{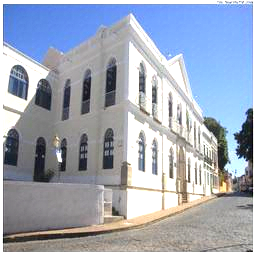}
\includegraphics[width=1.8cm]{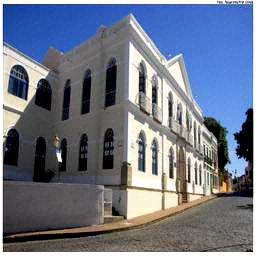}
\includegraphics[width=1.8cm]{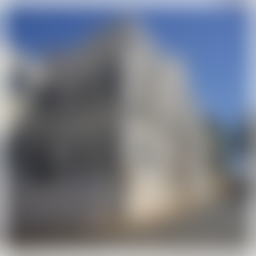}
\includegraphics[width=1.8cm]{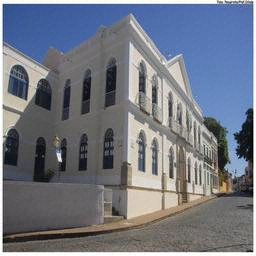}
\includegraphics[width=1.8cm]{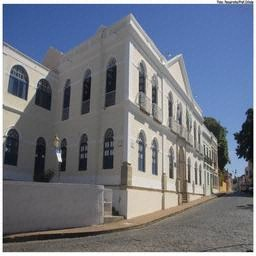}
\end{subfigure}
\hspace*{-1.ex}
\begin{subfigure}{0.107\textwidth}
\centering
\includegraphics[width=1.8cm]{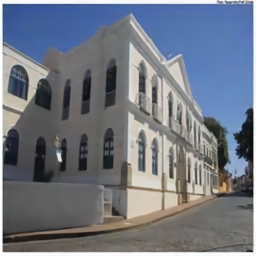}
\end{subfigure}
\hspace*{.0ex}
\begin{subfigure}{0.107\textwidth}
\centering
\includegraphics[width=1.8cm]{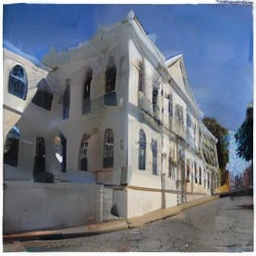}
\end{subfigure}
\hspace*{.0ex}
\begin{subfigure}{0.109\textwidth}
\centering
\includegraphics[width=1.8cm]{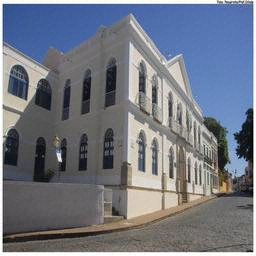}
\end{subfigure}

\begin{subfigure}{0.107\textwidth}
\centering
\includegraphics[width=1.8cm]{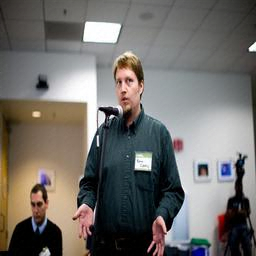}
\subcaption{WMs}
\end{subfigure}
\hspace*{-1.ex}
\begin{subfigure}{0.55\textwidth}
\centering
\includegraphics[width=1.8cm]{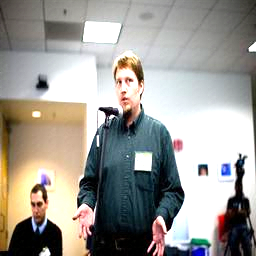}
\includegraphics[width=1.8cm]{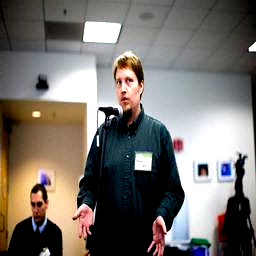}
\includegraphics[width=1.8cm]{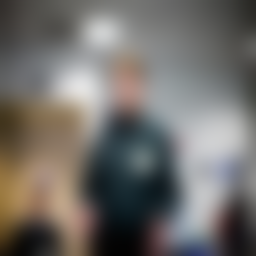}
\includegraphics[width=1.8cm]{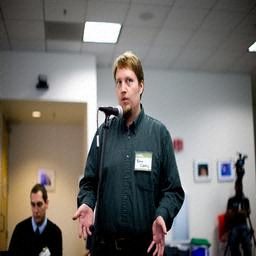}
\includegraphics[width=1.8cm]{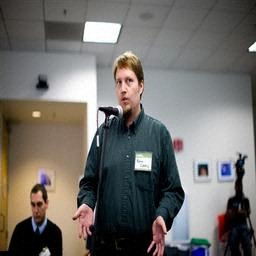}
\caption{\DistortionAttack}
\end{subfigure}
\hspace*{-1.ex}
\begin{subfigure}{0.107\textwidth}
\centering
\includegraphics[width=1.8cm]{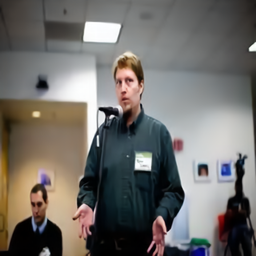}
\subcaption{\RegenVAE}
\end{subfigure}
\hspace*{.0ex}
\begin{subfigure}{0.107\textwidth}
\centering
\includegraphics[width=1.8cm]{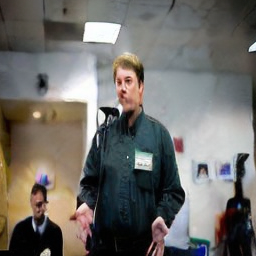}
\subcaption{\RegenDiffusion}
\end{subfigure}
\hspace*{.0ex}
\begin{subfigure}{0.109\textwidth}
\centering
\includegraphics[width=1.8cm]{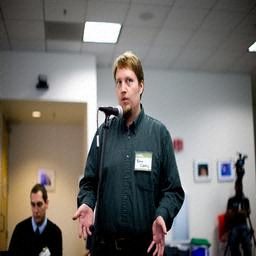}
\subcaption{\ComMark}
\end{subfigure}
\caption{Comparison of \ComMark, \DistortionAttack (brightness, contrast, blurring, Gaussian noise, and compression in that order), \RegenVAE, and \RegenDiffusion on image integrity using two
  \textbf{\MBRS-generated}
watermarked images (WMs)  from 
OpenImage.}
\label{fig:regenvae_diffusion_commark_mbrs}
\end{figure*}

\begin{figure*}[htbp]
\begin{subfigure}{0.107\textwidth}
\centering
\includegraphics[width=1.8cm]{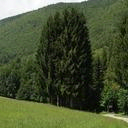}
\end{subfigure}
\hspace*{-1.ex}
\begin{subfigure}{0.55\textwidth}
\centering
\includegraphics[width=1.8cm]{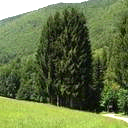}
\includegraphics[width=1.8cm]{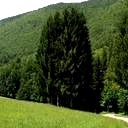}
\includegraphics[width=1.8cm]{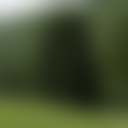}
\includegraphics[width=1.8cm]{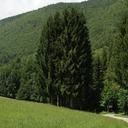}
\includegraphics[width=1.8cm]{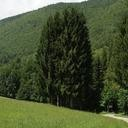}
\end{subfigure}
\hspace*{-1.ex}
\begin{subfigure}{0.107\textwidth}
\centering
\includegraphics[width=1.8cm]{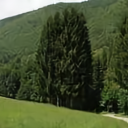}
\end{subfigure}
\hspace*{.0ex}
\begin{subfigure}{0.107\textwidth}
\centering
\includegraphics[width=1.8cm]{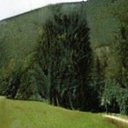}
\end{subfigure}
\hspace*{.0ex}
\begin{subfigure}{0.109\textwidth}
\centering
\includegraphics[width=1.8cm]{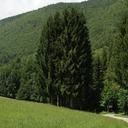}
\end{subfigure}

\begin{subfigure}{0.107\textwidth}
\centering
\includegraphics[width=1.8cm]{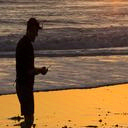}
\subcaption{WMs}
\end{subfigure}
\hspace*{-1.ex}
\begin{subfigure}{0.55\textwidth}
\centering
\includegraphics[width=1.8cm]{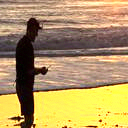}
\includegraphics[width=1.8cm]{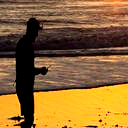}
\includegraphics[width=1.8cm]{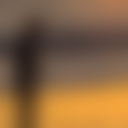}
\includegraphics[width=1.8cm]{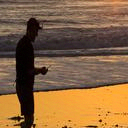}
\includegraphics[width=1.8cm]{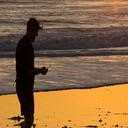}
\caption{\DistortionAttack}
\end{subfigure}
\hspace*{-1.ex}
\begin{subfigure}{0.107\textwidth}
\centering
\includegraphics[width=1.8cm]{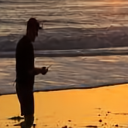}
\subcaption{\RegenVAE}
\end{subfigure}
\hspace*{.0ex}
\begin{subfigure}{0.107\textwidth}
\centering
\includegraphics[width=1.8cm]{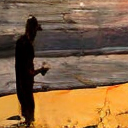}
\subcaption{\RegenDiffusion}
\end{subfigure}
\hspace*{.0ex}
\begin{subfigure}{0.109\textwidth}
\centering
\includegraphics[width=1.8cm]{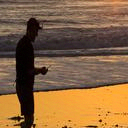}
\subcaption{\ComMark}
\end{subfigure}
\caption{Comparison of \ComMark,  \DistortionAttack (brightness, contrast, blurring, Gaussian noise, and compression in that order), \RegenVAE, and \RegenDiffusion on image integrity using two
  \textbf{\CIN-generated}
watermarked images (WMs)  from 
OpenImage.}
\label{fig:regenvae_diffusion_commark_CIN}
\end{figure*}

\begin{figure*}[ht]
\begin{subfigure}{0.107\textwidth}
\centering
\includegraphics[width=1.8cm]{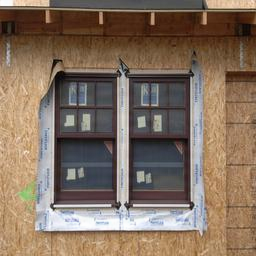}
\end{subfigure}
\hspace*{-1.ex}
\begin{subfigure}{0.55\textwidth}
\centering
\includegraphics[width=1.8cm]{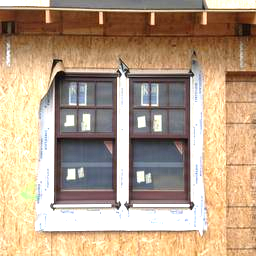}
\includegraphics[width=1.8cm]{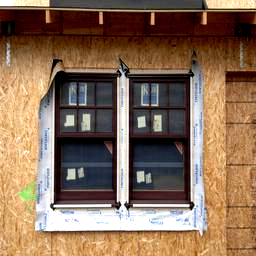}
\includegraphics[width=1.8cm]{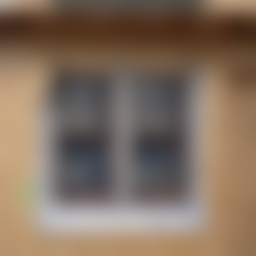}
\includegraphics[width=1.8cm]{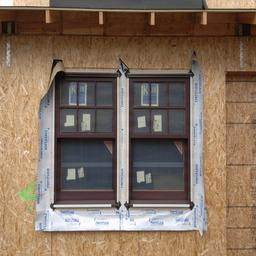}
\includegraphics[width=1.8cm]{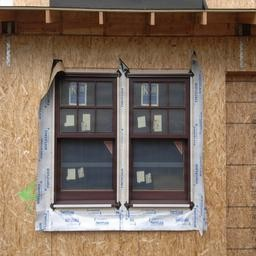}
\end{subfigure}
\hspace*{-1.ex}
\begin{subfigure}{0.107\textwidth}
\centering
\includegraphics[width=1.8cm]{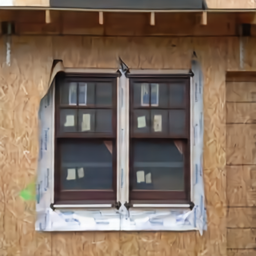}
\end{subfigure}
\hspace*{.0ex}
\begin{subfigure}{0.107\textwidth}
\centering
\includegraphics[width=1.8cm]{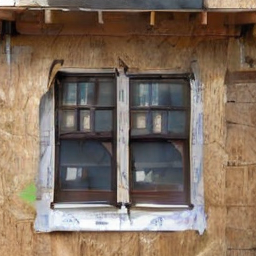}
\end{subfigure}
\hspace*{.0ex}
\begin{subfigure}{0.109\textwidth}
\centering
\includegraphics[width=1.8cm]{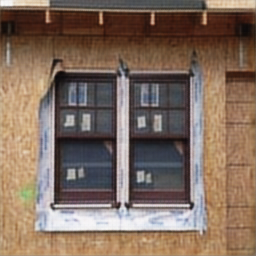}
\end{subfigure}

\begin{subfigure}{0.107\textwidth}
\centering
\includegraphics[width=1.8cm]{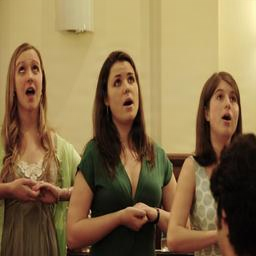}
\subcaption{WMs}
\end{subfigure}
\hspace*{-1.ex}
\begin{subfigure}{0.55\textwidth}
\centering
\includegraphics[width=1.8cm]{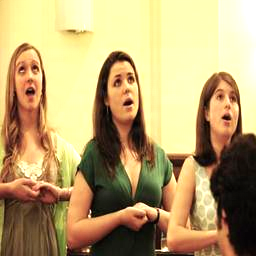}
\includegraphics[width=1.8cm]{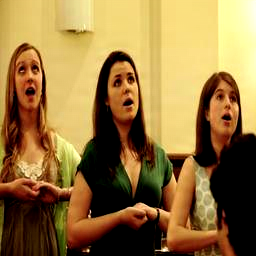}
\includegraphics[width=1.8cm]{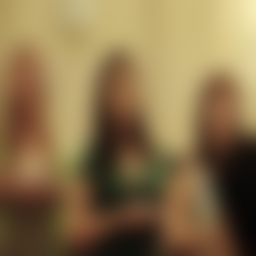}
\includegraphics[width=1.8cm]{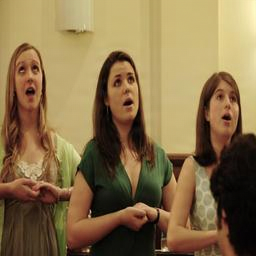}
\includegraphics[width=1.8cm]{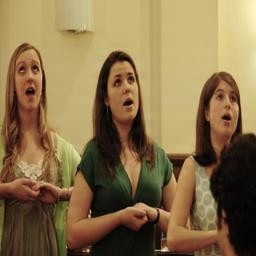}
\caption{\DistortionAttack}
\end{subfigure}
\hspace*{-1.ex}
\begin{subfigure}{0.107\textwidth}
\centering
\includegraphics[width=1.8cm]{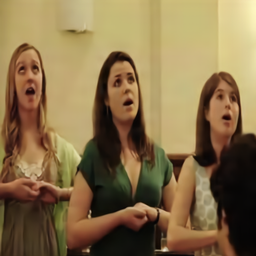}
\subcaption{\RegenVAE}
\end{subfigure}
\hspace*{.0ex}
\begin{subfigure}{0.107\textwidth}
\centering
\includegraphics[width=1.8cm]{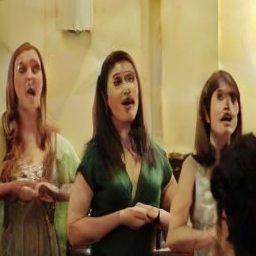}
\subcaption{\RegenDiffusion}
\end{subfigure}
\hspace*{.0ex}
\begin{subfigure}{0.109\textwidth}
\centering
\includegraphics[width=1.8cm]{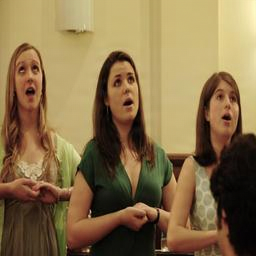}
\subcaption{\ComMark}
\end{subfigure}
\caption{Comparison of \ComMark,  \DistortionAttack (brightness, contrast, blurring, Gaussian noise, and compression in that order), \RegenVAE, and \RegenDiffusion on image integrity using two
  \textbf{\TrustMark-generated}
watermarked images (WMs)  from 
OpenImage.}
\label{fig:regenvae_diffusion_commark_trustmark}
\end{figure*}

\begin{figure*}[ht]
\begin{subfigure}{0.107\textwidth}
\centering
\includegraphics[width=1.8cm]{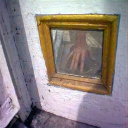}
\end{subfigure}
\hspace*{-1.ex}
\begin{subfigure}{0.55\textwidth}
\centering
\includegraphics[width=1.8cm]{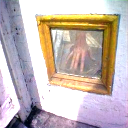}
\includegraphics[width=1.8cm]{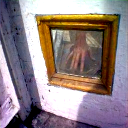}
\includegraphics[width=1.8cm]{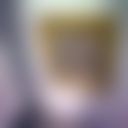}
\includegraphics[width=1.8cm]{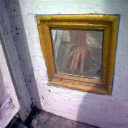}
\includegraphics[width=1.8cm]{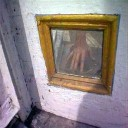}
\end{subfigure}
\hspace*{-1.ex}
\begin{subfigure}{0.107\textwidth}
\centering
\includegraphics[width=1.8cm]{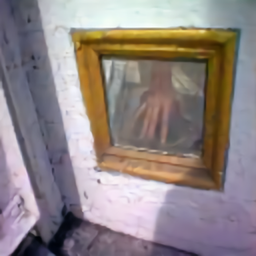}
\end{subfigure}
\hspace*{.0ex}
\begin{subfigure}{0.107\textwidth}
\centering
\includegraphics[width=1.8cm]{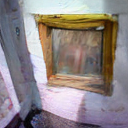}
\end{subfigure}
\hspace*{.0ex}
\begin{subfigure}{0.109\textwidth}
\centering
\includegraphics[width=1.8cm]{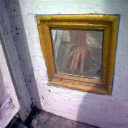}
\end{subfigure}

\begin{subfigure}{0.107\textwidth}
\centering
\includegraphics[width=1.8cm]{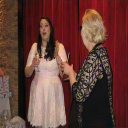}
\subcaption{WMs}
\end{subfigure}
\hspace*{-1.ex}
\begin{subfigure}{0.55\textwidth}
\centering
\includegraphics[width=1.8cm]{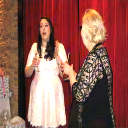}
\includegraphics[width=1.8cm]{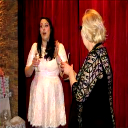}
\includegraphics[width=1.8cm]{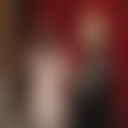}
\includegraphics[width=1.8cm]{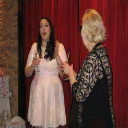}
\includegraphics[width=1.8cm]{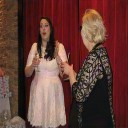}
\caption{\DistortionAttack}
\end{subfigure}
\hspace*{-1.ex}
\begin{subfigure}{0.107\textwidth}
\centering
\includegraphics[width=1.8cm]{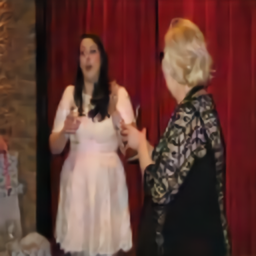}
\subcaption{\RegenVAE}
\end{subfigure}
\hspace*{.0ex}
\begin{subfigure}{0.107\textwidth}
\centering
\includegraphics[width=1.8cm]{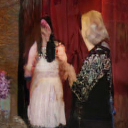}
\subcaption{\RegenDiffusion}
\end{subfigure}
\hspace*{.0ex}
\begin{subfigure}{0.109\textwidth}
\centering
\includegraphics[width=1.8cm]{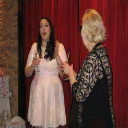}
\subcaption{\ComMark}
\end{subfigure}
\caption{Comparison of \ComMark,  \DistortionAttack (brightness, contrast, blurring, Gaussian noise, and compression in that order), \RegenVAE, and \RegenDiffusion on image integrity using two
  \textbf{\PIMoG-generated}
watermarked images (WMs)  from 
OpenImage.}
\label{fig:regenvae_diffusion_commark_PIMoG}
\end{figure*}

\begin{figure*}[ht]
\begin{subfigure}{0.107\textwidth}
\centering
\includegraphics[width=1.8cm]{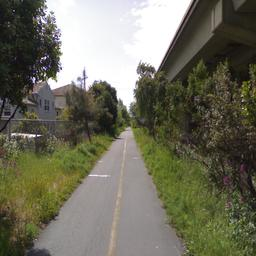}
\end{subfigure}
\hspace*{-1.ex}
\begin{subfigure}{0.55\textwidth}
\centering
\includegraphics[width=1.8cm]{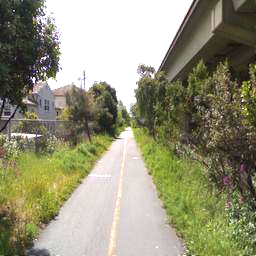}
\includegraphics[width=1.8cm]{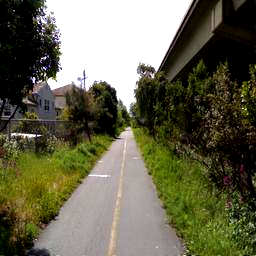}
\includegraphics[width=1.8cm]{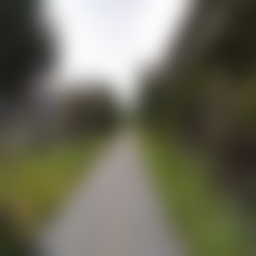}
\includegraphics[width=1.8cm]{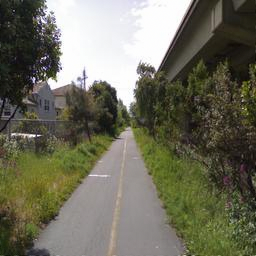}
\includegraphics[width=1.8cm]{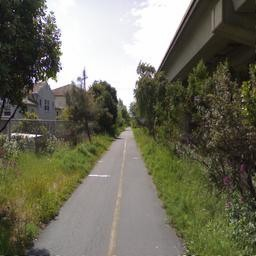}
\end{subfigure}
\hspace*{-1.ex}
\begin{subfigure}{0.107\textwidth}
\centering
\includegraphics[width=1.8cm]{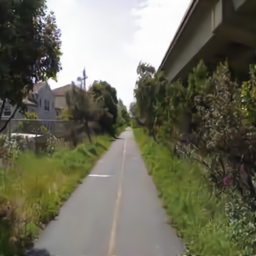}
\end{subfigure}
\hspace*{.0ex}
\begin{subfigure}{0.107\textwidth}
\centering
\includegraphics[width=1.8cm]{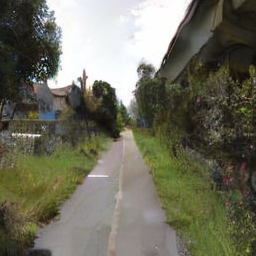}
\end{subfigure}
\hspace*{.0ex}
\begin{subfigure}{0.109\textwidth}
\centering
\includegraphics[width=1.8cm]{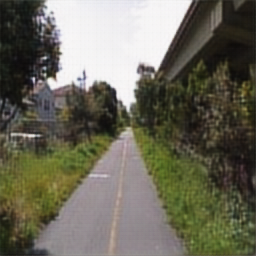}
\end{subfigure}

\begin{subfigure}{0.107\textwidth}
\centering
\includegraphics[width=1.8cm]{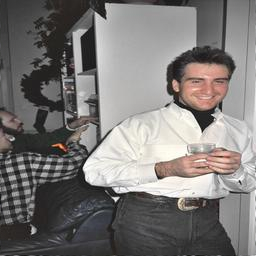}
\subcaption{WMs}
\end{subfigure}
\hspace*{-1.ex}
\begin{subfigure}{0.55\textwidth}
\centering
\includegraphics[width=1.8cm]{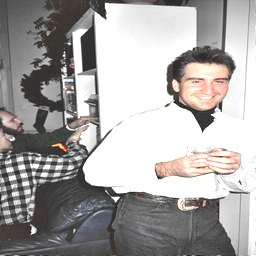}
\includegraphics[width=1.8cm]{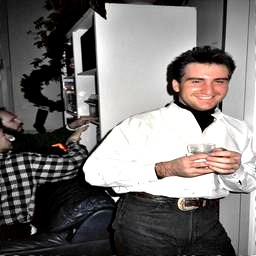}
\includegraphics[width=1.8cm]{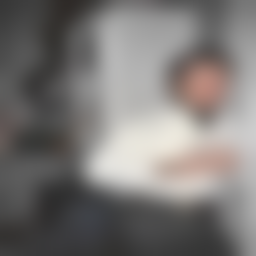}
\includegraphics[width=1.8cm]{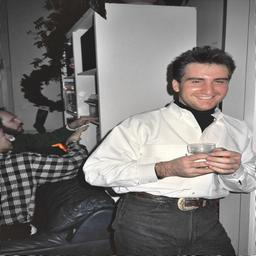}
\includegraphics[width=1.8cm]{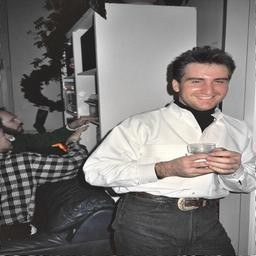}
\caption{\DistortionAttack}
\end{subfigure}
\hspace*{-1.ex}
\begin{subfigure}{0.107\textwidth}
\centering
\includegraphics[width=1.8cm]{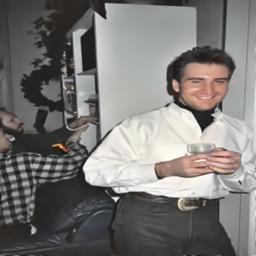}
\subcaption{\RegenVAE}
\end{subfigure}
\hspace*{.0ex}
\begin{subfigure}{0.107\textwidth}
\centering
\includegraphics[width=1.8cm]{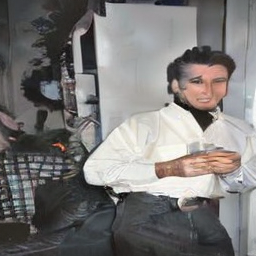}
\subcaption{\RegenDiffusion}
\end{subfigure}
\hspace*{.0ex}
\begin{subfigure}{0.109\textwidth}
\centering
\includegraphics[width=1.8cm]{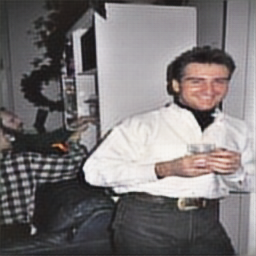}
\subcaption{\ComMark}
\end{subfigure}
\caption{Comparison of \ComMark,  \DistortionAttack (brightness, contrast, blurring, Gaussian noise, and compression in that order), \RegenVAE, and \RegenDiffusion on image integrity using two
  \textbf{\VINEB-generated}
watermarked images (WMs)  from 
OpenImage.}
\label{fig:regenvae_diffusion_commark_vinb}
\end{figure*}

\begin{figure*}[ht]
\begin{subfigure}{0.107\textwidth}
\centering
\includegraphics[width=1.8cm]{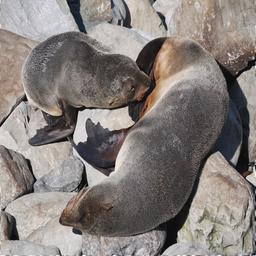}
\end{subfigure}
\hspace*{-1.ex}
\begin{subfigure}{0.55\textwidth}
\centering
\includegraphics[width=1.8cm]{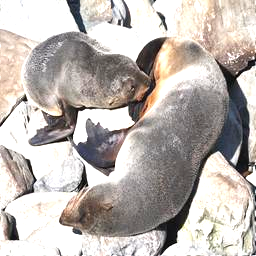}
\includegraphics[width=1.8cm]{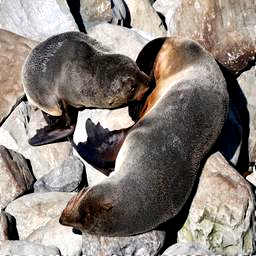}
\includegraphics[width=1.8cm]{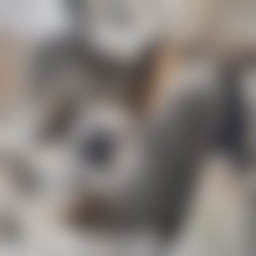}
\includegraphics[width=1.8cm]{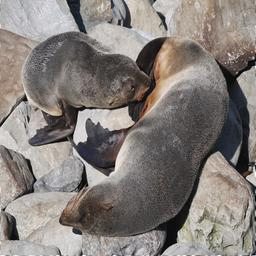}
\includegraphics[width=1.8cm]{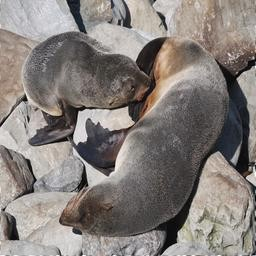}
\end{subfigure}
\hspace*{-1.ex}
\begin{subfigure}{0.107\textwidth}
\centering
\includegraphics[width=1.8cm]{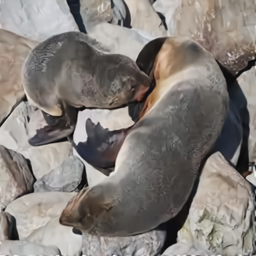}
\end{subfigure}
\hspace*{.0ex}
\begin{subfigure}{0.107\textwidth}
\centering
\includegraphics[width=1.8cm]{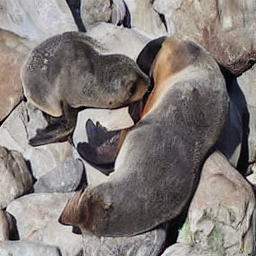}
\end{subfigure}
\hspace*{.0ex}
\begin{subfigure}{0.109\textwidth}
\centering
\includegraphics[width=1.8cm]{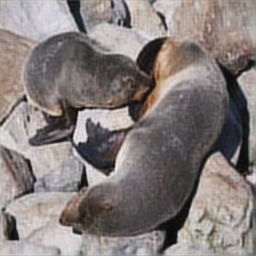}
\end{subfigure}

\begin{subfigure}{0.107\textwidth}
\centering
\includegraphics[width=1.8cm]{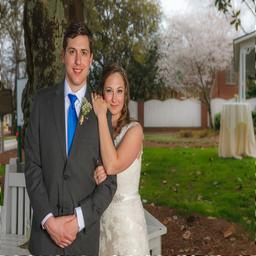}
\subcaption{WMs}
\end{subfigure}
\hspace*{-1.ex}
\begin{subfigure}{0.55\textwidth}
\centering
\includegraphics[width=1.8cm]{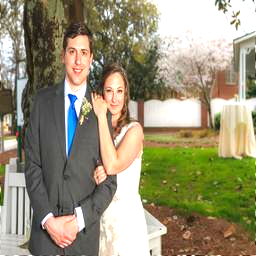}
\includegraphics[width=1.8cm]{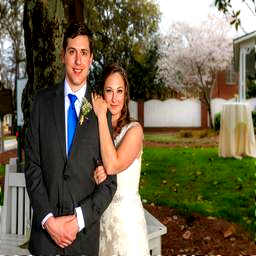}
\includegraphics[width=1.8cm]{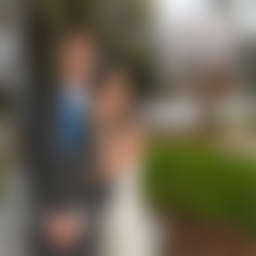}
\includegraphics[width=1.8cm]{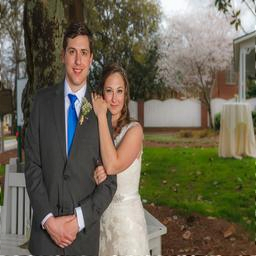}
\includegraphics[width=1.8cm]{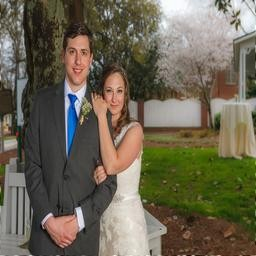}
\caption{\DistortionAttack}
\end{subfigure}
\hspace*{-1.ex}
\begin{subfigure}{0.107\textwidth}
\centering
\includegraphics[width=1.8cm]{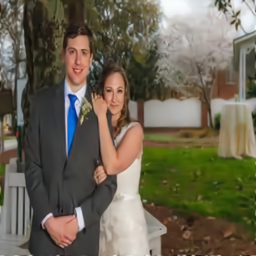}
\subcaption{\RegenVAE}
\end{subfigure}
\hspace*{.0ex}
\begin{subfigure}{0.107\textwidth}
\centering
\includegraphics[width=1.8cm]{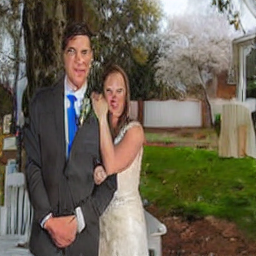}
\subcaption{\RegenDiffusion}
\end{subfigure}
\hspace*{.0ex}
\begin{subfigure}{0.109\textwidth}
\centering
\includegraphics[width=1.8cm]{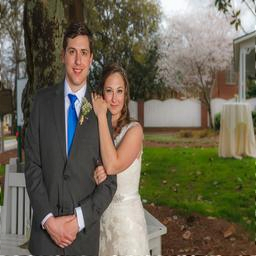}
\subcaption{\ComMark}
\end{subfigure}
\caption{Comparison of \ComMark,  \DistortionAttack (brightness, contrast, blurring, Gaussian noise, and compression in that order), \RegenVAE, and \RegenDiffusion on image integrity using two
  \textbf{\VINER-generated}
watermarked images (WMs)  from 
OpenImage.}
\label{fig:regenvae_diffusion_commark_vinr}
\end{figure*}

\begin{figure*}[ht]
\begin{subfigure}{0.107\textwidth}
\centering
\includegraphics[width=1.8cm]{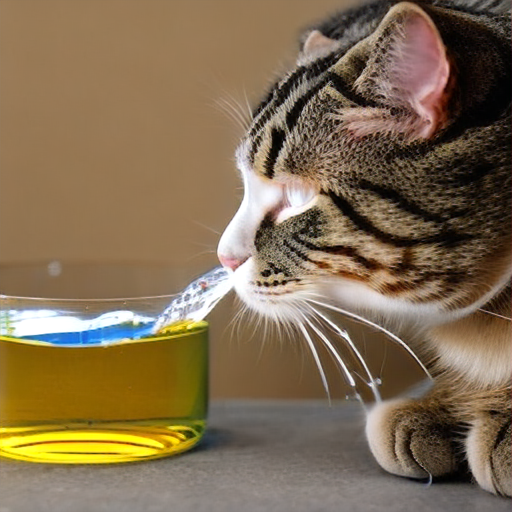}
\end{subfigure}
\hspace*{-1.ex}
\begin{subfigure}{0.55\textwidth}
\centering
\includegraphics[width=1.8cm]{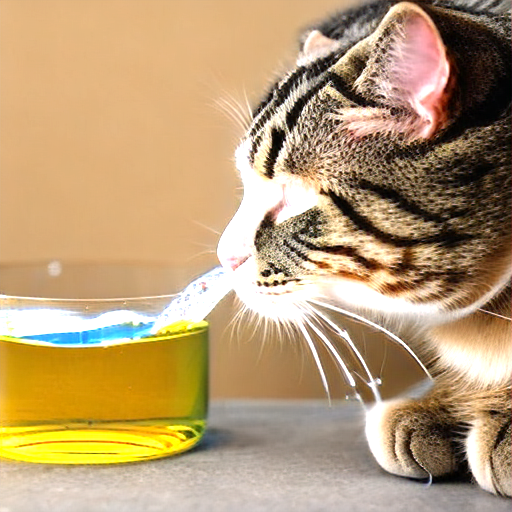}
\includegraphics[width=1.8cm]{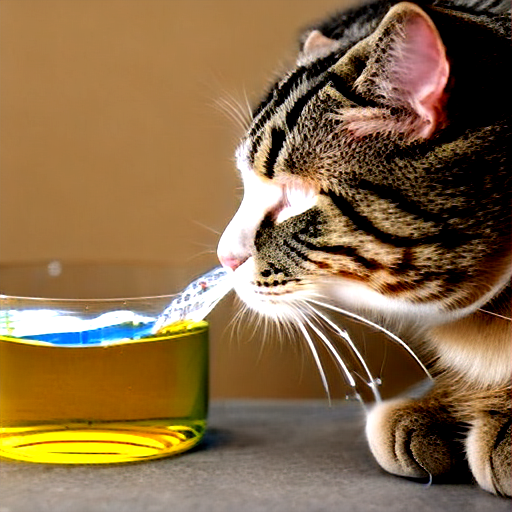}
\includegraphics[width=1.8cm]{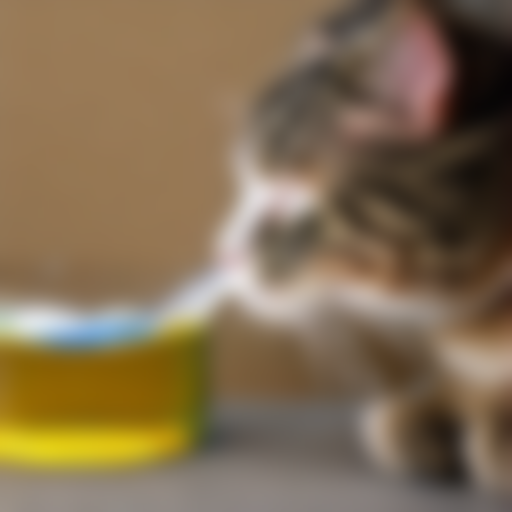}
\includegraphics[width=1.8cm]{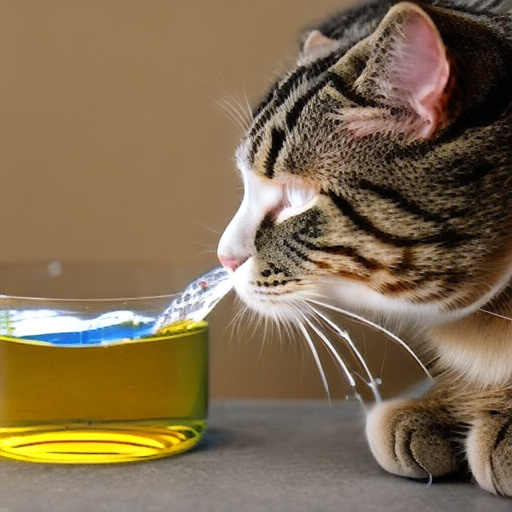}
\includegraphics[width=1.8cm]{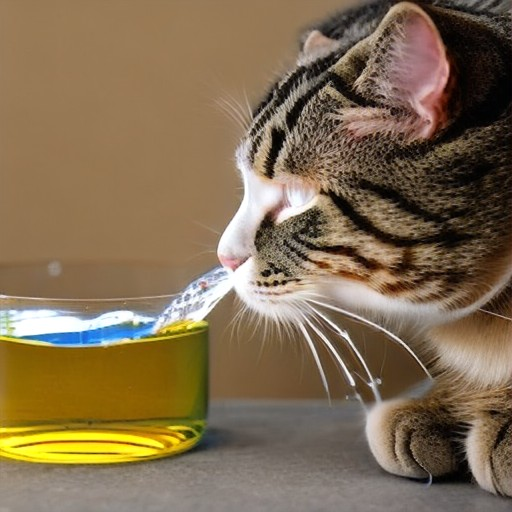}
\end{subfigure}
\hspace*{-1.ex}
\begin{subfigure}{0.107\textwidth}
\centering
\includegraphics[width=1.8cm]{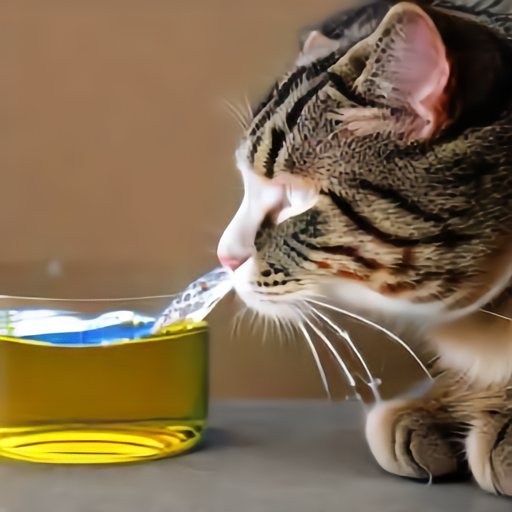}
\end{subfigure}
\hspace*{.0ex}
\begin{subfigure}{0.107\textwidth}
\centering
\includegraphics[width=1.8cm]{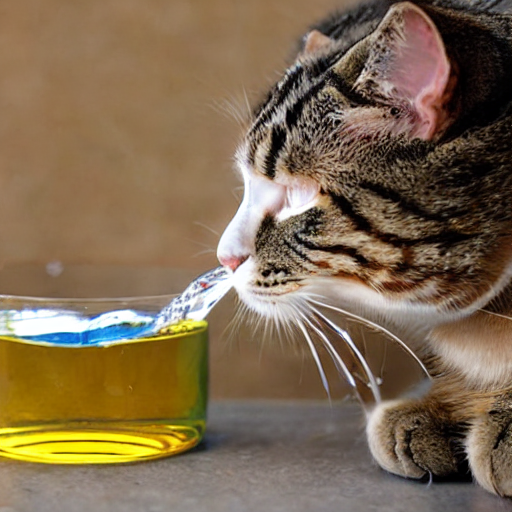}
\end{subfigure}
\hspace*{.0ex}
\begin{subfigure}{0.109\textwidth}
\centering
\includegraphics[width=1.8cm]{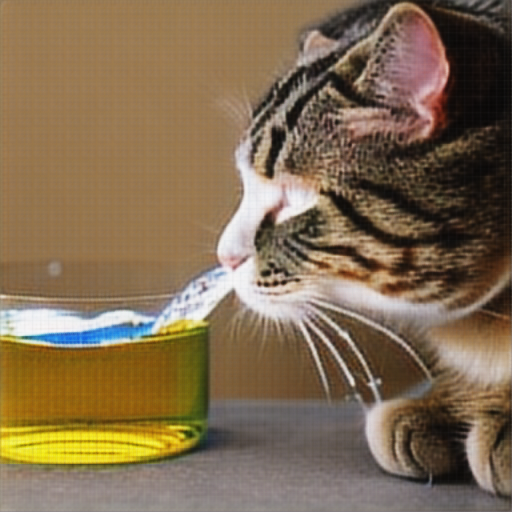}
\end{subfigure}

\begin{subfigure}{0.107\textwidth}
\centering
\includegraphics[width=1.8cm]{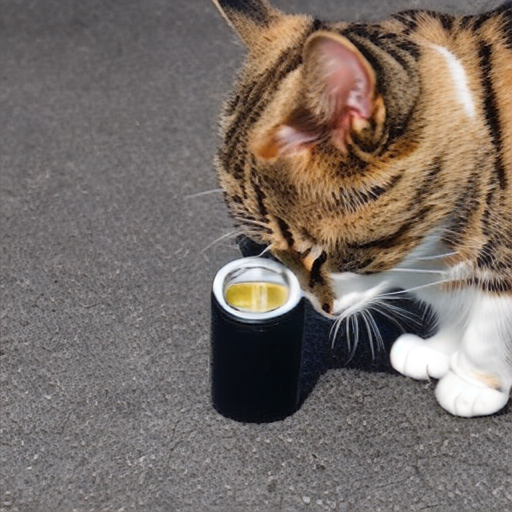}
\subcaption{WMs}
\end{subfigure}
\hspace*{-1.ex}
\begin{subfigure}{0.55\textwidth}
\centering
\includegraphics[width=1.8cm]{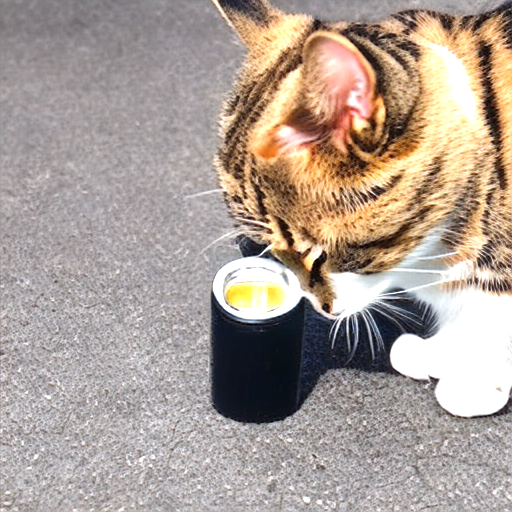}
\includegraphics[width=1.8cm]{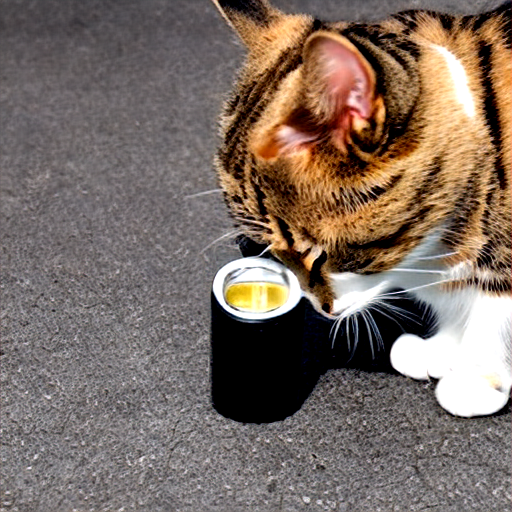}
\includegraphics[width=1.8cm]{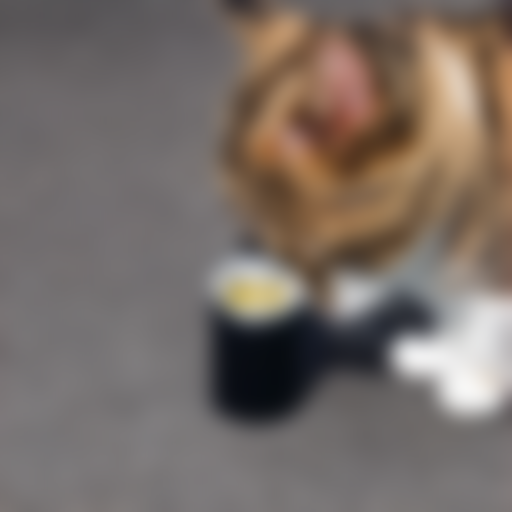}
\includegraphics[width=1.8cm]{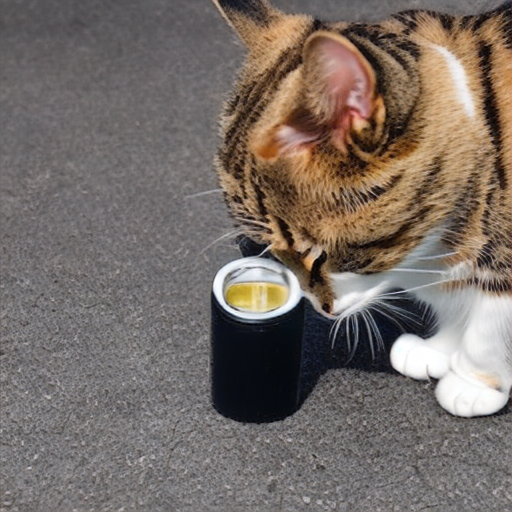}
\includegraphics[width=1.8cm]{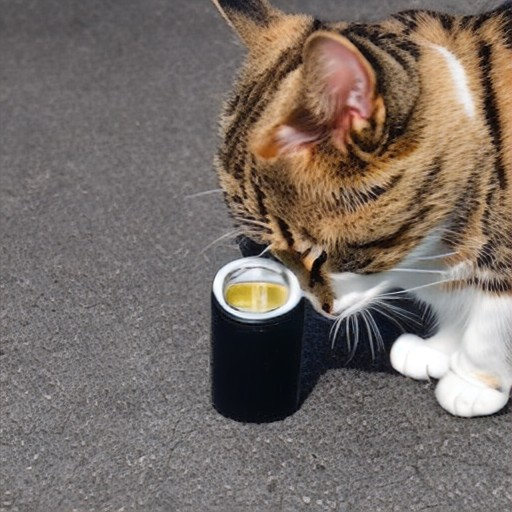}
\caption{\DistortionAttack}
\end{subfigure}
\hspace*{-1.ex}
\begin{subfigure}{0.107\textwidth}
\centering
\includegraphics[width=1.8cm]{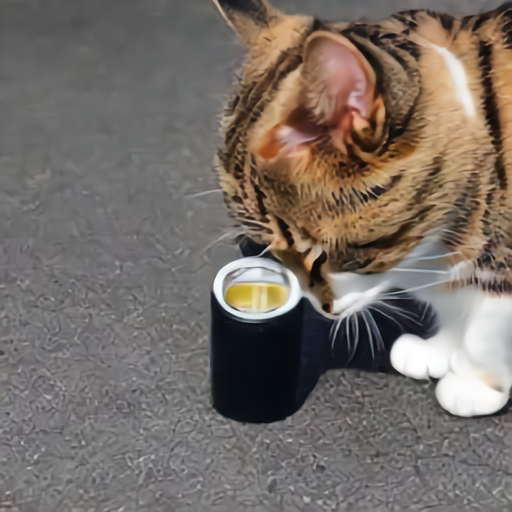}
\subcaption{\RegenVAE}
\end{subfigure}
\hspace*{.0ex}
\begin{subfigure}{0.107\textwidth}
\centering
\includegraphics[width=1.8cm]{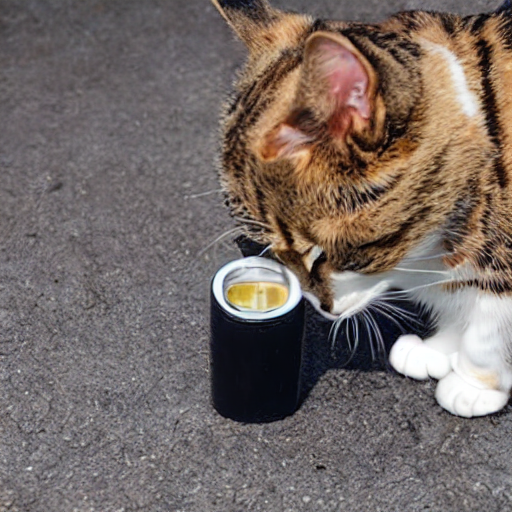}
\subcaption{\RegenDiffusion}
\end{subfigure}
\hspace*{.0ex}
\begin{subfigure}{0.109\textwidth}
\centering
\includegraphics[width=1.8cm]{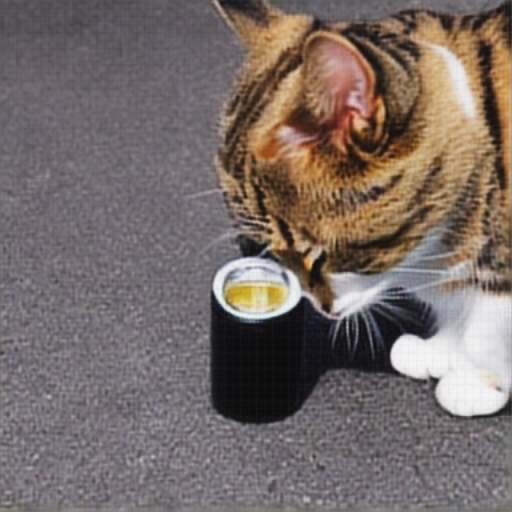}
\subcaption{\ComMark}
\end{subfigure}
\caption{Comparison of \ComMark, \DistortionAttack (brightness, contrast, blurring, Gaussian noise, and compression in that order), \RegenVAE, and \RegenDiffusion on image integrity using two
  \textbf{\StableSignature-generated}
watermarked images (WMs).}
\label{fig:regenvae_diffusion_commark_stablesignature}
\end{figure*}

\begin{figure*}[ht]
\begin{subfigure}{0.107\textwidth}
\centering
\includegraphics[width=1.8cm]{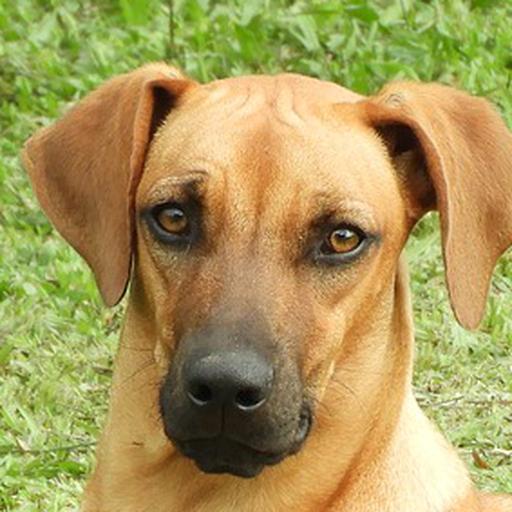}
\end{subfigure}
\hspace*{-1.ex}
\begin{subfigure}{0.55\textwidth}
\centering
\includegraphics[width=1.8cm]{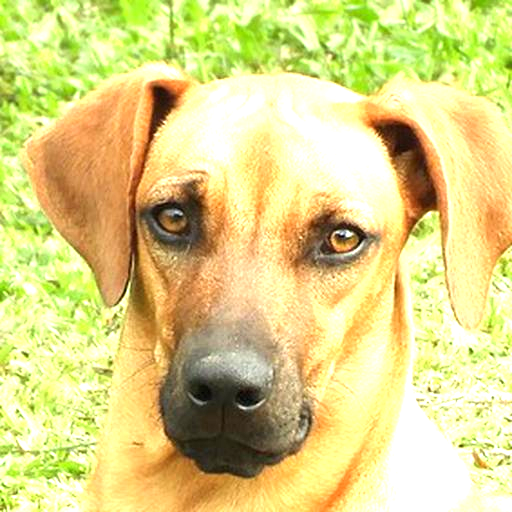}
\includegraphics[width=1.8cm]{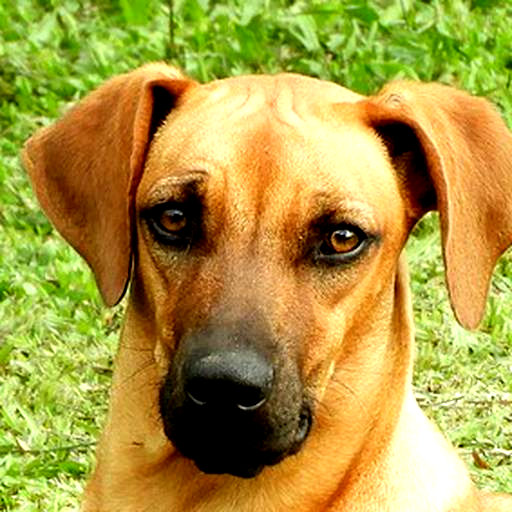}
\includegraphics[width=1.8cm]{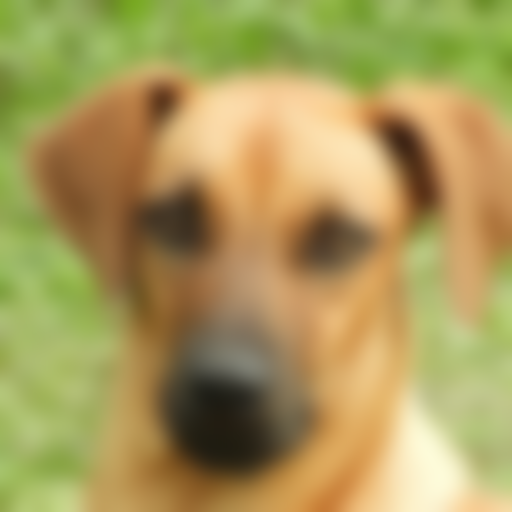}
\includegraphics[width=1.8cm]{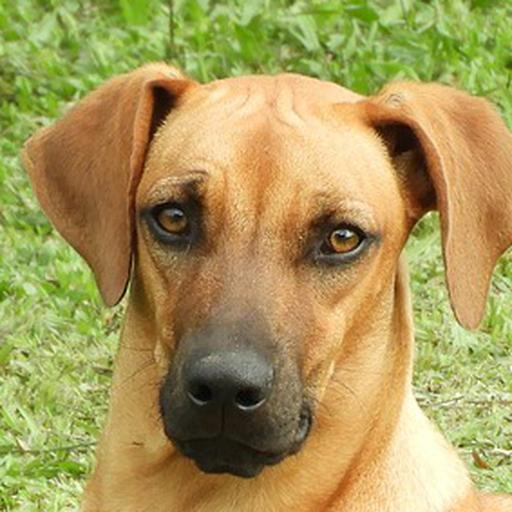}
\includegraphics[width=1.8cm]{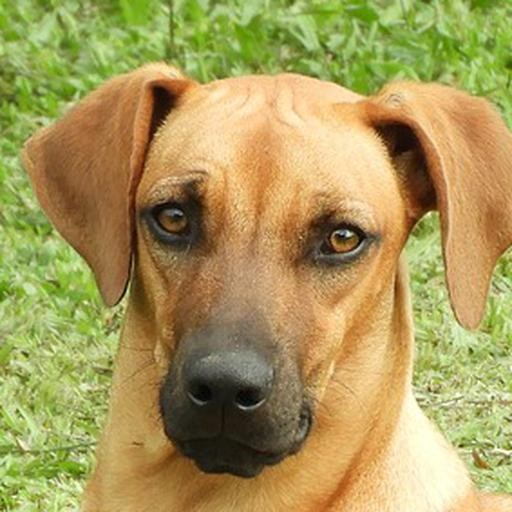}
\end{subfigure}
\hspace*{-1.ex}
\begin{subfigure}{0.107\textwidth}
\centering
\includegraphics[width=1.8cm]{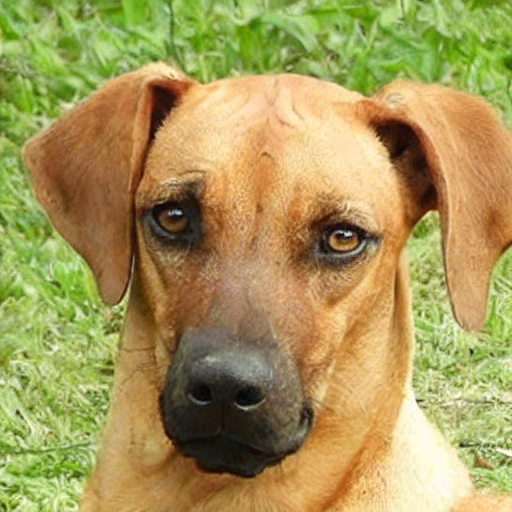}
\end{subfigure}
\hspace*{.0ex}
\begin{subfigure}{0.107\textwidth}
\centering
\includegraphics[width=1.8cm]{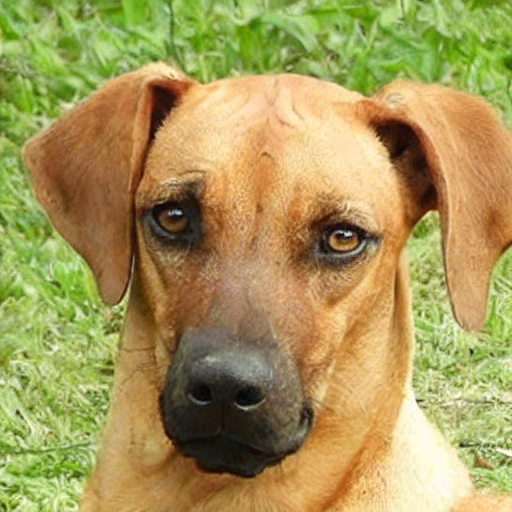}
\end{subfigure}
\hspace*{.0ex}
\begin{subfigure}{0.109\textwidth}
\centering
\includegraphics[width=1.8cm]{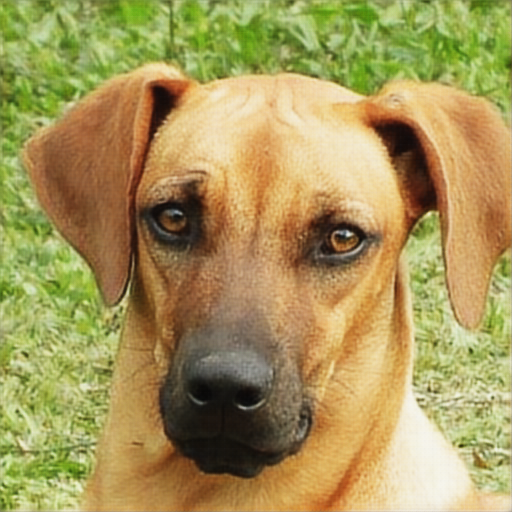}
\end{subfigure}

\begin{subfigure}{0.107\textwidth}
\centering
\includegraphics[width=1.8cm]{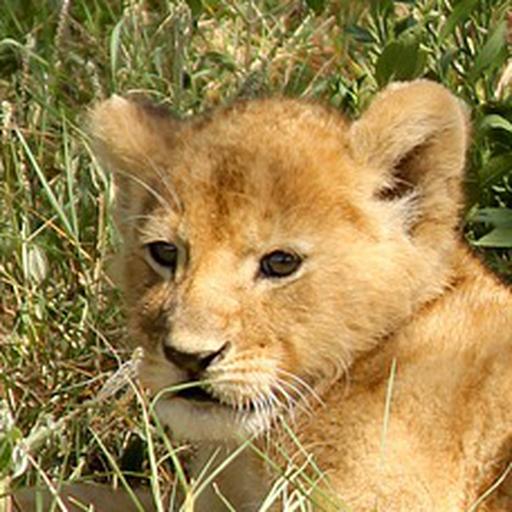}
\subcaption{WMs}
\end{subfigure}
\hspace*{-1.ex}
\begin{subfigure}{0.55\textwidth}
\centering
\includegraphics[width=1.8cm]{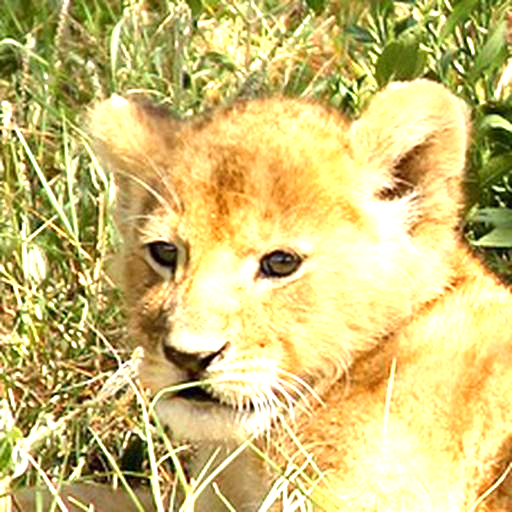}
\includegraphics[width=1.8cm]{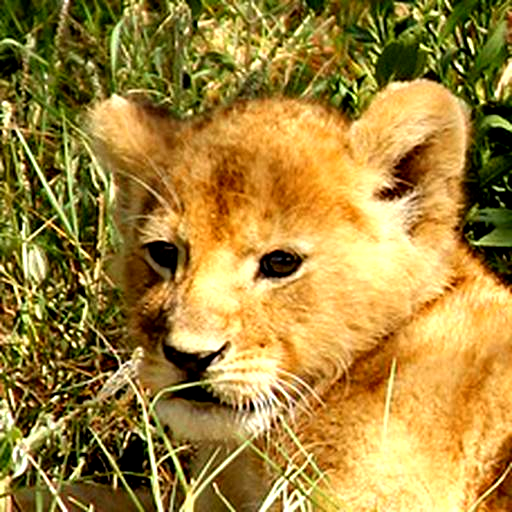}
\includegraphics[width=1.8cm]{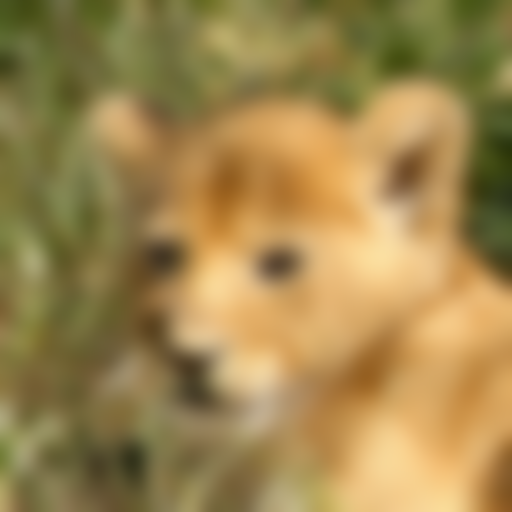}
\includegraphics[width=1.8cm]{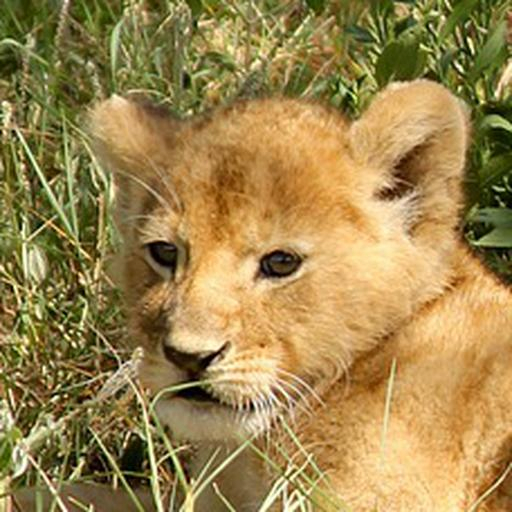}
\includegraphics[width=1.8cm]{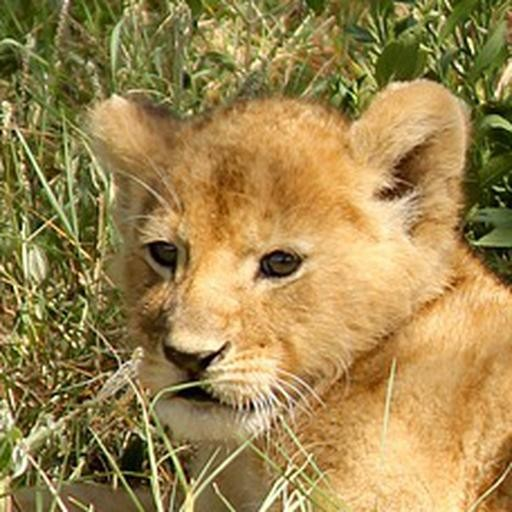}
\caption{\DistortionAttack}
\end{subfigure}
\hspace*{-1.ex}
\begin{subfigure}{0.107\textwidth}
\centering
\includegraphics[width=1.8cm]{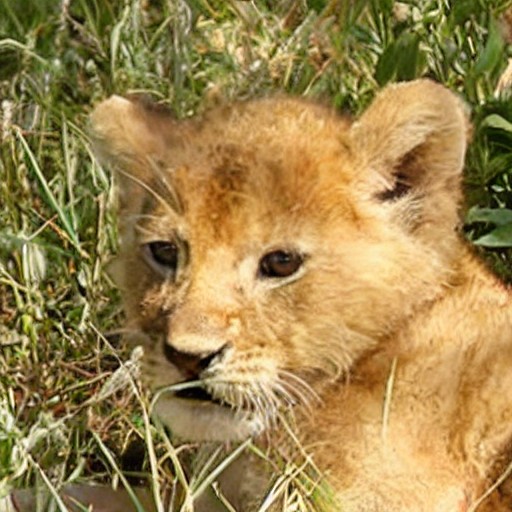}
\subcaption{\RegenVAE}
\end{subfigure}
\hspace*{.0ex}
\begin{subfigure}{0.107\textwidth}
\centering
\includegraphics[width=1.8cm]{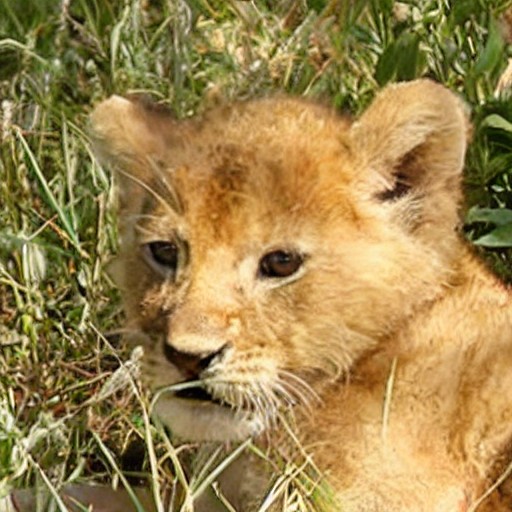}
\subcaption{\RegenDiffusion}
\end{subfigure}
\hspace*{.0ex}
\begin{subfigure}{0.109\textwidth}
\centering
\includegraphics[width=1.8cm]{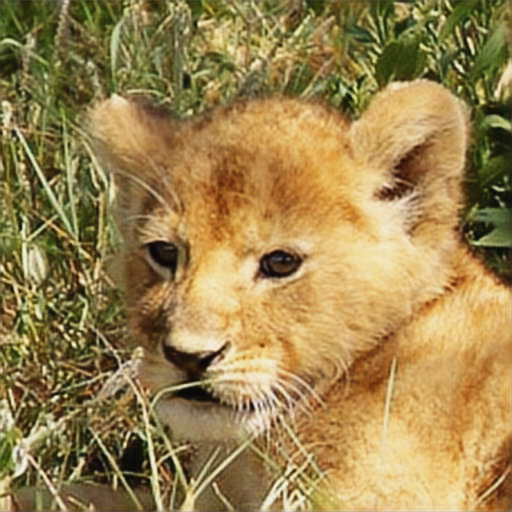}
\subcaption{\ComMark}
\end{subfigure}
\caption{Comparison of \ComMark,  \DistortionAttack (brightness, contrast, blurring, Gaussian noise, and compression in that order), \RegenVAE, and \RegenDiffusion on image integrity using two \textbf{\PTW-generated}
watermarked images (WMs).}
\label{fig:regenvae_diffusion_commark_PTW}
\end{figure*}

For each of the eight watermarking schemes—six post-processing schemes (\MBRS, \CIN, \TrustMark, \PIMoG, \VINEB, and \VINER) and two in-processing schemes (\StableSignature and \PTW)—evaluated in this paper, we present two watermarked images alongside their counterparts attacked by \DistortionAttack, \RegenVAE, \RegenDiffusion, and \ComMark in a separate figure below. This provides a visual comparison of the image integrity preserved by each attack method, complementing the quantitative metrics (\PSNR, \SSIM, \FID, and \LPIPS) reported in \Cref{table:quality_main_table,table:quality_main_table_in}.
Figures~\ref{fig:regenvae_diffusion_commark_mbrs} -- \ref{fig:regenvae_diffusion_commark_PTW} are referenced in \Cref{subsec:RQ2} as part of the analysis for RQ2.



\end{document}